\documentclass[12pt]{article}
\pdfoutput=1
\usepackage{amsmath,amssymb,amsthm,bm,graphicx,float,array,multirow,multicol,rotfloat,caption,subcaption,hyperref,cleveref,enumerate,geometry,mathdots,adjustbox,booktabs,parskip,mathtools,tikz,tikz-cd,pdflscape,csquotes,lscape,rotating,empheq,mathrsfs,longtable}
\usepackage[para]{threeparttable}
\usepackage[all]{xy}
\usepackage[normalem]{ulem}
\usepackage[aligntableaux=center]{ytableau}
\usepackage[numbers,sort&compress]{natbib}
\numberwithin{equation}{section}
\numberwithin{figure}{section}
\allowdisplaybreaks
\makeatletter

\usepackage{comment}

\theoremstyle{plain}
\newtheorem*{thm*}{Theorem}

\theoremstyle{definition}

\newtheorem*{defn*}{Definition}

\makeatother

\graphicspath{{figures/}}

\usepackage{hyzsyntax}

\tikzstyle{none} = []
\tikzstyle{A}=[fill=none, draw=black, shape=circle, radius=20]
\tikzstyle{B}=[fill=blue, draw=blue, shape=circle, size=0.3mm]
\tikzstyle{D}=[fill=green, draw=black, shape=rectangle]
\tikzstyle{C}=[fill=red, draw=red, shape=circle, textcolor=red]
\tikzstyle{X}=[fill=cyan, draw=black, shape=circle]
\tikzstyle{new style 0}=[fill=red, draw=red, shape=circle]
\tikzstyle{new style 1}=[fill=orange, draw=orange, shape=rectangle]
\tikzstyle{grayText}=[fill=none, draw=black, shape=circle, text=black]
\tikzstyle{redText}=[fill=none, draw=none, shape=circle, text=red]
\tikzstyle{blueText}=[fill=none, draw=none, shape=circle, text=blue]
\tikzstyle{magentaText}=[fill=none, draw=none, shape=circle, text=magenta]

\tikzstyle{black}=[-, line width=0.1mm]
\tikzstyle{black2}=[-, line width=0.2mm]
\tikzstyle{magenta}=[-, draw=magenta, line width=0.1mm]
\tikzstyle{magenta2}=[-, draw=magenta, line width=0.2mm]
\tikzstyle{densely dotted}=[-, densely dotted, draw=black]
\tikzstyle{red arrows}=[<->, solid, draw=red, line width=0.5mm]
\tikzstyle{green arrow}=[<->, draw=green, line width=0.5mm]
\tikzstyle{green dashed}=[<->, dashed, draw=green, line width=0.5mm]
\tikzstyle{red solid}=[-, draw=red]
\tikzstyle{red dashed}=[-, densely dotted, draw=red]
\tikzstyle{blue solid}=[-, dashed, draw=blue, line width=0.2mm]
\tikzstyle{blue solid 2}=[-, draw=blue, line width=0.2mm]
\tikzstyle{blue dashed}=[-, densely dotted, draw=blue]
\tikzstyle{magenta arrow}=[->, draw=magenta]
\tikzstyle{dotted}=[-, dotted, draw=black]
\tikzstyle{red wavy}=[-, dashed, draw=red]
\tikzstyle{blue wavy}=[-, dashed, draw=blue]
\tikzstyle{black wavy}=[-, draw=black, line width=0.5mm]
\tikzstyle{blue arrow}=[->, draw=blue]

\begin{document}

\begin{titlepage}
\vspace*{-3cm} 
\begin{flushright}
{\tt DESY-23-079}\\
{\tt UPR-1323-T}\\
\end{flushright}
\begin{center}
\vspace{1cm}
{\Large\bfseries Intermediate Defect Groups, Polarization Pairs, \vspace{2mm} \\
and Non-invertible Duality Defects \vspace{5mm} \\ 
}
\vspace{0.7cm}

{\large
Craig Lawrie$^{1}$, Xingyang Yu$^{2}$, and Hao Y. Zhang$^{3}$\\}
\vspace{.2cm}
{ $^1$ Deutsches Elektronen-Synchrotron DESY, }\par{Notkestr.~85, 22607 Hamburg, Germany}\par
\vspace{.1cm}
{ $^2$ Center for Cosmology and Particle Physics,}\par {New York University, New York, NY 10003, USA}\par
\vspace{.1cm}
{ $^3$ Department of Physics and Astronomy, University of Pennsylvania,}\par
{Philadelphia, PA 19104, USA}\par
\vspace{.1cm}

\vspace{.3cm}

\scalebox{.9}{\tt craig.lawrie1729@gmail.com, xy1038@nyu.edu, zhangphy@sas.upenn.edu}\par
\vspace{5mm}
\textbf{Abstract}
\end{center}

Within the framework of relative and absolute quantum field theories (QFTs), we present a general formalism for understanding polarizations of the intermediate defect group and constructing non-invertible duality defects in theories in $2k$ spacetime dimensions with self-dual gauge fields. We introduce the \emph{polarization pair}, which fully specifies absolute QFTs as far as their $(k-1)$-form defect groups are concerned, including their $(k-1)$-form symmetries, global structures (including discrete $\theta$-angle), and local counterterms. Using the associated symmetry TFT, we show that the polarization pair is capable of succinctly describing topological manipulations, e.g., gauging $(k-1)$-form global symmetries and stacking counterterms, of absolute QFTs. Furthermore, automorphisms of the $(k-1)$-form charge lattice naturally act on polarization pairs via their action on the defect group; they can be viewed as dualities between absolute QFTs descending from the same relative QFT. Using this formalism, we present a prescription for building non-invertible symmetries of absolute QFTs. A large class of known examples, e.g., non-invertible defects in 4D $\mathcal{N}=4$ super-Yang--Mills, can be reformulated via this prescription. As another class of examples, we identify and investigate in detail a family of non-invertible duality defects in 6D superconformal field theories (SCFTs), including from the perspective of the symmetry TFT derived from Type IIB string theory.

\vfill 
\end{titlepage}

\tableofcontents


\newpage

\section{Introduction}

Global symmetries play a fundamental role in the study of quantum field theories (QFTs). In particular, they provide an intrinsic property of the QFT which is independent of any specific description, such as via a Lagrangian, of the QFT. Importantly, symmetries have many applications in studying the low-energy dynamics of QFTs, their renormalization group (RG) flows, and other properties; symmetries provide an especially powerful technique for extracting physical features of theories without a (known) Lagrangian description.

Recently, there has been much ado about generalized global symmetries, spurred on by \cite{Gaiotto:2014kfa}, which proposes an extension of the usual notions of symmetry in such a way that the powerful consequences we are used to from ordinary symmetry are maintained.\footnote{For recent reviews on generalized global symmetries, see \cite{Cordova:2022ruw,Schafer-Nameki:2023jdn}; for a small selection of recent papers, see \cite{Gaiotto:2010be,Kapustin:2013qsa,Kapustin:2013uxa,Aharony:2013hda,DelZotto:2015isa,Sharpe:2015mja, Heckman:2017uxe, Tachikawa:2017gyf, Cordova:2018cvg,Benini:2018reh,Hsin:2018vcg,Wan:2018bns, Thorngren:2019iar,GarciaEtxebarria:2019caf,Eckhard:2019jgg,Wan:2019soo,Bergman:2020ifi,Morrison:2020ool, Albertini:2020mdx,Hsin:2020nts,Bah:2020uev,DelZotto:2020esg,Hason:2020yqf,Bhardwaj:2020phs, Apruzzi:2020zot,Cordova:2020tij,Thorngren:2020aph,DelZotto:2020sop,BenettiGenolini:2020doj,Yu:2020twi,Bhardwaj:2020ymp,DeWolfe:2020uzb,Gukov:2020btk,Iqbal:2020lrt,Hidaka:2020izy,Brennan:2020ehu,Komargodski:2020mxz,Closset:2020afy,Thorngren:2020yht,Closset:2020scj,Bhardwaj:2021pfz,Nguyen:2021naa,Heidenreich:2021xpr,Apruzzi:2021phx,Apruzzi:2021vcu,Hosseini:2021ged,Cvetic:2021sxm,Buican:2021xhs,Bhardwaj:2021zrt,Iqbal:2021rkn,Braun:2021sex,Closset:2021lhd,Thorngren:2021yso,Sharpe:2021srf,Bhardwaj:2021wif,Hidaka:2021mml,Lee:2021obi,Lee:2021crt,Hidaka:2021kkf,Koide:2021zxj,Kaidi:2021xfk,Choi:2021kmx,Gukov:2021swm,Closset:2021lwy,Yu:2021zmu,Apruzzi:2021nmk,Beratto:2021xmn,Bhardwaj:2021mzl,Debray:2021vob, Wang:2021vki,Cvetic:2021maf,Cvetic:2022uuu,DelZotto:2022fnw,Cvetic:2022imb,DelZotto:2022joo,DelZotto:2022ras,Bhardwaj:2022yxj,Hayashi:2022fkw,Kaidi:2022uux,Roumpedakis:2022aik,Choi:2022jqy,Choi:2022zal,Arias-Tamargo:2022nlf,Cordova:2022ieu, Bhardwaj:2022dyt,Benedetti:2022zbb, Bhardwaj:2022scy,Antinucci:2022eat,Carta:2022spy,Apruzzi:2022dlm, Heckman:2022suy, Choi:2022rfe,Bhardwaj:2022lsg, Lin:2022xod, Bartsch:2022mpm, Apruzzi:2022rei,GarciaEtxebarria:2022vzq, Heckman:2022muc,Cherman:2022eml, Lu:2022ver, Niro:2022ctq, Kaidi:2022cpf,Mekareeya:2022spm, vanBeest:2022fss, Antinucci:2022vyk, Giaccari:2022xgs, Bashmakov:2022uek,Cordova:2022fhg,GarciaEtxebarria:2022jky, Choi:2022fgx, Robbins:2022wlr, Bhardwaj:2022kot, Bhardwaj:2022maz, Bartsch:2022ytj, Gaiotto:2020iye,Robbins:2021ibx, Robbins:2021xce,Huang:2021zvu,Inamura:2021szw, Cherman:2021nox,Sharpe:2022ene,Bashmakov:2022jtl, Inamura:2022lun, Damia:2022bcd, Lin:2022dhv,Burbano:2021loy,Damia:2022rxw,Apte:2022xtu,Kaidi:2023maf,Etheredge:2023ler,Carta:2023bqn,Koide:2023rqd,Bhardwaj:2023wzd,Cao:2023doz,Inamura:2023qzl,Dierigl:2023jdp,Chen:2023qnv,Cvetic:2023plv,Bhardwaj:2023ayw,Bashmakov:2023kwo,Copetti:2023mcq,Garcia-Valdecasas:2023mis}. We refer to the cited reviews for comprehensive references to this vast literature.} In this generalized perspective, a QFT with an ordinary global symmetry with symmetry group $G$ is viewed as possessing codimension-one topological operators $U_g(M_{d-1})$, for each $g \in G$ and where $M_{d-1}$ is any $(d-1)$-dimensional submanifold of spacetime.\footnote{In this paper, the terms ``operator'' and ``defect'' are used interchangeably since the distinction between the spatial and temporal directions is not essential for our analysis.} These operators are such that when $U_g(M_{d-1})$ crosses a charged local operator/excitation, it exerts an action via the group element $g$. The group-structure comes from the fusion rule of the topological defects: $U_g(M_{d-1}) \times U_{g'}(M_{d-1}) = U_{g'g}(M_{d-1})$. This formulation suggests several generalizations. First of all, we can consider higher-codimension topological operators that act on charged extended operators/excitations; these operators generate higher-form symmetries. In another direction, we can relax the condition that the topological defects obey a group-like fusion rule; in this case the associated symmetries are called non-invertible symmetries.\footnote{Non-invertible symmetries are familiar in two dimensions, such as the Verlinde lines in rational conformal field theory \cite{Verlinde:1988sn}.}

The consequences of the existence of generalized global symmetries apply to QFTs that realize such symmetries. Therefore, to take advantage of this new symmetry toolkit, it is necessary to first construct QFTs with such symmetries. In this paper, we focus on theories with non-invertible symmetries associated to codimension-one topological operators. In \cite{Choi:2021kmx,Kaidi:2022uux} the authors consider a generalization of the Kramers--Wannier duality of the Ising model to construct 4D gauge theories with non-invertible symmetries. The key feature is the presence of self-dual one-form gauge fields: such theories possess an associated discrete one-form global symmetry. If the gauge theory also realizes a zero-form global symmetry that acts (in a certain precise sense that we elucidate later) on the one-form gauge sector, then a non-invertible symmetry can be observed by considering the gauging of the one-form symmetry. In this paper, we consider the explicit generalization of this ``duality defect'' construction to QFTs in $2k$ dimensions which involve self-dual $(k-1)$-form gauge fields.

Theories of self-dual higher-form gauge fields are often plagued by subtleties. In particular, they do not a priori admit a scalar-valued partition function on an arbitrary closed spacetime manifold; instead they have a partition \emph{vector}; this property is a feature of a so-called \emph{relative} quantum field theory \cite{Witten:1996hc,Aharony:1998qu,Witten:1998wy,Moore:2004jv,Belov:2006jd,Freed:2006yc,Witten:2007ct,Henningson:2010rc,Freed:2012bs,Tachikawa:2013hya}.\footnote{A relative quantum field theory in $2k$ dimensions should be viewed as living on the boundary of a $2k+1$-dimensional topological quantum field theory, whose Hilbert space contains states which are the partition vector of the relative $2k$-dimensional theory.} A $2k$-dimensional theory involving self-dual $(k-1)$-form gauge fields involves both ``light'' $(k-1)$-dimensional excitations and ``heavy'' $(k-1)$-dimensional defects. The charges of the light objects take values in the lattice $\Lambda$, whereas the charges of the heavy objects are valued in the dual lattice: $\Lambda^*$, which is a $\bbQ$-refinement of $\Lambda$.\footnote{See, e.g., \cite{Deser:1997se} for a discussion of the charges of extended objects in self-dual Abelian $p$-form gauge theory.} The intermediate defect group \cite{tHooft:1977nqb,tHooft:1979rtg}
\begin{equation}
    \bbD = \Lambda^*/\Lambda \,,
\end{equation}
measures the failure of the Dirac pairing between $(k-1)$-dimensional objects to be integral.\footnote{The definition of the intermediate defect group as $\bbD = \Lambda^*/\Lambda$ shows that the pairing on the lattice $\Lambda^*$, the Dirac pairing $\langle - \,, - \rangle$, descends to a pairing on $\bbD$. The intermediate defect ``group'' is the Abelian group together with this inherited pairing, though we often leave the pairing implicit and just write a group $\bbD$.} A consistent quantum field theory with a well-defined partition function on an arbitrary closed spacetime manifold requires a choice of sublattice of charges for the $(k-1)$-dimensional objects that are all mutually integer under the Dirac pairing: this corresponds to a choice of Lagrangian subgroup $L$ of $\bbD$, often referred to as a choice of polarization.\footnote{We remark that one could in principle go beyond the intermediate defect group to consider the defect groups of all form degrees, which we leave for future analysis. Heuristically, one then needs to pick a ``Lagrangian subcategory'' of the ``defect category'' (see, e.g., \cite{Naidu_2008}). We leave this for future analysis.} In summary: to have a well-defined QFT for a $2k$-dimensional theory involving self-dual $(k-1)$-form gauge fields, it is necessary to also prescribe a Lagrangian subgroup $L$ of the intermediate defect group $\bbD$.\footnote{In $4s$-dimensions, the choice of polarization is often left implicit, since the intermediate defect group decomposes into a sum of ``electric'' and ``magnetic'' Lagrangian subgroups: $\bbD = \bbD_e \oplus \bbD_m$. Thus, a polarization can always be chosen, however the different choices of polarization lead to differing spectra of extended operators in the absolute theory. See, for example, \cite{Aharony:2013hda}.}  Given a choice of $L$, the resulting absolute theory has a $(k-1)$-form global symmetry group $L^\vee = \bbD/L$.

We are interested in cases where the $(k-1)$-form symmetry $L^\vee$ is gaugable. This occurs when $L$ is a splittable polarization of $\bbD$, i.e., where 
\begin{equation}
    \bbD = L \oplus \ovL \,,
\end{equation}
for $\ovL$ another Lagrangian subgroup of $\bbD$.\footnote{When the polarization is splittable the $(k-1)$-form symmetry $L^\vee = \bbD/L$ can always be uplifted to $\ovL \subset \bbD$, in such a way that the $(k-1)$-form symmetry is non-anomalous.} As groups $\ovL \cong L^\vee$, however, recalling that the defect group also includes the information of the pairing, there can be distinct uplifts of $L^\vee$ to $\bbD$, and this is captured by the different choices of $\ovL$. As we can see, to explicitly specify a splittable polarization of $\bbD$, it is insufficient to specify $L$, but we must also provide the choice of uplift of $L^\vee$ to $\ovL$. As explained in \cite{Gukov:2020btk} (see also Appendix \ref{app:absorb counterterm}), different choices of $\ovL$ correspond to different SPT phase descriptions of the vacuum; as such they describe a physically observable property of the QFT.

To accommodate the information of a possible non-trivial SPT phase, we extend the notion of a choice of polarization to a choice of polarization pair.\footnote{In 6D, the polarization pair overlaps with the ``refined polarization'' of \cite{Gukov:2020btk}.} We emphasize that our discussion only applies to the case when the $(k-1)$-form global symmetry is gaugable. The discussion splits into the two scenarios of $2k = 4s$ and $2k = 4s+2$, with the key distinction that the $\mathbb{Z}$-valued Dirac pairing on the dynamical charge lattice (and thus the inherited $\bbQ/\mathbb{Z}$-valued pairing on the defect group) is \textit{antisymmetric} in $4s$ spacetime dimensions, while it is \textit{symmetric} in $4s+2$ spacetime dimensions. In $4s$ dimensions it is always possible to find such a splittable polarization due to the antisymmetry of the pairing, whereas this is not necessarily the case in $4s+2$ dimensions.

With these preparations in place, we now introduce our general, explicit description of the polarization for such theories. For clarity in this introduction, we focus on the case where 
\begin{equation}\label{eqn:introsimp}
    \bbD = \bbZ_N \oplus \bbZ_N \,,
\end{equation}
with $N$ a prime number.\footnote{In general, without making the assumption in equation \eqref{eqn:introsimp}, one needs to specify a polarization pair by providing a complete set of generators of $L$ and $\ovL$, respectively. See Section \ref{sec:polarizationPair} for more details.} In particular, $L \cong \ovL \cong \bbZ_N$. In this case, a \textbf{polarization pair} is an \textit{ordered pair} given by a generator of $L$, $\ell$, and a generator of $\ovL$, $\ovl$:
\begin{equation}\label{eqn:introPP}
(\ell, \ovl), \quad \text{ s.t. } \quad \langle \ell, \ovl \rangle = \frac{1}{N} \,,
\end{equation}
where the last identity leads to nice behavior of the polarization pair under gauging, as described in more detail in Section \ref{sec:polarizationPair}. Physically, $\ell, \ovl$ labels the pair of $\bbZ_N$ fields for which we give Neumann and Dirichlet boundary conditions, respectively. Once the polarization pair is defined as in equation \eqref{eqn:introPP}, the ``topological manipulations'' of gauging $L^\vee$, stacking an SPT phase, or acting by an automorphism of the charge lattice admit a succinct algebraic characterization. These actions are determined in Section \ref{sec:polarizationPair}, and we summarize them briefly here.

\paragraph{Gauging $(k-1)$-form symmetries} In $4s+2$ dimensions, gauging $L^\vee$ simply corresponds to flipping the polarization pair:
\begin{equation}
    (\ell, \ovl)\ \rightarrow (\ovl, \ell) \,,
\end{equation}
whereas in $4s$ dimensions, we not only need to flip the elements of the polarization pair but we also need to ensure that the antisymmetric Dirac pairing is unchanged. Thus, either $\ell$ or $\ovl$ needs to be multiplied by $-1$. Without loss of generality, this ``symplectic'' flip can be realized as:
\begin{equation}
    (\ell, \ovl)\ \rightarrow (-\ovl, \ell) \,.
\end{equation}

\paragraph{Stacking an SPT phase/adding a local counterterm} Stacking a local counterterm amounts to fixing $\ell$ and only changing $\ovl$:
\begin{equation}
    (\ell, \ovl) \rightarrow (\ell, \ovl') = (\ell, \ovl + r\ell), \quad r = 0, 1,\cdots, N-1 \,,
\end{equation}
which ensures that the Dirac pairing holds fixed. We remark that in $4s+2$ dimensions the possibility of stacking a counterterm is much more restricted than in $4s$ dimensions.

\paragraph{Automorphism action on the charge lattice} An automorphism of the charge lattice $\hat{a} \in \text{Aut}(\Lambda^*)$ can always be descended to an automorphism of the defect group: $a \in \text{Aut}(\bbD)$. In fortuitous circumstances, the automorphism of the charge lattice $\hat{a}$ can be uplifted to a full duality of the local operator content of the theory; i.e., to being a duality of the relative theory. The chosen set of generators $\ell_i, \ovl_i$ of the Lagrangian subgroups $L, \ovL \subset \bbD$ in the polarization pair \textit{inherits} the action of $\hat{a}$ via $a \in \text{Aut}(\bbD)$. This way, we naturally know how the charge lattice automorphisms act on the set of polarizations.  

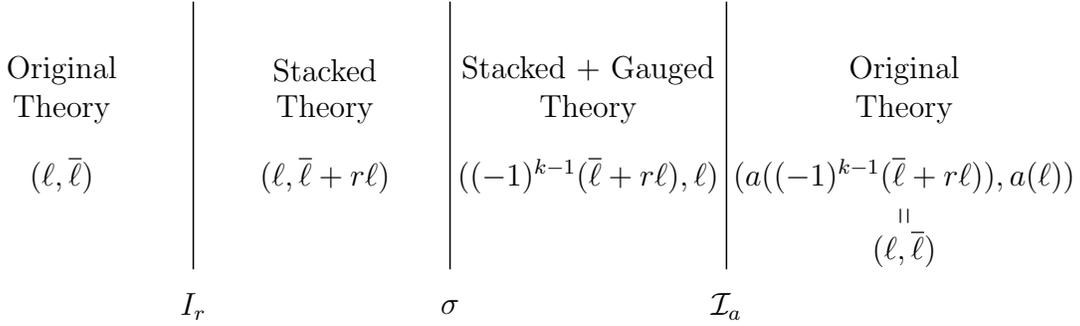
\begin{figure}
\centering
\begin{tikzpicture}[scale= .35]
	\node [style=none] (1) at (-5, -6){};
	\node [style=none] (2) at (-5, 5) {};
 	\node [style=none] (3) at (4.75, -6) {};
    \node [style=none] (4) at (4.75, 5) {};
    \node [style=none] (n1) at (15.25, -6) {};
    \node [style=none] (n2) at (15.25, 5) {};
    \node [style=none] (5) at (-10, 2) {Original};
    \node [style=none] (5a) at (-10, 0.5) {Theory};
    \node [style=none] (5b) at (-10, -2) {$(\ell, \overline{\ell})$};
    \node [style=none] (6) at (0, 2) {Stacked};
    \node [style=none] (6a) at (0, 0.5) {Theory};
    \node [style=none] (6b) at (0, -2){$(\ell, \overline{\ell} + r\ell)$};
    \node [style=none] (7) at (22, 2) {Original};
    \node [style=none] (7) at (22, 0.5) {Theory};
    \node [style=none] (7b) at (22, -2) {$(a((-1)^{k-1}(\overline{\ell} + r\ell)), a(\ell))$};
    \node [style=none] (7b) at (22, -3.5) {$\shortparallel$};
    \node [style=none] (7b) at (22, -4.8) {$(\ell, \overline{\ell})$};

    \node [style=none] (l) at (10, 2) {Stacked + Gauged};
    \node [style=none] (la) at (10, 0.5) {Theory};
    \node [style=none] (7b) at (10, -2) {$((-1)^{k-1}(\overline{\ell} + r\ell), \ell)$};

	\node [style=none] (8) at (-5, -7){$I_r$};
    \node [style=none] (9) at (4.75, -7){$\sigma$};
    \node [style=none] (9) at (15.25, -7){$\mathcal{I}_a$};
    
    \draw [style=black2] (1) to (2);
    \draw [style=black2] (3) to (4);
    \draw [style=black2] (n1) to (n2);

\end{tikzpicture}
\caption{An illustration of the general construction of non-invertible duality defects by combining the half-space gauging interface $\sigma$, the stacking-counterterm interface $I_r$, and the interface $\calI_a$ implementing a charge lattice automorphism. If the absolute theories on the left and the right are the same, for which it is required that the polarization pairs are identical, then the combined defect $\mathcal{N} =  I_r \cdot \sigma \cdot \mathcal{I}_a$ is a non-invertible topological duality defect.}
\label{fig:CombiningInterfaces}
\end{figure}

We now describe how non-invertible symmetry defects can be built out of these fundamental operations. Consider a $2k$-dimensional absolute theory of self-dual $(k-1)$-form gauge fields that is specified via a polarization pair $(\ell, \overline{\ell})$. We may introduce a domain wall in the theory which implements the stacking of an SPT phase, $I_r$ where $r \in \mathbb{Z}_N$. Next, we can introduce a domain wall $\sigma$ which implements the gauging of the $(k-1)$-form discrete global symmetry: this domain wall functions as an interface, with Dirichlet boundary conditions, between two distinct QFTs. Finally, introduce another domain wall, labeled by $\mathcal{I}_a$ for $a \in \mathrm{Aut}(\mathbb{D})$, in the stacked + gauged theory which implements the action of an automorphism of the charge-lattice (that uplifts to a duality of the QFT). Let us assume that QFT on the other side of this automorphism domain wall is the same absolute QFT as the original theory we started with. We depict this sequence of interfaces in Figure \ref{fig:CombiningInterfaces}. In fact, one can include an arbitrary number of gauging/stacking/automorphism interfaces in the middle, as long as the QFTs at the far left and the far right are the same absolute QFT. Collapsing all of the interfaces on top of each other leads to a codimension-one topological defect in the absolute QFT; when this combined defect involves an odd number of gauging interfaces, this so-called duality defect will have a non-invertible fusion rule.  

\paragraph{Example: reformulating non-invertible duality defects in 4D $\calN = 4$ SYM} To show the unifying power of our description, we start by reformulating the well-studied case of 4D $\calN = 4$ SYM. We explain how the duality defects in these theories based on $SL(2, \bbZ)$ transformations at special values of $\tau_{YM}$, half-space gauging of center 1-form symmetry, and stacking SPT phases in \cite{Choi:2021kmx,Kaidi:2022uux} are captured by our formulation in full generality. 

The 4D defect group is well-known to split into an electric component and a magnetic component,
\begin{equation}
    \bbD = \bbD^{(e)} \times \bbD^{(m)} \,,
\end{equation}
with the standard Dirac pairing:
\begin{equation}
    \bbD \times \bbD \rightarrow \bbZ, \quad  (e, m), (e', m') \rightarrow em' - e'm \in \bbZ \,.
\end{equation}
Therefore, at least the electric polarization $L = \bbD^{(m)}$ and the magnetic polarization $L = \bbD^{(e)}$ always exist. Various $SL(2, \bbZ)$ duality transformations can be viewed as  automorphisms of the charge lattice at special $\tau_{YM}$ values, e.g.,
\begin{equation}
  \begin{aligned}
    S \in \text{Aut}(\bbD) &\text{ for } \tau_{YM} = i \,, \\ 
    \ \ ST \in \text{Aut}(\bbD) &\text{ for }\tau_{YM} = e^{\frac{2\pi i}{3}} \,, \\
    \ \ -1 \in \text{Aut}(\bbD) &\text{ for any } \tau_{YM} \,.
  \end{aligned}
\end{equation}
For example, if we take $\ksu (N)$ SYM at $\tau_{YM} = i$, then an S-transformation acts on the polarization pair as:
\begin{equation}
    (\ell, \ovl) \rightarrow (S \ell, S \ovl) \,,
\end{equation}
where $\ell = (\ell_e, \ell_m)$ and $S\ell = S(l_e, l_m) = (l_m, -l_e)$, and the same for $S\ovl$.

Now, we can completely reformulate the duality defects found in \cite{Kaidi:2022uux} algebraically in terms of polarization pairs. Continuing with the same example, take $\ksu(N)$ at $\tau_{YM} = i$ with the choice of polarization pair 
\begin{equation}
    (\ell, \ovl) = ((0,1) (-1, 0)) \,.
\end{equation}
This is usually known as the theory with ``electric polarization''. Combine the half-space gauging interface $\sigma$ and the $S$-duality interface (viewed as an order-4 lattice automorphism) to build the duality defect 
\begin{equation}
\calN(M_3) = \sigma \cdot \calI_S \,,
\end{equation}
with the action on polarization pairs as:
\begin{equation}
 ((0,1) (-1, 0)) \overset{\sigma}{\rightarrow}  ((0, 1) (1, 0)) \overset{\calI_S}{\rightarrow} ((0, 1), (-1,0) \,,
\end{equation}
and thus we find the original absolute theory. We then recover the duality defect $\mathcal{N}(M_3)$ in $\calN = 4$ SYM satisfying the non-invertible fusion rule as in \cite{Choi:2021kmx,Kaidi:2022uux}. We remark that, in four dimensions, the extra data of specifying generators beyond merely picking $(L, \ovL)$ is important (specifically when $\bbD \neq \bbZ_2^n$ for some $n$). 

Physically, this extra data amounts to picking a specific generator of the background field of the 1-form global symmetry for each of its cyclic generators, modulo suitable equivalence relations. For example, as we will see explicitly in the $\mathfrak{su}(3)$ theory, this extra data can be seen as charge conjugation, which indeed will be modded out if one merely specifies $(L, \ovL)$ as the polarization pair.  

\paragraph{Example: novel non-invertible duality defects in 6D SCFTs}

Of course, the abstract construction of non-invertible symmetries in self-dual higher gauge theories in arbitrary dimensions is complemented by a connection to explicit, known and well-studied, quantum field theories where such non-invertible duality defects are present. We revisit known examples in 2D and 4D, and we make a detailed analysis of six-dimensional field theories with self-dual two-form gauge fields as a demonstration of our construction. Such 6D QFTs are particularly challenging as the self-dual two-forms preclude a simple Lagrangian description, while at the same time, with enhanced superconformal symmetry, they are ubiquitous in string theory, and they function as parent theories shedding light on lower-dimensional QFT.\footnote{For a recent summary of the power of 6D superconformal theories for the understanding of lower-dimensional quantum field theory, see \cite{Argyres:2022mnu}.} It is thus especially pressing to take advantage of all possible tools and techniques to understand the physical properties of this important class of theories. In particular, we give a general construction of non-invertible duality defects in 6D $(2,0)$ SCFTs, which we illustrate exhaustively for the $D_4$ and $A_4 + A_4$ theories. To support our analysis of 6D SCFTs, we also study the symmetry TFT as derived from Type IIB string theory directly.

The rest of this paper is organized as follows. In Section \ref{sec:polarizationPair}, we review the relevant concepts, including relative and absolute QFTs, the intermediate defect group, polarizations, and Heisenberg flux non-commutativity. We then establish the general formulation of the polarization of a quantum field theory in terms of a \textit{polarization pair}. With the language of a polarization pair for a $2k$ dimensional QFT, we explain how one can conveniently implement the operations of gauging $(k-1)$-form symmetry, stacking SPT phases and implementing automorphisms of the charge lattice. In Section \ref{sec:nonInvGeneral}, we give the general construction of duality defects in $2k$ dimensional QFTs using polarization pairs and revisit the 2D Ising model and 4D $\calN = 4$ SYM as warm-up examples. In Section \ref{sec:6DSCFTs}, we take the hitherto unexplored example of six dimensions and construct non-invertible duality defects in a number of 6D SCFTs, for which we give concrete examples by 6D $(2,0)$ and $(1,0)$ SCFTs. Then in Section \ref{sec:symTFTs}, we determine the symmetry TFTs for 6D $(2,0)$ SCFTs via Type IIB string compactification, from which we do a detailed study of the interplay between our non-invertible duality symmetry and other symmetries in 6D SCFTs. Finally, in Section \ref{sec:conclusion}, we discuss a variety of consequences and future directions. In Appendix \ref{apdx:relative}, we give more technical reviews and treatments regarding intermediate defect groups and polarizations. In Appendix \ref{app:absorb counterterm}, we demonstrate that polarization pairs intrinsically incorporate the information of SPT phases/local counterterms.

\section{Polarization Pairs on the Intermediate Defect Group} \label{sec:polarizationPair}

In this section, we present the general formulation of a polarization pair for an even-dimensional QFT. The motivation is to review and refine the concept of a polarization, which specifies an absolute QFT from a given relative QFT \cite{Freed:2012bs}. Such a refinement allows us also to incorporate the data of an SPT phase or discrete counterterm, which removes the ambiguity involved in the outcome of gauging the $(k-1)$-form global symmetry in $2k$-dimensional spacetime.

A polarization pair succinctly captures the algebraic data of relevant coefficients in the partition function, which can be further viewed as coming from the topological boundary condition of the associated symmetry TFT. It has the advantage that all polarizations (e.g., what is known in 4D as electric, magnetic, and dyonic polarizations) with all possible choices of SPT phase are treated on an equal footing. In addition, manipulations like gauging, stacking counterterms, and duality transformations are straightforward to handle in this formalism. 

We begin by reviewing the intermediate defect group $\bbD$ for heavy defects with spacetime dimension $k-1$, on which a Dirac pairing is defined. Then, motivated by the objective of constructing duality defects, we specialize to the QFTs whose $(k-1)$-form symmetry is non-anomalous and thus gaugable. This holds as long as the defect group splits into a pair of Lagrangian subgroups, $\bbD = L \oplus \ovL$. In such a case, the corresponding $(2k+1)$-dimensional symmetry TFT admits a finite gauge theory (sometimes also referred to as a BF theory) description. At a later stage, we will also specialize to the simplest case of a defect group for the sake of clarity:
\begin{equation}
    \bbD = \bbZ_N \oplus \bbZ_N, \quad N \text{ prime } \,,
\end{equation}
so that we are not distracted by the additional subtlety of gauging or stacking counterterms with respect to a proper subgroup of the $(k-1)$-form symmetry group. 

In this section, we emphasize the key concepts, however some of the more formal or technical explanations are gathered in Appendix \ref{apdx:relative}.

\subsection{Intermediate Defect Groups and the Dirac Pairing}

In $2k$ dimensions, one can define the Dirac pairing between a pair of dynamical objects both of spacetime dimension $k-1$ (see Appendix \ref{subapdx:Dirac} for more details). Crucially, the parity of such pairing depends on the parity of $k$:
\begin{itemize}
    \item For $k = 2s$ even, we have a $4s$-dimensional spacetime equipped with an \textit{anti-symmetric} Dirac pairing. In these dimensions (starting from 4D), any charged element has a trivial self-pairing; therefore a non-degenerate Dirac pairing forces the simultaneous existence of electric and magnetic objects.
    \item For $k = 2s+1$ odd, we have a $4s+2$-dimensional spacetime which is equipped with a \textit{symmetric} Dirac pairing. \textit{A priori}, there is no electric-magnetic splitting of $(k-1)$-dimensional states, and thus these objects can be referred to as intrinsically dyonic.
\end{itemize}

In $2k$ dimensions, in addition to dynamical objects, the Dirac pairing also involves $(k-1)$-dimensional heavy defects in spacetime. The dynamical objects carry charges valued in a charge lattice $\Lambda$, but the heavy defects carry charges valued in a refined charge lattice $\Lambda^* \supset \Lambda$, whose equivalence classes under screening of dynamical objects (via the usual 't Hooft screening argument \cite{THOOFT19781}) are labeled by the defect group: 
\begin{equation}
    \mathbb{D} = \Lambda^*/\Lambda \,.
\end{equation}
By definition, the free lattice $\Lambda^* = \mathbb{Z}^r$ with rank $r$ comes with a bilinear pairing given by the integer-coefficient matrix $K$
\begin{equation}
    b_{\Lambda^*}(\lambda_1, \lambda_2) = \lambda_1^T K^{-1} \lambda_2, \in \mathbb{Q}, \quad \lambda_1, \lambda_2 \in \mathbb{Z} \,, \label{eqn:LatticeBilinear}
\end{equation}
with the associated quadratic form $q_{\Lambda^*}(\lambda) = \frac{1}{2} \lambda^T K^{-1} \lambda$ (sometimes called a \textit{quadratic refinement} of the bilinear form.\footnote{We have $b_{\Lambda^*}(\lambda_1, \lambda_2) = q_{\Lambda^*}(\lambda_1 + \lambda_2) - q_{\Lambda^*}(\lambda_1) - q_{\Lambda^*}(\lambda_2)$ and thus $ b_{\Lambda^*}(\lambda, \lambda) = 2q_{\Lambda^*}(\lambda)$} The dual lattice $\Lambda^*$ is a torsional refinement of the original lattice $\Lambda$ given by $\Lambda^* = K^{-1} \Lambda$, on which one has an inherited pairing $q_{\Lambda^*}(\hat{\lambda}) \in \mathbb{Q}$ for $\hat{\lambda} \in \Lambda^*$. By definition, for elements in the original lattice $\Lambda$, the pairings $q_{\Lambda}$ and $q_{\Lambda^*}$ are the same. Therefore, such a $\mathbb{Q}$-valued quadratic form $q_{\Lambda^*}(\hat{\lambda})$ on $\Lambda^*$ consistently descends onto a $\mathbb{Q}/\mathbb{Z}$-valued pairing on the defect group $\bbD$:
\begin{equation}\label{eq:quadratic pairing on defect group}
    q(\mu) \in \bbQ/\bbZ, \ \mu \in \mathbb{D} \,.
\end{equation}
Similarly, the $\bbQ$-valued bilinear pairing $b_{\Lambda^*}(\lambda_1, \lambda_2)$ on $\Lambda^*$ descends to a $\bbQ/\bbZ$ bilinear form on $\bbD$:
\begin{equation}
    b(\mu_1, \mu_2) \in \bbQ/\bbZ, \ \ \mu_1, \mu_2 \in \bbD \,.\label{eq:bilinear pairing on defect group}
\end{equation}
In this paper, whenever we talk about a defect group $\mathbb{D}$, we always implicitly assume that it comes with a $\mathbb{Q}/\mathbb{Z}$-valued bilinear form $b(\mu_1, \mu_2)$ and an associated quadratic form $q(\mu)$, both of which are inherited from an underlying dual charge lattice $\Lambda^*$. 

We can now define an isotropic subgroup of the defect group
\begin{equation}
    G \subset \bbD \,,
\end{equation}
as one for which any pair of elements trivializes the bilinear form:
\begin{equation}
    b(\mu_1, \mu_2) = 0 \in \bbQ/\bbZ \,, \qquad \forall \mu_1, \mu_2 \in G \subset \bbD \,.
\end{equation}
A key type of isotropic subgroup of $\bbD$ for our purpose are maximal isotropic subgroups, also known as \textit{Lagrangian subgroups}, which we usually denote by 
\begin{equation}
L \subset \bbD, \quad \text{ satisfying } |L|^2 = |\bbD|\,, \,\,\text{ and } \,\,b(\mu_1, \mu_2) = 0 \in \bbQ/\bbZ, \ \ \forall \mu_1, \mu_2 \in L \subset \bbD \,.
\end{equation}
See \cite{Heckman:2015bfa,Gukov:2020btk} and references therein for more details of Lagrangian subgroups, which for convenience we also review in Appendix \ref{subapdx:defectGroup}).

\subsection{Polarizations and a Phase Ambiguity}\label{subsec:polar to polar pair}

After reviewing the intermediate defect group and the Dirac pairing, we explain the notion of polarization of a QFT. We start with reviewing the conventional approach of specifying a polarization by specifying a Lagrangian subgroups of the intermediate defect group to obtain absolute QFTs from relative QFTs. Then, under the motivation of gauging the $(k-1)$-form symmetry, we will guide ourselves to the point where the notion of the polarization faces its limitation. At that point, we are be forced to work with the more refined notion of polarization pair, which will be the topic of the next subsection.

According to the standard story, the starting point of picking a polarization is to notice that the heavy defects valued in $\Lambda^*$ do not have integer-valued Dirac pairing among themselves! This signals an inconsistency of the theory upon quantization, resulting in a \emph{relative} QFT \cite{Freed:2012bs}. A relative QFT cannot be consistently defined on its own but has to be defined as the boundary theory of a $2k+1$ dimensional \textit{bulk} QFT. This relative $2k$ dimensional theory has an \textit{anomalous} $(k-1)$-form global symmetry - this anomaly is precisely captured by braiding relations of $k$-dimensional operators in the $2k+1$ dimensional bulk QFT \cite{GarciaEtxebarria:2019caf,Gukov:2020btk}. Such fractional pairing is also visible at the level of flux operators as the Heisenberg flux \textit{non-commutativity} $\hat{X}\hat{Y} = \hat{Y}\hat{X} e^{2\pi i \phi}$ with $\phi \neq 0$ precisely the non-integer Dirac pairing on the defect group $\mathbb{D}$. 

Therefore, in order to restore a consistent quantum theory, we need to restrict ourselves to a \textit{maximum commuting subset} of flux observables out of the full set. At the level of defect groups, the commutation condition dictates that the Dirac pairing between the states on such a maximal commuting subset has to be integer-valued. So we need to pick a maximal subset of heavy defects that have integer-valued mutual Dirac pairing. Mathematically, this choice of maximal commuting observables amounts to choosing a \textit{maximal isotropic} sublattice $\Lambda^L$ of the defect charge lattice $\Lambda^*$ such that 
\begin{equation}
    \Lambda \subset \Lambda^L \subset \Lambda^* \,,    
\end{equation}
where the isotropic condition is defined as $b_{\Lambda^*}(\lambda^L_1, \lambda^L_2) \in \bbZ$ for any $\lambda^L_1, \lambda^L_2 \in \Lambda^L$.

After quotienting every individual entry by $\Lambda$, this choice of $\Lambda^L$ is equivalent to choosing a Lagrangian subgroup (i.e., maximal isotropic subgroup) of the defect group: 
\begin{equation}
    L = \Lambda^L/\Lambda \subset \Lambda^*/\Lambda = \mathbb{D} \,.
\end{equation}
In this way, one specifies an \emph{absolute} QFT in the conventional sense, which has a $(L^\vee = \mathbb{D}/L)$-valued $(k-1)$-form global symmetry. This process is also referred to as picking a \emph{polarization}. \footnote{$L^{\vee} \cong L$ since $L^\vee$ is the Pontryagin dual of $L$ as induced by the Dirac pairing on $\bbD$.} 

The existence of such a polarization $L \subset \bbD$ depends on the parity of $k$: 
\begin{itemize}
    \item For $k = 2s$ even, the defect group automatically comes with two copies $\mathbb{D} = \mathbb{D}_e \oplus \mathbb{D}_m$ where $|\mathbb{D}_e| = |\mathbb{D}_m|$ due to the antisymmetric Dirac pairing.\footnote{Any finite dimensional symplectic vector space $H$ can be written as a decomposition $H=V\oplus V^\star$.} Thus, a choice of $L$ is always possible. For example, one can always pick the electric polarization $L = \mathbb{D}_m, L^\vee = \mathbb{D}_e$, or the magnetic polarization $L = \mathbb{D}_e, L^\vee = \mathbb{D}_m$, so that the remaining global symmetry is the electric and magnetic global symmetry, respectively.
    A simple example is 4D $\mathfrak{su}(N)$ $\mathcal{N}=4$ SYM with defect group $\mathbb{D}=\mathbb{Z}_N^{(e)} \oplus \mathbb{Z}_N^{(m)}$. Choosing $\mathbb{Z}_N^{(e)}$ (resp. $\mathbb{Z}_N^{(m)}$) as the Lagrangian subgroup results in the magnetic (resp. electric) 1-form symmetry, whose genuine line defects are Wilson (resp. 't Hooft) lines. Indeed, when the pairing is anti-symmetric, any cyclic subgroup is automatically isotropic, as can be seen by examining the generator. We hope to comment further on the specifics of the $s > 1$ case in the future. 
    \item For $k=2s+1$ odd, \textit{a priori}, the defect group $\mathbb{D}$ does not always decompose, so there is no guarantee that a given relative theory allows a polarization to an absolute theory by picking a Lagrangian subgroup $L \subset \mathbb{D}$. In particular, such a choice is impossible when the order $|\mathbb{D}|$ is not a complete square. When $|\mathbb{D}|$ is a complete square, there are cases where such an isotropic subgroup $L \subset \mathbb{D}$ exists, and the corresponding polarization can be picked. As we will see in Section \ref{sec:6DSCFTs}, there are many examples in 6D that polarizations can be picked and one arrives at 6D absolute theories.
\end{itemize}

\paragraph{Partition functions for absolute theories and a phase ambiguity}

Given an relative theory with defect group $\bbD = \bbZ_N \oplus \bbZ_N$, specifying a polarization $L \subset \bbD$ would specify an absolute theory whose partition function coupled to a background gauge field $B$ is denoted as:
\be
    \text{``}Z[B], B \in L^\vee \text{''} \,,
\ee
where $B$ takes values in the global $(k-1)$-form symmetry $L^\vee \cong \bbZ_N$, and the splittings of $\bbD = L \oplus \ovL$ ensures that the $L^\vee$ symmetry is gaugable. 

However, we will soon see that the meaning of ``$Z[B]$'' in the above equation suffers from a phase ambiguity. Correspondingly, in an absolute QFT with only $L \subset \bbD$ specified, the resulting theory after gauging the $(k-1)$-form symmetry is ambiguous, as was also pointed out in \cite{Gukov:2020btk}. As will soon see, resolving such an ambiguity precisely requires us to refine the notion of a polarization into a polarization pair. Such a resolution of this ambiguity is closely related to that of specifying an SPT phase, namely a quadratic counterterm, which we will also take into account.

\subsection{Polarization Pair via Heisenberg Group } \label{subsec:symTFTpolpair}

In this part, we present a refined analysis of partition functions and topological boundary conditions of symmetry TFT via the representation space of the Heisenberg group. The six-dimensional case of such a refinement has already been treated in great detail in \cite{Gukov:2020btk}.

\paragraph{Basis of Partition Vector Space via Heisenberg Group} To go one step further and define the topological boundary states associated with polarization pairs (from which we can build the well-defined partition functions), we need to go deeper into the partition vector space of relative QFTs and the quantization of the corresponding symmetry TFTs. 

A relative QFT no longer has a scalar-valued partition function, but it has a partition \textit{vector} instead. It turns out that the partition vector space of a relative QFT can be regarded as the Hilbert space from the quantization of the corresponding symmetry TFT \cite{Witten:1998wy}, which is succinctly captured by the Heisenberg group $\underline{H}^{k}(M_{2k}, \mathbb{D})$ with coefficients in the defect group $\mathbb{D}$ \cite{Witten:1998wy,Tachikawa:2013hya,DelZotto:2015isa,Gukov:2020btk}, following \cite{mumford2006tata}. The Heisenberg group $\underline{H}^{k}(M_{2k}, \mathbb{D})$ is defined by the following extension:
\begin{equation}
    1 \rightarrow U(1) \rightarrow \underline{H}^{k}(M_{2k}, \mathbb{D}) \rightarrow H^{k}(M_{2k}, \mathbb{D}) \rightarrow 1 \,.
\end{equation}
The partition \textit{vector} space of a relative QFT carries a representation of $\underline{H}^{k}(M_{2k}, \mathbb{D})$. Associated with any cohomology class $A \in H^{k}(M_{2k}, \mathbb{D})$, we denote the corresponding flux operators as $\Phi(A) \in \underline{H}^{k}(M_{2k}, \mathbb{D})$. \footnote{As pointed out in \cite{Tachikawa:2013hya,GarciaEtxebarria:2019caf,Gukov:2020btk}, $\Phi(A)$ is not a global section of $H^{k}(M_{2k}, \mathbb{D})$ in $\underline{H}^{k}(M_{2k}, \mathbb{D})$ due to flux non-commutativity, but we still treat $\Phi(A)$ as a physically-defined flux operator.}

A polarization of the defect group $\mathbb{D}$ always induces a polarization of the group of fluxes $H^k(M_{2k}, \mathbb{D})$ via the following long exact sequence \cite{Witten:1998wy,Gukov:2020btk,GarciaEtxebarria:2019caf}:
\begin{equation}
    \dots \rightarrow H^{k-1}(M_{2k}, L^\vee) \overset{\beta_{k-1}}{\rightarrow} H^k(M_{2k}, L) \overset{\imath_*}{\rightarrow} H^k(M_{2k}, \mathbb{D}) \rightarrow H^k(M_{2k}, L^\vee) \overset{\beta_{k}}{\rightarrow} \dots \,. \label{eqn:LES}
\end{equation}
Then one sees that $H^k(M_{2k}, L) \subset H^k(M_{2k}, \mathbb{D})$ \footnote{Technically $\imath_*(H^k(M_{2k}, L)) \subset H^k(M_{2k}, \mathbb{D})$.} is an isotropic subgroup of $H^k(M_{2k}, \mathbb{D})$. So one can write down the group of physical fluxes $H^k(M_{2k}, \mathbb{D})/H^k(M_{2k}, L)$. When $\bbD = L \oplus \ovL$ splits, $H^k(M_{2k}, \mathbb{D})/H^k(M_{2k}, L)$ can similarly be uplifted to $H^k(M_{2k}, \ovL)$. \footnote{In particular, when $\mathbb{D} = L \oplus \overline{L}$, the Bockstein homomorphisms $\beta_l$ in the above sequence are always trivial.} The dimension of the vector space is then given by $|H^k(M_{2k}, L)| = |H^k(M_{2k}, \ovL)^\vee| = \sqrt{|H^k(M_{2k}, \mathbb{D})|}$.\footnote{That the order of $|H^k(M_{2k}, \mathbb{D})|$ is a complete square follows as can be found in \texttt{https://mathoverflow.net/questions/58825/non-degenerate-\\alternating-bilinear-form-on-a-finite-abelian-group/58828\#58828}, which makes use of the non-degenerate antisymmetric pairing on $\bbD$ for $2k = 4s$, and of the non-degenerate antisymmetric pairing on $H^k(M_{2k}, \mathbb{D})$ for $2k = 4s+2$.}

Given such a decomposition $\mathbb{D} = L \oplus \overline{L}$ which induces the split of $H^k(M_{2k}, \bbD) = H^k(M_{2k}, L) \oplus H^k(M_{2k}, \ovL)$, the standard procedure of constructing a basis of topological boundary states for the symmetry TFT involves the flux operators of $\Phi(A), \Phi(B)$ (where $A \in H^k(M_{2k}, L), B \in H^k(M_{2k}, \ovL)$), which satisfies the well-known flux non-commutativity relation:
\be
    \Phi(A) \Phi(B)\ = \exp\left(\frac{2\pi i}{N} \int_{M_{2k}} A \cup B \right) \ \Phi(B) \Phi(A) \,. 
\ee

The existence of the partition space (mathematically the representation space of $H^k(M_{2k}, \bbZ)$) requires the splitness of $\bbD$ into $L \oplus \ovL$. In this situation, we introduce the formalism of a \textbf{polarization pair}. Given a polarization associated with the Lagrangian subgroup $L$, we specify a certain $\ovL$ to form a ordered pair of Lagrangian subgroups $(\ell, \ovl)$. A polarization pair is defined by a chosen set of generators of $L$ and $\ovL$ as 
\begin{equation}
    (\{\ell_1,\ell_2,\cdots \ell_n\},\  \{\ovl_1,\ovl_2,\cdots \ovl_n\}),  \quad \text{ s.t. } \ell_i, \ovl_i \in \bbD \ \ \langle \ell_i, \ovl_i \rangle = \frac{1}{N_i} \, \forall i \,,
\end{equation}
where $\ell_i$ (resp. $\ovl_i$) denote the $i$-th generator of $L$ (resp. $\ovL$) with order $N_i$. To illustrate the idea more explicitly, we focus on the case of $\bbD = \bbZ_N \oplus \bbZ_N$, i.e., $L \cong \ovL \cong \bbZ_N$. The polarization pair thus takes a simple form:
\be
    (\ell, \ovl) \quad \text{ s.t. }\ell, \ovl \in \bbD, \ \ L = \langle \ell \rangle, \ovL = \langle \ovl \rangle, \ \ \langle \ell, \ovl \rangle = \frac{1}{N} \,.
\ee

More concretely, the basis of partition functions (namely a basis of the partition vector space) can be written as
\be
Z_{(\ell, \ovl)}[B] \,,
\ee
where we emphasize that the reason why we specify a pair of generators $(\ell, \ovl)$ on top of $L, \ovL$ is to keep track of the relabeling of the background fields, such that for any non-trivial element $t \neq 0 \in \bbZ_N$ (with its inverse element $t^\vee$ such that $t t^\vee \equiv 1 \mod \bbZ_N$)
\be
    Z_{(\ell, \ovl)}[{\color{red}t}B] \equiv Z_{\color{red} (t^\vee \ell, t \ovl)}[B] \,.
\ee

By definition, such a basis has the nice property that the $\Phi(A)$ behave as clock operators under this basis, and the $\Phi(B)$ behave as shift operators \cite{Witten:1998wy,Tachikawa:2013hya,GarciaEtxebarria:2019caf,Gukov:2020btk}:
\be
    \Phi(A) Z_{(\ell, \ovl)}[B]  = \exp\left(\frac{2\pi i}{N} \int_{M_{2k}} A \cup B \right) Z_{(\ell, \ovl)}[B], \quad \Phi(B_2) Z_{(\ell, \ovl)}[B_1] =  Z_{(\ell, \ovl)}[B_1 + B_2] \,.
\ee
We remark that only the state with zero boundary field value $Z_{(\ell, \ovl)}[B = 0]$ has the property that its phase factor does not depend on the choice of $\ovl$ \cite{Gukov:2020btk}, and that a $\Phi(A)$ acts trivially onto it. Therefore, operationally, we begin with this basis vector $Z_{(\ell, \ovl)}[B = 0]$, and then acts on it with shift operator $\Phi(B)$ to generate the remaining basis vectors with $B \neq 0$, all of whose phases will depend on the choice of $\ovl$ and thus on $\ovL$. 

In summary, the partition functions $Z_{(\ell, \ovl)}[B]$ specified by the polarization pair $(\ell, \ovl)$ and the background field value $B$ is capable of completely capturing the information of topological boundary conditions. 

We remark that in the special case of 6D, our phrasing largely overlaps with that of \cite{Gukov:2020btk}. The full data for them involves specifying $H^3(M_6, L)$ and $H^3(M_6, \ovL)$ for given $M_6$, whereas our polarization pair $(\ell, \ovl)$ is the minimal version of their data that one already need to specify before specifying a spacetime manifold $M_{2k}$. In addition, specifying $(\ell, \ovl)$ on top of $(L, \ovL)$ further specifies the data of background field relabeling (which incorporates charge conjugation).\footnote{The definition of \cite{Gukov:2020btk} in 6D also involves a quadratic refinement of the bilinear pairing on $H^3(M_6, \bbD)$, which we address in Appendix \ref{app:absorb counterterm}.}

In particular, we stress that the conventional notion of ``quadratic counterterm'' in the literature can be absorbed when we change the basis of topological boundaries of a symmetry TFT (and thus the basis of partition functions) by changing the second element $\ovl$ in the polarization pair.

\paragraph{Symmetry TFT and Topological Boundary Conditions} 

Equivalently, the information associated with a polarization pair can be recast in terms of a
symmetry TFT \cite{Freed:2012bs,Freed:2022qnc} (see also \cite{Apruzzi:2021nmk,Kaidi:2022cpf} for discussion in various specific contexts). Symmetry TFTs are topological theories in $(d+1)$ spacetime dimensions living on $N_{d+1}$, with boundary $\p N_{d+1} = M_d$, which capture the information of global symmetries in $d$-dimensional QFTs living on $M_d$. 

In the case we are currently interested in, namely $2k$-dimensional QFTs with intermediate defect groups $\mathbb{D}$, the symmetry TFT elegantly encodes the structure of polarizations via its topological boundary conditions. They have been intensively studied in various systems. Here we present a treatment of polarizations via symmetry TFTs, which applies to all QFTs with a split intermediate defect group $\bbD = L \oplus \ovL$ where $L, \ovL \subset \bbD$ are both Lagrangian subgroups of $\bbD$. To illustrate our main idea, our presentation focuses on the case with $L \cong \ovL \cong \bbZ_N$, and we leave the treatment of generic non-cyclic $L$ to Appendix \ref{subapdx:symTFTnoncyclic}. 

Denoting the defect group decomposition as $\bbD = \bbZ_N^{(x)} \oplus \bbZ_N^{(y)}$, we have $k$-form gauge field $b_i \in H^k(N_{2k+1}, \bbZ_N^{(i)}) , \ \ i \in \{x, y\}$ where $H^k(N_{2k+1}, \bbD) = H^k(N_{2k+1}, \bbZ_N^{(x)}) \oplus H^k(N_{2k+1}, \bbZ_N^{(y)})$. Namely, the splitness of the defect group induces a splitness of the group of fluxes. Keeping the relative theory in mind, every flux will eventually be rewritten as valued in the defect group $\bbD$. \footnote{We avoid labeling the defect group as $\bbD = \bbZ_N^{(e)} \oplus \bbZ_N^{(m)}$, since such a notation is only natural in $4s$ dimensions while misleading in $4s+2$ dimensions, as discussed in Section \ref{subsec:polar to polar pair}.} 

The symmetry TFT then has an action of the generic form 
\begin{equation}
    S_{\text{symTFT}}[b_x,b_y]=\int_{M_{2k+1}} \frac{1}{2} b_i Q_{ij}\delta b_j \quad (i, j \in \{x, y\}) \,,
\end{equation}
where the $2\times 2$ matrix $Q_{ij}$ is the coefficient of the bilinear pairing $b(\mu_1, \mu_2)$ on the defect group $\mathbb{D}$, defined in equation (\ref{eq:bilinear pairing on defect group}).\footnote{If we were to not restrict ourselves to fields associated with the intermediate defect group, then there are more terms in the bulk symmetry TFT beyond the quadratic Chern--Simons terms. For all such fields involved, one needs to specify a basis with respect to the involved commutation relations for the quantization of such extra terms in the symmetry TFT. In our paper, we restrict to the situation where the background fields beyond the intermediate defect group are not turned on, so that such complications do not arise. We thank J.~J.~Heckman for comments on this point.} In this case, each element 
$\ell, \ovl \in \bbD$ of our polarization pair $(\ell, \ovl)$ comes with two components each corresponding to a $\bbZ_N$ subgroup of $\bbD$:
\be
    \ell = (\ell_x, \ell_y), \ \ \ovl = (\ovl_x, \ovl_y) \,.
\ee

For the $(2k+1)$-dimensional symmetry TFT, a dynamical boundary always exists, on which the $2k$-dimensional relative QFT lives. A topological (gapped) boundary, on the other hand, is specified by a Lagrangian subgroup $L\subset \mathbb{D}$. There is a set of well-defined topological boundary conditions on the topological boundary, including the Dirichlet boundary conditions
\begin{equation}\label{eq:dirichlet boundary condition}
    \delta(d(b_i) - \ovl_x D_x - \ovl_y D_y) = 0 \Big|_{M_{2k}}, \quad D_x \in H^{k}(M_{2k}, \bbZ_{N}^{(x)}), \ D_y \in H^{k}(M_{2k}, \bbZ_{N}^{(y)}) \,.
\end{equation}
As a linear combination of $b_i$, we have $d(b_i) \in H^k(N_{2k+1}, \ovL)$ are $\ovL$-valued components of bulk fluxes, whose boundary profiles are determined by $D_x$ and $D_y$. Here $d$ stands for \textit{Dirichlet} boundary conditions with boundary value $D$. In particular, $\ovl$ specifies the embedding of the generator of $\ovL$ into $\bbD$:
\be
    1 \in \ovL \cong \bbZ_N\ \  \hookrightarrow\ \  (\ovl_x, \ovl_y) \in \bbD \cong \bbZ_N^{(x)} \oplus \bbZ_N^{(y)} \,.
\ee
$\ovl$ labels the boundary profiles of the $\ovL$-valued bulk fields. In other words, the above boundary condition in equation (\ref{eq:dirichlet boundary condition}) can be derived by expressing the $\ovL$-valued $B$ in terms of the embedding of $\ovL$ in the full defect group  $\mathbb{D}$, whose components in $\bbZ_N \times \bbZ_N$ can be denoted as $\ovl=(\ovl_x, \ovl_y), \ \ \ovl_x, \ovl_y \in \bbZ_N$.

The canonical dual of $d(b_i)$ is another linear combination of bulk fields, denoted as $n(b_i)$ (where $n$ stands for \textit{Neumann}). This $n(b_i)$ is the background field for the gauged symmetry $L$ that has Neumann boundary conditions on the topological boundary.

More concretely, the basis of boundary states, written as,
\be
|\ell, \ovl, B \rangle \,,
\ee
are such that we can stack the topological boundary onto the dynamical boundary $\langle \calR |$ to get the partition function of the absolute theory specified by the polarization pair $(\ell, \ovl)$:
\be
    Z_{(\ell, \ovl)}[B] = \langle \calR |\ell, \ovl, B\rangle \,,
\ee
which is just the projection of the partition vector $|\mathcal{R}\rangle$ onto a topological boundary state under a given basis. 

Therefore, $\Phi(A), \Phi(B)$ also act as clock-shift operators on the boundary states:
\be
    \Phi(A) |\ell, \ovl, B \rangle  = \exp\left(\frac{2\pi i}{N} \int_{M_{2k}} A \cup B \right) |\ell, \ovl, B \rangle, \quad \Phi(B_2) |\ell, \ovl, B_1 \rangle =  |\ell, \ovl, B_1 + B_2 \rangle \,.
\ee

\subsection{Topological Manipulations via Polarization Pairs} 

In the remainder of this section, we explain that polarization pairs are particularly convenient for the purpose of understanding the gauging of the $(k-1)$-form global symmetries. By working with polarization pairs, we no longer need to do detailed computations involved in gauging the global symmetry in the presence of quadratic counter terms (sometimes known as ``twist gauging''). Since conceptually, a twist gauging is always decomposed into the following two steps: (1) transforming into a polarization pair $(\ell, \ovl) \rightarrow (\ell, \ovl')$ where the counterterm disappears, and (2) doing a direct gauging of $(k-1)$-form symmetry via implementing a single discrete Fourier transformation.

\subsubsection{Stacking counterterms as changing \texorpdfstring{$\ovl$}{lbar}}

Recall the step when introducing polarization pairs where for a given $L^{\vee}$, the uplift to $\overline{L}$ is not unique. The choice of $(\ell, \ovl)$ for the split $\bbD = L \oplus \ovL$ leads to the split of the group of fluxes. As a consequence, we will get a specific basis by examining the Heisenberg group and its representation, as explained in Section \ref{subsec:symTFTpolpair}

Now, it is natural to ask the explicit consequence of changing from one option $\ovL$ to another $\ovL$. In short, such a difference will result in a phase shift of the partition function $Z_{(\ell, \ovl)}[B]$ and the topological boundary state $|\ell, \ovl, B\rangle$. In Appendix \ref{app:absorb counterterm}, we explain in detail that such a redefinition can generate quadratic counterterms as the integral of $\epsilon(B)$. At the same time, there always exists such a redefinition that precisely cancels the phase associated with the quadratic counterterm. Therefore, we say that the choice of SPT phase has been captured / incorporated in the choice of the polarization pair $(\ell, \ovl)$.

We now briefly discuss the possible form of quadratic counterterms in $4s$ and $4s+2$ dimensions:

\begin{itemize}
    \item For $4s$ dimensions, a quadratic counterterm (an SPT phase) should be described by a Pontryagin square $\calP(B_{2k})$, which is defined as 
    \be
    B_k \in H^{2s}(M_{4s}, \bbZ_{n}) \rightarrow \left\{ \begin{array}{cc}
       \calP(B_k) \in H^{4s}(M_{4s}, \bbZ_{2n}) & n \text{ even,}     \\
       \calP(B_k) \in H^{4s}(M_{4s}, \bbZ_{n}) & n \text{ odd.}
    \end{array} \right. \,
    \ee
    For $n$ odd $\calP(B_k)$ coincides with $B_k \cup B_k$, while for $n$ even, $\calP(B_k)$ takes valued in a ``quadratic-refined'' coefficient $\bbZ_{2n}$, and its reduction mod $\bbZ_n$ coincides with $B_k \cup B_k$. 
    \item Whereas for $4s+2$ dimensions, the only possibility of a quadratic counterterm can be expressed as (e.g., see Appendix \ref{app:absorb counterterm} of \cite{Gukov:2020btk})
    \be
        Q(x): H^{2l+1}(M_{4s+2}, \bbZ_2) \rightarrow \bbZ_2 \,,
    \ee
    which is a quadratic refinement of the bilinear pairing $\langle \cdot, \cdot \rangle$ on $H^{2l+1}(M_{4s+2}, \bbZ_2)$ defined via integration, such that for $B_x, B_y \in H^{2l+1}(M_{4s+2}, \bbZ_2)$,
    \be
        \int_{M_{4s+2}} B_x \cup B_y = Q(B_x + B_y) - Q(B_x) - Q(B_x) \mod 2 \,.
    \ee
    To understand why the above quadratic form is the only possibility, we remark that usually for an odd-degree form, we say they have trivial self-pairing due to anticommutativity $B_x \cup B_x = - B_x \cup B_x$ so $2 B_x \cup B_x = 0$. This almost always gives $B_x \cup B_x = 0$, unless if they take the coefficients in $\bbZ_2$.
\end{itemize}

We conclude by remarking that, shifting $\ovl$ to absorb the counterterm also holds for more general cases. Indeed, when $L$ has $\bbZ_2^s$ subgroup with $s > 1$, there will be more ways to stack counterterms corresponding to more $\bbZ_2$ generators, but at the same time, there are equally many ways to cancel these counterterms by shifting $\ovl$ with these generators. For example, if $\bbD = \bbZ_2^4$ and $L = \bbZ_2^{(a)} \oplus \bbZ_2^{(b)}$, then there are three order $3$ elements so that any one of the three possible counterterms 
\be
Q(B_a),\ \  Q(B_b),\ \  Q(B_a + B_b) \,,
\ee
can be canceled by shifting the generators in $\ovL$ via generators of $L$ accordingly.

\subsubsection{Gauging as Flipping the Polarization Pair}

We next consider gauging the $(k-1)$-form symmetry $L^{\vee}$ in a given absolute theory $\mathcal{T}$, which is only possible when $L^{\vee}$ is non-anomalous. The anomaly-free condition is equivalent to the existence of uplift of $L^{\vee}$ to $\overline{L}$, which gives rise to a direct sum decomposition of the defect group $\mathbb{D} = L \oplus \overline{L}$ in a pair of Lagrangian subgroups $(L, \overline{L})$ \cite{Gukov:2020btk} (also see detailed discussion in Appendix \ref{subapdx:k-1Anomaly}) so that one can fix a pair of generators $(\ell, \ovl)$ of $(L, \ovL)$ as the polarization pair. 

Then, under this particular basis, gauging a $\ovL$-valued $(k-1)$-form symmetry amounts to summing in the partition function $Z_{B}$ over its all possible $k$-form background field values $B \in H^k(M_{2k}, \ovL)$. Such a summation would turn the background field $B$ into a dynamical field of the gauged $(k-1)$-form symmetry, but the dual $L$-valued gauge field $A \in H^k(M_{2k}, L)$ would instead turn into a global symmetry.

In the symmetry TFT language, such a gauging amounts to doing the following Fourier transformation on the boundary state. Here $B = B_0$ the boundary value for the Dirichlet boundary condition of the $\ovL$-valued field, and $A$ is the dual $L$-valued field which acquires a Dirichlet boundary condition after gauging:
\be
    |\ell, \ovl, B= B_0 \rangle \rightarrow \sum_{B_0 \in H^k(M_{2k}, \ovL)} \exp \left(\frac{2\pi i}{N}\int_{M_{2k}} A \cup B\right) |\ell, \ovl, B= B_0 \rangle \,.
\ee
By definition, we introduce $\ell$ with $\langle \ell, \ovl \rangle = \tfrac{1}{N}$ to keep track of the to-appear generator of the dual global symmetry after gauging. Therefore, we have:
\be
     \sum_{B_0 \in H^k(M_{2k}, \ovL)} \exp \left(\frac{2\pi i}{N}\int_{M_{2k}} A \cup B\right) |\ell, \ovl, B= B_0 \rangle =  \boxed{|(-1)^{k-1}\ovl, l, A = A_0\rangle } \,.
\ee

Under this description, gauging $(k-1)$-form symmetry can be succinctly described as:
\begin{itemize}

\item{\textbf{In $4s+2$ dimensions}} \textit{exchanging} $\ell$ and $\overline{\ell}$ \cite{Gukov:2020btk}: 
\begin{equation}\label{eq:gauging as exchanging}
    \mathcal{T}~\text{associated with}~ (\ell,\ovl) \xrightarrow{\text{gauging}~L^{\vee}} \mathcal{T}/\overline{L}~\text{associated with}~(\ovl, \ell) \,. 
\end{equation}
\item{\textbf{in $4s$ dimensions}} \textit{exchanging} $\ell$ and $\overline{\ell}$ and then put an extra minus sign on $\ovl$
\be
    \calT\text{ associated with }(\ell,\ovl) \xrightarrow{\text{gauging }L^\vee} \mathcal{T}/\overline{L}~\text{associated with}~ (-\ovl, \ell) \,.
\ee
The information in the two elements of $(\ell, \ovl)$ is, in fact, correlated with each other under the Dirac pairing constraint. But as we have just seen keeping both of them explicitly is very helpful in making our notation well-behaved under gauging. For these spacetime dimensions, the gauging is a symplectic transformation on the defect group $\bbD = \bbD^{(e)} \oplus \bbD^{(m)}$, since the Dirac form is antisymmetric in $4s$ dimensions.
\end{itemize}

Therefore, after shrinking the symmetry TFT slab, the above discrete Fourier transformation would thus be implemented on the partition function $Z[B]$:
\begin{equation}
    Z[B] \rightarrow \widehat{Z}[A] = \sum_{B \in H^k(M_{2k}, \ovL)} \exp \left(\frac{2\pi i}{N}\int_{M_{2k}} A \cup B\right)\ Z[B], \ \ (A \in H^k(M_{2k}, L)) \,.
\end{equation}

As we have seen, the $N$-valued field $B_x$ which were previously thought of as a dynamical field, now become the background field, with coefficients in $L$, i.e., the emergent $(k-1)$-form global symmetry after gauging. Instead, the $\overline{N}$ valued field $B_y$ as background field of the original theory now becomes the dynamical field of the new theory. 

We end this part with two concluding remarks:
\begin{itemize}
    \item It is natural to examine the consequence of gauging twice. Indeed, gauging twice in $4s+2$ dimensions gives us back the original theory, while gauging twice in $4s$ dimensions gives us the charge-conjugated version of the original theory.
    \item The above treatment of gauging seems very basic at first glance. But the power of our formulation comes from the fact that this is all we need to do for gauging. Indeed, the complication of ``twist gauging'', namely of gauging in the presence of counterterms, has been simplified by decomposing into two smaller steps: changing $\ovl$ to $\ovl'$ and then doing a (symplectic) pair flip.
\end{itemize}

\section{Non-invertible Duality Defects via Polarization Pairs} \label{sec:nonInvGeneral}

In this section, we reformulate the half-space gauging construction of non-invertible duality defects in $2k$ dimensions, as building defects separating dual absolute QFTs arising from the same relative QFT. Our construction involves stacking an interface implementing a discrete automorphism of the charge lattice for $(k-1)$-dimensional charged operators, with another interface implementing a half-space gauging of the $(k-1)$-form global symmetry.

\paragraph{Review of Existing Formulations} We begin by reviewing how to construct non-invertible symmetry defects in even-dimensional QFTs via half-space gauging and refer the reader to \cite{Choi:2021kmx, Choi:2022zal} for more details. Consider a $2k$-dimensional QFT $\mathcal{T}$ with a non-anomalous $(k-1)$-form global symmetry $\mathbb{Z}_N^{(k-1)}$. Gauge the $\mathbb{Z}_N^{(k-1)}$ in half of the $2k$-dimensional spacetime, and then impose Dirichlet boundary conditions for the associated $k$-form gauge field $B_k$ on the resulting interface. Such a gauging corresponds to summing over the background field $B_k$ in the partition function; this is a \textit{topological manipulation} denoted as $\sigma$. If the original theory $\mathcal{T}$ and the gauged one $\mathcal{T}/\mathbb{Z}_N^{(k-1)}$ are dual to each other
\be
    \calT/\bbZ_N \cong \calT \,,
\ee
either trivially or via a duality transformation, then the interface becomes a symmetry defect,  corresponding to a non-invertible 0-form symmetry of the theory $\mathcal{T}$ \cite{Choi:2021kmx}. 

In addition to directly gauging the $\mathbb{Z}_N^{(k-1)}$, one can consider other topological manipulations combined with the gauging. If performing the resulting action on half of the spacetime again gives rise to a topological interface between $\mathcal{T}$ and its dual theory, then one can end up with a higher-order non-invertible duality defect, which is also referred to as an $n$-ality defect in the literature (see, e.g., \cite{Kaidi:2022uux, Choi:2022zal}).\footnote{In this paper we will also use ``duality defect" for whatever order of the associated duality.} For example, define the topological manipulation $\tau$ as stacking a quadratic counterterm (i.e., an SPT phase) $\epsilon(B_{k})$ on the theory $\mathcal{T}$ \cite{Witten:2003ya,Gaiotto:2014kfa}
\begin{equation}\label{eq:stacking SPT face}
\tau: \mathcal{Z}_{\mathcal{T}}[B_{k}]\rightarrow \mathcal{Z}_{\tau\mathcal{T}}\equiv \mathcal{Z}_{\mathcal{T}}[B_{k}]\exp{ \left(i \int \epsilon(B_{k})\right)} \,.
\end{equation}
One can then perform a twisted gauging (i.e., stacking a counterterm and then gauging the $\mathbb{Z}_N^{(k-1)}$ symmetry of the resulting theory $\tau\mathcal{T}$) of the $\mathbb{Z}_N^{(k-1)}$ via the interface $\mathcal{Z}_{(\sigma \cdot \tau)\mathcal{T}}\equiv \mathcal{Z}_{(\tau \mathcal{T})/\mathbb{Z}_N^{(k-1)}}$.
If the resulting theory is dual to the original one, i.e.,
\be
    \sigma \cdot \tau \calT = (\tau \calT)/\bbZ_N \cong \calT \,,
\ee 
a non-invertible duality defect can be constructed via performing a twisted gauging $\sigma \cdot \tau$ on half of the spacetime.\footnote{Gauging without twist can lead to a defect with order $> 2$, however, the non-invertible defect is still commonly referred to as a duality defect. See \cite{Choi:2022zal,Kaidi:2022uux} for discussions on this point.}

In this section, we show that all the above operations can naturally be reformulated in terms of polarization pairs, which inevitably leads to higher dimensional generalizations. We begin by explaining how any discrete automorphism acts on the polarization pair via its action on the defect group $\bbD$. We then combine all ingredients to give the general construction of non-invertible duality defects in the language of polarization pairs.

\subsection{Automorphisms as Dualities among Absolute Theories} 

In addition to the gauging, we also focus on cases where there exist discrete automorphisms  
\begin{equation}
    a \in \text{Aut}(\Lambda^*) \,,
\end{equation}
acting on the charge lattice of $k-1$ dimensional objects, such that under the action of $a$ the charge lattice of dynamical objects, $\Lambda$, is mapped to itself. These automorphisms thus descend to automorphisms of the defect group $\mathbb{D}$ 
\begin{equation}
    \text{Aut}(\Lambda^*) \rightarrow \text{Aut}(\mathbb{D}) \,.
\end{equation}
These can be regarded as discrete automorphisms for the relative QFT; these can be either discrete global symmetries or dualities, depending on the particular setup (see, e.g., \cite{Apruzzi:2017iqe} for applications to six dimensional theories). 

However, once we descend to an absolute theory by picking a polarization pair, then these ``automorphisms" are no longer global symmetries. However, they may become dualities mapping absolute QFTs which are equivalent locally (i.e., having the same local operators and their correlation functions) but different global structures (e.g., extended operators). Take 4D $\mathfrak{su}(N)$ $\mathcal{N}=4$ SYM as a simple example. Consider the Montonen--Olive S-duality for $\mathcal{N}=4$ SYM. For the $\mathfrak{su}(N)$ relative theory, just considering the local operator spectrum, it is an automorphism that can be regarded as a ``discrete global symmetry''. For absolute theories with well-defined global forms, it becomes a duality transformation between different absolute theories, e.g., with gauge groups $SU(N)$ and $PSU(N)$, at certain points on the conformal manifold \cite{Kapustin:2006pk} (i.e, $\tau_{YM} = i$). For our purposes, such a statement will be refined by including discrete $\theta$ angles and SPT phases, for which we give a detailed illustration in Section \ref{sec:4Dreformulated}.

For $2k$-dimensional QFTs, using polarization pairs, one can immediately obtain how the $\text{Aut}(\Lambda^*)$ acts on the set of polarizations, thus read off the possible dualities between absolute theories. Since $\text{Aut}(\Lambda^*)$ acts on the defect group, it acts on the components $(\ell, \ovl)$ of the polarization pair which are simply elements of the defect group. Therefore, implementing an action of a discrete automorphism of the defect group $a \in \text{Aut}(\bbD)$ is determined by having $a$ acting on $\ell$ and $\ovl$ separately:
\be
    (\ell, \ovl) \rightarrow (a(\ell), a(\ovl)) \,.
\ee

\subsection{Constructing Non-invertible Duality Defects}\label{sec:duality defects from polar pairs}

Having discussed gauging and dualities of absolute theories as various manipulations on the defect group, we are now ready to reformulate the half-space gauging construction of non-invertible duality defects. The construction has the following steps, which is also illustrated in Figure \ref{fig:CombiningInterfaces}.

\begin{itemize}
    \item Step 1. Start with an absolute theory $\mathcal{T}$ associated with the Lagrangian subgroup pair $(L,\ovL)$, which directly decomposes the defect group as $\mathbb{D}=L\oplus \overline{L}$. In addition, we need to also specify a pair of generators $(\ell, \ovl)$ for $(L, \ovL)$.
    \item Step 2. Stack quadratic counterterms onto the partition function of $\calT$. This effect can be described by shifting a basis of the partition vector space, which amounts to shifting $(\ell, \ovl) \rightarrow (\ell, \ovl')$, where $\ovl' = \ovl + r \ell$.
    \item Step 3. Gauge the non-anomalous $(k-1)$-form symmetry $L^\vee$ in half of the spacetime $M_{2k}^{x \geq 0}$ with Dirichlet boundary conditions for the corresponding $k$-form background gauge field. According to equation (\ref{eq:gauging as exchanging}), the resulting topological interface $\sigma(M_{2k-1})$ separates $\mathcal{T}$ and its gauged absolute theory $\mathcal{T}/L^\vee$ associated with $(\overline{L},L)$. Specifically, the new polarization pair one gets is $(\ovl', \ell)$ for $4s+2$ dimensions and $(-\ovl', \ell)$ for $4s$ dimensions.
    \item Step 4. Assume that there exists an automorphism element $a\in \text{Aut}(\mathbb{D})$ exchanging the two Lagrangian subgroups $L$ and $\overline{L}$, thus the two absolute theories $\mathcal{T}$ and $\mathcal{T}/L^\vee$ are dual to each other under this automorphism. Concretely, this automorphism takes the new polarization pair $(\pm \ovl', \ell)$ and restores the old one $(\ell, \ovl) = (a(\pm \ovl'), a(\ell))$ before stacking the counterterms and doing the gauging. Introduce the topological interface $\mathcal{I}_a(M_{2k-1})$ implementing this automorphism $a$. 
    \item Step 5. Stack the above sequence of topological interfaces together. The resulting codimension-1 operator 
    \begin{equation}
        \mathcal{N}(M_{2k-1})=I_r(M_{2k-1}) \cdot \sigma(M_{2k-1})\cdot \mathcal{I}_a(M_{2k-1}) \,,
    \end{equation}
    is a non-invertible duality defect for the theory $\mathcal{T}$.
\end{itemize}
In some literature, e.g., \cite{Choi:2022zal, Heckman:2022xgu}, the half-space gauging interface $\sigma(M_{2k-1})$ is promoted as the non-invertible duality defect itself if  $\mathcal{T}/\overline{L}$ is dual to $\mathcal{T}$, without writing down $\mathcal{I}_a(M_{2k-1})$ explicitly. At the level of fusion rules, this is equivalent to the above construction in the sense that $\mathcal{I}_a(M_{2k-1})$ is invertible and does not affect the non-trivial fusion rules. 
 
The fusion rules for the non-invertible defect $\mathcal{N}$ and the $L^\vee$ symmetry defect $\eta$ are 
\begin{equation}
\begin{split}
    \mathcal{N}(M_{2k-1})\times \mathcal{N}^\dagger(M_{2k-1}) &= \frac{1}{|H^0(M_{2k-1}, \mathbb{D})|}\sum_{\Sigma_k \in H_{k}(M_{2k-1}; \overline{L})} \eta(\Sigma_{k})\,,\\
    \mathcal{N}(M_{2k-1})\times \eta(\Sigma_k)&=\mathcal{N}(M_{2k-1})\,,
\end{split}
\end{equation}
where $\mathcal{N} (M_{2k-1})^\dagger$ is the orientation reversal of $\calN (M_{2k-1})$. The RHS of the first fusion rule is summing over all $L^\vee$ symmetry defects $\eta$ along the codimension-1 manifold $M_{2k-1}$, which is known as the condensation defect via higher gauging \cite{Kaidi:2021xfk,Roumpedakis:2022aik}.\footnote{The coefficient $\frac{1}{|H^0(M_{2k+1}, \mathbb{D})|}$ can be derived by summing over all gauge configurations in the symmetry TFT slab $M_{2k+1} \times I$, and then convert to homology by using the Poincare--Lefshetz duality of a tubular neighborhood $M_{2k+1} \times I$ of $M_{2k+1}$ relative to its boundary. See \cite{Choi:2021kmx,Kaidi:2022cpf} for more details.}

\paragraph{Action on states with $L^\vee$-valued charges.} Let us comment on how non-invertible duality defects act on $(k-1)$-dimensional charged defects. Starting from the relative theory, a defect charged under $\overline{L}\in \mathbb{D}$ would descend to a genuine $(k-1)$-dimensional defect charged under the non-anomalous global symmetry $L^\vee$ in the absolute theory $\mathcal{T}$ associated with $(\ell,\ovl)$. If we instead consider a defect whose charge is valued in $L \subset \mathbb{D}$, then it descends to a non-genuine $(k-1)$-dimensional defect in the absolute theory $\mathcal{T}$ associated with $(\ell,\ovl)$, which is only well-defined when attached to a $k$-dimensional topological operator. 

\begin{figure}[H]
    \centering
    \includegraphics[width=7.5cm]{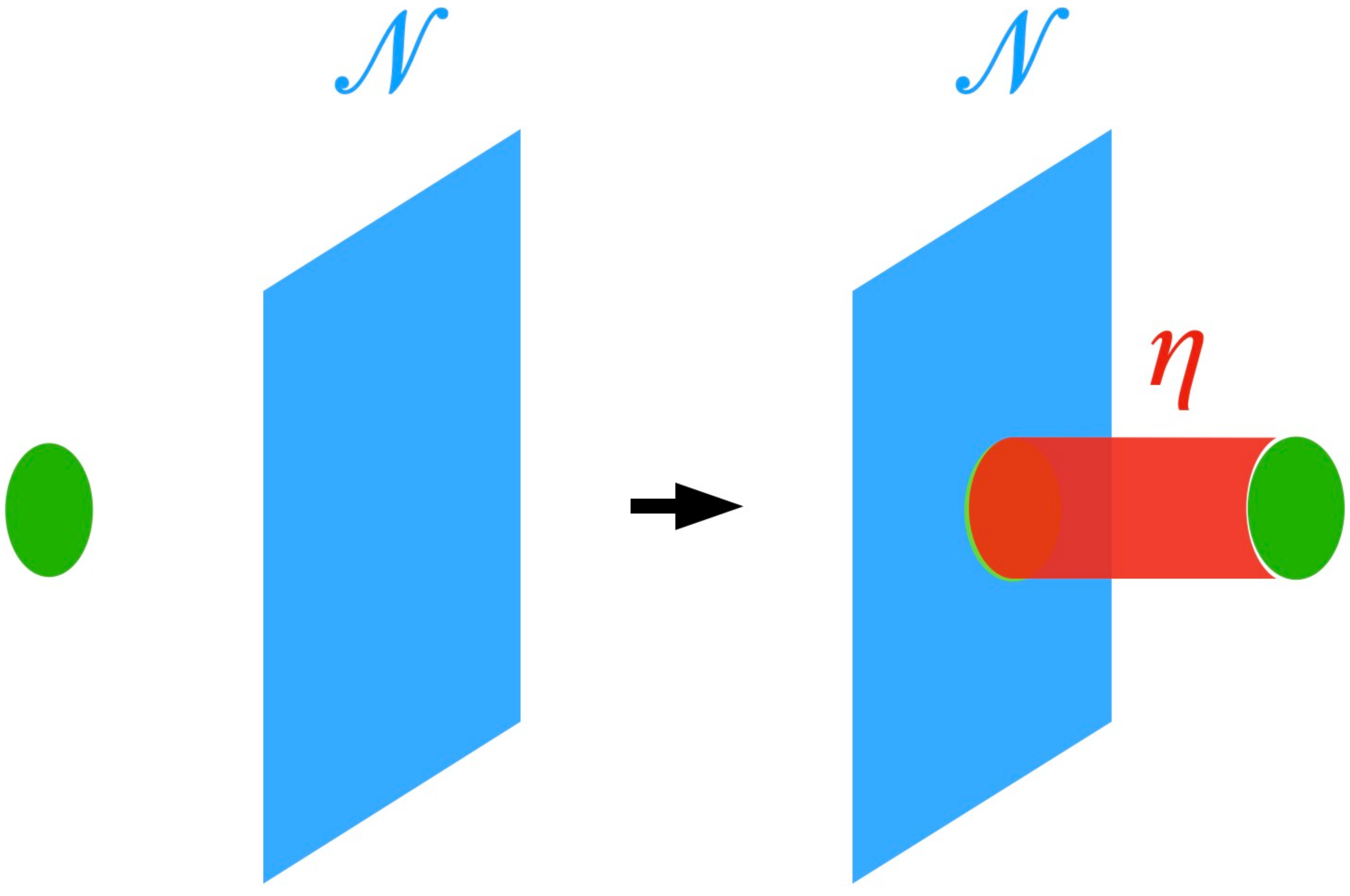}
    \caption{The action of non-invertible duality defects in 6D on charged surface operator. $\eta$ is a 3-dimensional topological operator generating the $\ovL$ global symmetry.}
    \label{fig:6D transition}
\end{figure}

Therefore, when the non-invertible duality defect $\mathcal{N}$ is swept past a $(k-1)$-dimensional $L^\vee$-charged defect, it becomes gauged, i.e., non-genuine and is now attached to $k$-dimensional $L^\vee$ symmetry operator $\eta$
which intersects with $\mathcal{N}$. This non-trivial transition has been investigated in 2D and 4D QFTs (see, e.g., \cite{Frohlich:2004ef, Choi:2021kmx}).\footnote{In some 4D examples, this non-trivial transition enjoys a string theory implementation as the Hanany--Witten transition \cite{Apruzzi:2022rei,Heckman:2022xgu}.} Figure \ref{fig:6D transition} illustrates this transition for $k=3$, i.e., in 6D. 

\section{Warm-up Examples: 2D and 4D Revisited} \label{sec:4Dreformulated}

In the previous sections, we gave a general discussion of how to realize various topological manipulations and how to build duality defects in $2k$-dimensional QFTs in terms of the polarization pair. In this section, we present a comprehensive discussion of how this reproduces the known results in the literature for the 2D Ising CFT and 4D $\mathcal{N}=4$ SYM theories. By doing so, we not only clarify our notations and conventions but also familiarize the reader with the language of polarization pairs in order to better understand our generic construction and the more exotic 6D examples in the following sections. Our notations in this section essentially follow \cite{Choi:2021kmx, Kaidi:2022uux}.

\subsection{2D Ising CFT}

The simplest example of a non-invertible duality defect is the Kramers--Wannier line in the 2D critical Ising CFT \cite{Frohlich:2004ef}. This duality defect can be realized via the half-space gauging construction \cite{Choi:2021kmx}, which can also be viewed from the symmetry TFT perspective \cite{Kaidi:2022cpf}. In this subsection, we reproduce the Kramers--Wannier duality defect from the polarization pair perspective. Furthermore, we argue that bosonization/fermionization among $c=\frac{1}{2}$ CFTs is also nicely unified via this description. 

The relative 2D theory that we start with has intermediate defect group $\mathbb{D}=\mathbb{Z}_2^{(x)}\times \mathbb{Z}_2^{(y)}$, with the symmetric (since we are in $4s+2$ dimensions) Dirac pairing given by 
\begin{equation}\label{eq:Dirac pairing of Ising CFT}
    \left(\begin{array}{cc}
        0 & \frac{1}{2} \\ 
        \frac{1}{2} & 0
    \end{array}\right) \,.
\end{equation}
We label each generator of $\mathbb{Z}_2^{(x)}\times \mathbb{Z}_2^{(y)}$ as 
\begin{equation}
    (n_x,n_y)\in \mathbb{Z}_2^{(x)}\times \mathbb{Z}_2^{(y)} \,,
\end{equation}
and the subgroup generated as $\langle (n_x,n_y) \rangle$. Lagrangian subgroups $L\in \mathbb{D}$ are then those generated by $(n_x,n_y)$ with the trivial pairing 
\begin{equation}
    (n_x,n_y)\left(\begin{array}{cc}
        0 & \frac{1}{2} \\ 
        \frac{1}{2} & 0
    \end{array}\right)(n_x,n_y)^T=0~\text{mod}~1 \,,
\end{equation}
from which it is straightforward to compute that $\mathbb{Z}_2^{(x)}$, $\mathbb{Z}_2^{(y)}$ and the diagonal subgroup $\mathbb{Z}_2^{(\text{diag})}\subset \mathbb{D}$ are the three Lagrangian subgroups, generated by $\langle (1,0)\rangle$, $\langle (1,0)\rangle$, and $\langle (1,1)\rangle$, respectively. There are thus six absolute theories, labeled by $\mathcal{T}^{c=\frac{1}{2}}_{\ell,\ovl}$, given by the following polarization pairs \begin{equation}\label{eq:list of polar pairs for c=1/2 CFT}
    \calT^{c=\frac{1}{2}}_{(1,0), (0,1)}, \ \ \calT^{c=\frac{1}{2}}_{(1,0), (1,1)}, \ \ \calT^{c=\frac{1}{2}}_{(0,1), (1,0)}, \ \ \calT^{c=\frac{1}{2}}_{(0,1), (1,1)}, \ \ \calT^{c=\frac{1}{2}}_{(1,1), (1,0)}, \ \ \calT^{c=\frac{1}{2}}_{(1,1), (0,1)} \,,
\end{equation}
where ``$c=\frac{1}{2}$'' indicates that these theories are not all the Ising CFT but all have central charge $c=\frac{1}{2}$. 

\paragraph{Kramers--Wannier Duality Defect} We take the convention that the Ising CFT is associated with the polarization pair as follows:
\begin{equation}
    \calT^{c=\frac{1}{2}}_{(1,0), (0,1)} \leftrightarrow \text{Ising} \,.
\end{equation}
Then, gauging the $\mathbb{Z}_2$ zero-form symmetry of the Ising model leads to the absolute theory:
\begin{equation}
     \calT^{c=\frac{1}{2}}_{(0,1), (1,0)} \leftrightarrow \text{Ising}/\mathbb{Z}_2 \,,
\end{equation}
as gauging is simply the flipping of the pair of generators in the polarization pair:
\begin{equation}
    (\ell,\ovl)=((1,0),(0,1))\rightarrow (\ell^\prime,\ovl^\prime)=((0,1),(1,0)) \,.
\end{equation}

The isomorphism of Ising and Ising$/\mathbb{Z}_2$ is implemented by an automorphism on the parent relative theory with the action on the polarization pair as
\begin{equation}
    a:((0,1),(1,0))\rightarrow ((1,0),(0,1)) \,.
\end{equation}
The Kramers--Wannier duality defect in the critical Ising CFT can then be simply realized as
\begin{equation}
\mathcal{N}=\sigma \cdots \mathcal{I}_a \,,
\end{equation}
where $\sigma$ is the half-space gauging interface and $\mathcal{I}_a$ is the invertible operator implementing the automorphism $a$.

\paragraph{Fermionization/Bosonization} Let us now interpret the other four polarization pairs in equation (\ref{eq:list of polar pairs for c=1/2 CFT}). Based on our general discussion in previous sections, theories with the same $\ell$ but different $\ovl$ should be regarded as distinguished via SPT phases/counterterms. However, there is no nontrivial SPT phase for the $\mathbb{Z}_N$ symmetry in 2D pure bosonic systems, since $H^2(\mathbb{Z}_N,U(1))=0$. Then how the theory $\calT^{c=\frac{1}{2}}_{(1,0), (1,1)}$ distinguished from the Ising CFT $\calT^{c=\frac{1}{2}}_{(1,0), (0,1)}$? The answer is that $\calT^{c=\frac{1}{2}}_{(1,0), (1,1)}$ is derived from stacking fermionic SPT phases, given by Arf invariants (see, e.g., \cite{Karch:2019lnn,Ji:2019ugf,Gukov:2020btk}), on the Ising CFT $\calT^{c=\frac{1}{2}}_{(1,0), (0,1)}$. Gauging the $\mathbb{Z}_2$ symmetry of $\calT^{c=\frac{1}{2}}_{(1,0), (1,1)}$, one flips the polarization pair and ends up with the theory 
\begin{equation}
    \calT^{c=\frac{1}{2}}_{(1,1), (1,0)} \leftrightarrow \text{CFT for a Majorana Fermion} \,.
\end{equation}
This reproduces the celebrated fermionization of the bosonic Ising CFT to a Majorana fermion. The absolute theory $\calT^{c=\frac{1}{2}}_{(1,1), (0,1)}$ is also the fermionic CFT for Majorana spinor, but differed from the $\calT^{c=\frac{1}{2}}_{(1,1), (1,0)}$ theory by the SPT phase/Arf invariant. Gauging the $\mathbb{Z}_2$ of the fermionic CFT gives rise back to the bosonic Ising CFT, which is exactly the bosonization process via summing over the spin structure appropriately.\footnote{We refer the reader to \cite{Gaiotto:2015zta,Ji:2019ugf} for a detailed discussion on 2D fermionization/bosonization from a modern perspective.} 

We close this subsection by emphasizing that though one can write automorphisms on the defect group in order to connect theories under fermionization/bosonization, these do not give rise to (non-invertible) duality defects but rather \emph{maps} between inequivalent bosonic and fermionic CFTs \cite{Lin:2019hks}; i.e., this is an example where the assumption that the charge lattice automorphism uplifts to a good duality of the full theory does not necessarily hold.

\subsection{4D \texorpdfstring{$\mathcal{N}=4$ $\mathfrak{su}(N)$}{N=4 su(N)} SYM}

For simple examples like the 2D critical Ising CFT, the polarization pair language for building duality defects might look unnecessarily abstract and formal. However for relative QFTs that possess a rich structure of associated absolute QFTs, the polarization pair is a powerful method for their investigation. With this in mind, let us now revisit non-invertible duality defects in 4D $\calN=4$ SYM theories via the polarization pair. 

For simplicity we focus on the $\mathfrak{su}(N)$ case for $N$ prime, whose defect group is $\bbZ_N^{(e)} \oplus \bbZ_N^{(m)}$ with the anti-symmetric Dirac pairing:
\begin{equation}
    \left(\begin{array}{cc}
        0 & \frac{1}{N} \\ 
        -\frac{1}{N} & 0
    \end{array}\right) \,.
\end{equation}
We label each generator via the following notation:
\be
    (n_e, n_m) \in \bbZ_N^{(e)} \oplus \bbZ_N^{(m)} \,,
\ee
and the cyclic subgroup generated as $\langle (n_e, n_m) \rangle$. In the language of polarization pairs, the total number of absolute theories is given by (recalling that $N$ is prime):
\be
    (N+1)N(N-1) = N^3 - N \,.
\ee
This can be counted by first counting the number of pairs of generators $(\ell, \ovl)$ with 
\begin{equation}
\begin{split}
    \ell=(n_e,n_m), ~\ovl=(\overline{n}_e,\overline{n}_m)\,, \quad \text{ such that } \quad 
    \langle \ell, \ovl \rangle=\frac{1}{N}(n_e\overline{n}_m-n_m\overline{n}_e) \,.
\end{split}
\end{equation}
 We will later see that these are exactly all global structures that the conventional approach covers, thereby supporting the validity of our formulation.\footnote{If $N$ is not prime, then we get a complication from the possibility of gauging proper subgroups of $\bbZ_N$, which, though somewhat involved, is also captured by the polarization pair.} 
 
 A standard way of defining all polarizations is to start from the partition function of the electric polarization $Z_{SU(N)_0}[\tau_{YM}, B]$, and do various manipulations on it to reach all to the remaining polarizations. But instead of going through all the technical details of the original approach, we will take the same steps and walk through these manipulations in the notion of polarization pairs. Our notation in 4D will follow that of \cite{Kaidi:2022uux}.

It is conventional to start from the electric polarization, which is usually denoted as $Z_{SU(N)_0}[B]$. For us, taking the electric polarization means that the global symmetry $L^\vee$ is the electric center symmetry. We have:
\be
   Z_{SU(N)_0}: L = \langle (0,1) \rangle, \ovL = \langle (1,0) \rangle \,.
\ee
We first consider the theory where the background field of the global symmetry carries a single unit of $B$; this amounts to fixing the generator to be $\ovl = (1,0)$, and then the Dirac pairing requires us to pick $\ell = (0,-1)$. Therefore, the above theory is refined into the polarization pair:
\be
    Z_{SU(N)_0}[B]: (\ell, \ovl) \text{  with  }\ell = (0, -1),\ \ovl = (1, 0) \,.
\ee
Now we take the following three-step procedure to get all absolute theories which correspond to the $\ksu (N)$ SYM as a relative theory.
\begin{itemize}
    \item $Z_{SU(N)_0}[tB]$. This theory is obtained by changing the multiplicity of the background field, i.e., implementing ``generalized charge conjugation" (named after the charged conjugation  $B\rightarrow -B$ with $t=N-1$). This is done by using different generators:
    \be
        Z_{SU(N)_0}[B]: (\ell, \ovl) \text{  with  }\ell = (0, -t),\ \ \ovl = (t^\vee, 0), \ \ t \in \bbZ_N \,,
    \ee
    where $t^\vee$ is the inverse of $t$ in $\mathbb{Z}_N$, i.e., such that $t t^\vee = 1 \mod N$. In particular, $t = -1$ implements charge conjugation. This gives $N-1$ absolute theories.
    \item $Z_{SU(N)_m}[tB]$. This theory is obtained by stacking $m$ units of counterterms onto the $Z_{SU(N)_0}[tB]$ theory, which is implemented by the shift of $\ovl \rightarrow \ovl + m \ell$. The resulting polarization pair is
    \be
        Z_{SU(N)_m}[tB]: (\ell, \ovl) \text{  with  }\ell = (0, -t),\ \ \ovl = (t^\vee, -mt) \,,
    \ee
    which gives $(N-1)^2$ more absolute theories.
    \item $Z_{PSU(N)_{n, m}}[tB]$. This theory is obtained by taking $Z_{SU(N)_0}[tB]$, first stacking $n$ units of counterterms to get $l = (0, -t), \ovl = (t^\vee, -nt)$, and then implementing the gauging via the symplectic flipping $(\ell,\ovl)\rightarrow (-\ovl, \ell)$ to get
    \be
    \ell = (-t^\vee, nt),\ \ \ovl = (0, -t) \,,
    \ee
    and finally stack $m$ units of counterterms once more to get:
    \be
        Z_{PSU(N)_{n, m}}[tB]: (\ell, \ovl) \text{  with  }\ell = (-t^\vee, nt),\ \ \ovl = (-m t^\vee, (mn-1)t) \,.
    \ee
    This gives $N \cdot N \cdot (N-1) = N^3 - N^2$ additional absolute theories, since $m, n, t \in \bbZ_N,\ t \neq 0$.
\end{itemize}
Adding up all three situations, we reproduce all $N^3 - N$ possible absolute theories via the polarization pairs $(\ell, \ovl)$. 

To summarize, the topological manipulations relating the different absolute theories translate into simple operations on polarization pairs $(\ell, \ovl)$, which provides an elegant and universal way to capture all absolute theories associated with the same relative theory. Instead of starting with a particular theory and exhaustively exploring all possible topological manipulations, one merely needs to enumerate the polarization pairs.

We now go into further details of $\ksu (2)$ and $\ksu (3)$ SYM theories to review the construction of duality defects via polarization pairs. As we will see, in the $\ksu (3)$ example where the charge conjugation is non-trivial, the polarization pair $(\ell, \ovl)$ serves as a powerful tool to fully specify the absolute theory and its possible duality defects.

\paragraph{$\ksu(2)$ Example}

The intermediate defect group is $\bbD = \bbZ_2 \oplus \bbZ_2$. There are 6 polarization pairs $(\ell, \ovl)$ (in this case reduced to only specifying $(\ell, \ovl)$ for $L = \ovL = \bbZ_2$) which are given by 
\begin{align}
    SU(2)_0: ((0,1), (1,0)),\ \ SO(3)_{+,0}: ((1,0), (0,1)),\ \  SO(3)_{-,0}:((1,1), (0,1)) \,,\\
    SU(2)_1: ((0,1), (1,1)), \ \ SO(3)_{+,1}: ((1,0), (1,1)),\ \ SO(3)_{-,1}:((1,1), (1,0)) \,.
\end{align}
To recapitulate, the topological manipulations are generated by:
\begin{equation}\label{eq:su(2) topological manipulation}
    \sigma: (\ell, \ovl) \rightarrow (-\ovl, \ell)=(\ovl, \ell), \quad \tau: (\ell, \ovl) \rightarrow (\ell, \ovl + \ell) \,,
\end{equation}
while the $SL(2, \bbZ)$ automorphisms/dualities act separately on $\ell$ and $\ovl$ as:
\be\label{eq:su(2) automorphisms}
    S: (1,0) \leftrightarrow (0,1), \ \ (1, 1)\text{ fixed,}\quad T: (1,0) \leftrightarrow (1,1), \ \ (0,1) \text{ fixed.}
\ee
Thus we reproduce the expected transformations for 4D $\ksu(2)$ SYM. There is only one non-trivial generator of $SU(2)$, so charge conjugation is completely trivial in this case. Based on our discussion in Section \ref{sec:duality defects from polar pairs}, the non-invertible duality defects for a given $(\ell,\ovl)$ are those corresponding to $\sigma$ combined with other actions in equations (\ref{eq:su(2) topological manipulation}) and (\ref{eq:su(2) automorphisms}) such that $(\ell,\ovl)$ is eventually mapped back to itself. It is straightforward to check this reproduces the result in \cite{Kaidi:2022uux}. We depict the polarization pairs and their connections via 

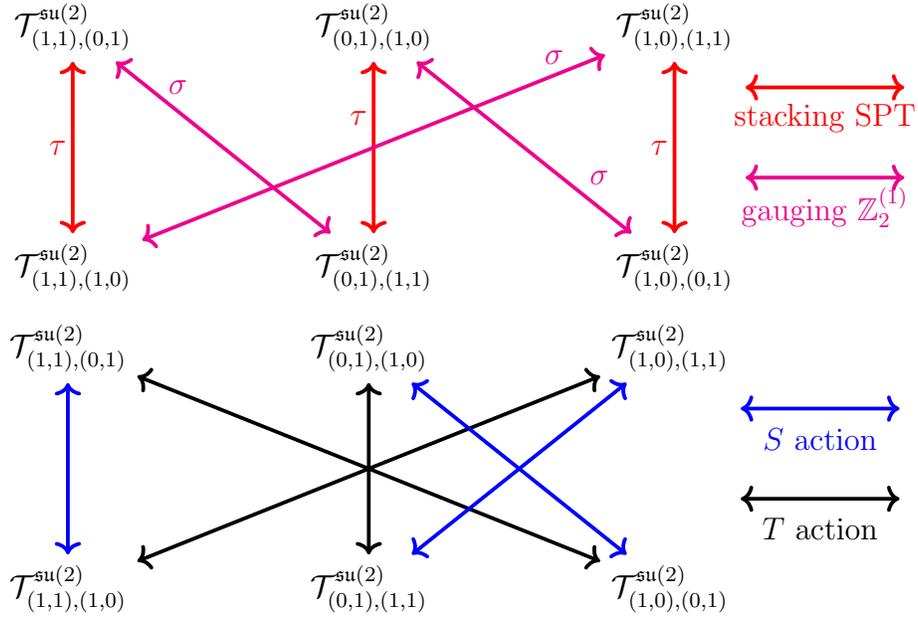
\begin{figure}
    \centering
    \begin{tabular}{c}
    \begin{tikzpicture}[scale= .4]
	\node [style=none] (6) at (-10, 4) {$\mathcal{T}^{\ksu(2)}_{(1,1), (0,1)}$};
	\node [style=none] (7) at (0, 4) {$\mathcal{T}^{\ksu(2)}_{(0, 1), (1, 0)}$};
	\node [style=none] (8) at (-10, -4) {$\mathcal{T}^{\ksu(2)}_{(1, 1), (1, 0)}$};
	\node [style=none] (9) at (0, -4) {$\mathcal{T}^{\ksu(2)}_{(0, 1), (1, 1)}$};
	\node [style=none] (10) at (10, -4) {$\mathcal{T}^{\ksu(2)}_{(1, 0), (0,1)}$};
	\node [style=none] (11) at (10, 4) {$\mathcal{T}^{\ksu(2)}_{(1, 0), (1,1)}$};
    \node [style=none] (12) at (-10.5, 0) {{\color{red} $\tau$}};
    \node [style=none] (13) at (-0.5, 1) {{\color{red} $\tau$}};
    \node [style=none] (14) at (9.5, 0) {{\color{red} $\tau$}};
    \node [style=none] (15) at (-6.5, 2) {{\color{magenta} $\sigma$}};
    \node [style=none] (16) at (7.5, -1) {{\color{magenta} $\sigma$}};
    \node [style=none] (17) at (6, 3) {{\color{magenta} $\sigma$}};





	\draw [style=red arrows] (6) to (8);
	\draw [style=red arrows] (9) to (7);
	\draw [style=red arrows] (10) to (11);
	\draw [style=magenta arrows] (6) to (9);
	\draw [style=magenta arrows] (7) to (10);
	\draw [style=magenta arrows] (8) to (11);

    \node [style=none] (x1) at (12, 2) {};
    \node [style=none] (x2) at (18, 2) {};
    \draw [style=red arrows] (x1) to (x2);
    
    \node [style=none] (y1) at (12, -1) {};
    \node [style=none] (y2) at (18, -1) {};
    \draw [style=magenta arrows] (y1) to (y2);  

    \node [style=none,text=red] (x3) at (15, 1) {stacking SPT};
    \node [style=none,text=magenta] (y3) at (15, -2) {gauging $\bbZ_2^{(1)}$};
 
\end{tikzpicture}\\
    \begin{tikzpicture}[scale= .4]
	\node [style=none] (1) at (-10, -4){$\mathcal{T}^{\ksu(2)}_{(1,1), (1,0)}$};
	\node [style=none] (2) at (0, -4) {$\mathcal{T}^{\ksu(2)}_{(0, 1), (1, 1)}$};
 	\node [style=none] (3) at (10, -4) {$\mathcal{T}^{\ksu(2)}_{(1, 0), (0,1)}$};
	\node [style=none] (4) at (-10, 4) {$\mathcal{T}^{\ksu(2)}_{(1, 1), (0,1)}$};
	\node [style=none] (5) at (0, 4) {$\mathcal{T}^{\ksu(2)}_{(0, 1), (1, 0)}$};
	\node [style=none] (6) at (10, 4) {$\mathcal{T}^{\ksu(2)}_{(1, 0), (1, 1)}$};

    \draw [style=black arrows] (1) to (6);
    \draw [style=black arrows] (2) to (5);
    \draw [style=black arrows] (3) to (4);

    \draw [style=blue arrows] (1) to (4);
    \draw [style=blue arrows] (2) to (6);
    \draw [style=blue arrows] (3) to (5);

    \node [style=none] (7) at (12, 2) {};
    \node [style=none] (8) at (18, 2) {};
    \draw [style=blue arrows] (7) to (8);
    
    \node [style=none] (9) at (12, -1) {};
    \node [style=none] (10) at (18, -1) {};
    \draw [style=black arrows] (9) to (10);  

    \node [style=none,text=blue] (11) at (15, 1) {$S$ action};
    \node [style=none] (11) at (15, -2) {$T$ action};
    
\end{tikzpicture}
    \end{tabular}
    \caption{We show the six different polarization pairs for the relative theory of $\mathcal{N}=4$ SYM with $\mathfrak{su}(2)$ gauge algebra, together with how the topological manipulations and dualities act on the polarization pairs. Any closed loop containing an odd number of topological manipulations associated to gauging the one-form symmetry, i.e., $\sigma$, gives rise to a non-invertible symmetry defect in the theory at the start/end of the loop.}
    \label{fig:su2N4}
\end{figure}

\paragraph{$\ksu(3)$ Example}

In this case, the polarization pair consists of a pair of generators inside the defect group $\mathbb{D}=\bbZ_3 \times \bbZ_3$. In contrast to $\mathfrak{su}(2)$ case, the charge conjugation plays a non-trivial role. For the convenience of the reader, 
we present all global forms of $\calN = 4$ $\ksu (3)$ SYM, enumerating the full list of $3^3 - 3 = 24$ polarization pairs $(\ell, \ovl)$. Since we only have two choices of the value of the background field $ \pm B$, we denote the $Z_G[B]$ as $G$ and the charge conjugated version $Z_G[-B]$ as $\overline{G}$ following \cite{Kaidi:2022uux}, where $G$ is $SU(3)$ or $PSU(3)$ with certain discrete parameters. Then the full list of polarization pairs $(\ell, \ovl)$ is given by the following:
\renewcommand{\arraystretch}{1.4}
\begin{equation}
\centering
{\small
\begin{array}{c|c|c|c}
    SU(3)_0: (0,2), (1,0) & PSU(3)_{0, 0}: (2,0), (0,2) &  PSU(3)_{1, 0}:(2,1), (0,2) & PSU(3)_{2, 0}:(2,2), (0,2)\\ \hline
    SU(3)_1: (0,2), (1,2) & PSU(3)_{0, 1}: (2,0), (2,2) &  PSU(3)_{1, 1}:(2,1), (2,0) & {\color{red} PSU(3)_{2, 1}:(2,2), (2,1)}\\ \hline
    SU(3)_2: (0,2), (1,1) & PSU(3)_{0, 2}: (2,0), (1,2) &  {\color{red} PSU(3)_{1, 2}:(2,1), (1,1)} & PSU(3)_{2, 2}:(2,2), (1,0)\\ \hline
    \overline{SU(3)}_0: (0,1), (2,0) & \overline{PSU(3)}_{0, 0}: (1,0), (0,1) &  \overline{PSU(3)}_{1, 0}:(1,2), (0,1) & \overline{PSU(3)}_{2, 0}:(1,1), (0,1)\\ \hline
    \overline{SU(3)}_1: (0,1), (2,1) & \overline{PSU(3)}_{0, 1}: (1,0), (1,1) &  \overline{PSU(3)}_{1, 1}:(1,2), (2,0) & \overline{PSU(3)}_{2, 1}:(1,1), (1,2)\\ \hline
    \overline{SU(3)}_2: (0,1), (2,2) & \overline{PSU(3)}_{0, 2}: (1,0), (2,1) &  \overline{PSU(3)}_{1, 2}:(1,2), (2,2) & \overline{PSU(3)}_{2, 2}:(1,1), (2,0)\\
\end{array}\nonumber
}
\end{equation}
\renewcommand{\arraystretch}{1.0}

The procedure of obtaining the full list of duality defects in \cite{Kaidi:2022uux} via polarization pairs is again followed our discussion in Section \ref{sec:duality defects from polar pairs}. Write down the 24 polarization pairs above, together with all of their connections via topological manipulation of duality; then, any closed loop involving an odd number of one-form symmetry gauging manipulations implies the presence of a non-invertible duality defect. For example, consider the theory with polarization pair
\be
    PSU(3)_{1, 2}: ((2,1), (1,1)) \,.
\ee
Gauging the $\mathbb{Z}_3$ one-form symmetry leads to the following absolute theory:
\be
((2,1), (1,1))\overset{\sigma}{\rightarrow} ((-1,-1),(2,1))=((2,2),(2,1)) \,,
\ee
which is written as $PSU(3)_{2,1}$. We mark these two theories in the table in red for visual clarity. One can express the $SL(2,\mathbb{Z})$ automorphisms (at $\tau_{YM}=i$) of the $\mathfrak{su}(3)$ theory as actions on $\ell$ and $\ovl$, similarly to equation (\ref{eq:su(2) automorphisms}), and then realize there is indeed an automorphism which is the duality transformation 
\begin{equation}
    ((2,2),(2,1))\overset{S}{\rightarrow} ((2,1), (1,1)) \,.
\end{equation}
Therefore, this leads to a non-invertible duality defect in the $PSU(3)_{1,2}$ theory via half-space gauging associated with $\sigma$ and the duality transformation $S$.

We close this section by emphasizing that the polarization pair construction in 2D and 4D is not limited to the Ising CFT and $\mathcal{N}=4$ SYM theories. One can revisit other theories with non-invertible duality defects, e.g., 2D $c=1$ CFTs \cite{Thorngren:2021yso}, or 4D class $\mathcal{S}$ theories \cite{Bashmakov:2022uek,Antinucci:2022cdi}. Namely, once the defect group and its pairing rule are derived for these theories, one can follow our generic construction to build polarization pairs and non-invertible duality defects.\footnote{In practice, looking for the defect group and its pairing can be non-trivial. For 2D $c=1$ rational CFTs, this translates into investigating the Lagrangian subalgebra for the conformal blocks (which give rise to the space of partition vectors) and defect braiding in the associated 3D TFT. For class $\mathcal{S}$ theories associated with 6D $\mathcal{N}=(2,0)$ theories compactified on Riemann surfaces, the relative theory to start with is the 6D $\mathcal{N}=(2,0)$ theory itself. Lagrangian subgroups are then given by maximally isotropic sublattices of 1-cycles of the Riemann surface and their linking.}

\section{Non-invertible Duality Defects in 6D (S)CFTs} \label{sec:6DSCFTs}

In this section, we apply our general constructions of duality defects to QFTs in 6 dimensions. Even though our general construction does not depend on supersymmetry, most of the known constructions of QFTs in 6D rely on string theory, therefore we focus on examples in 6D SCFTs with $(2,0)$ and $(1,0)$ supersymmetry, see \cite{Heckman:2018jxk} for a review. 

However, this section will not assume any background in 6D SCFTs. Even though we will express each 6D SCFT we discuss in terms of the effective field theory description on the tensor branch, following the notation in \cite{Heckman:2013pva,Heckman:2015bfa}, we immediately specify the field-theoretic data of 2-form charge lattices, their intermediate defect groups, and the associated bilinear pairing. As the latter three objects are all that is necessary for our discussion, together with the datum that a relevant automorphism uplifts to a duality of the relative theory, the tensor branch description is only provided as an aid to readers who are familiar with that description.

A crucial object is the charge lattice automorphisms of 6D SCFTs. They are introduced and exhaustively studied in \cite{Apruzzi:2017iqe} for 6D SCFTs viewed as relative theories. However, down to the level of absolute 6D theories, such Green--Schwarz automorphisms should be viewed as dualities between different absolute theories descending from the same relative theory. Their role is highly analogous to that played by $SL(2, \bbZ)$ duality at special $\tau_{YM}$ values in 4D SYM theories.

By combining the Green--Schwarz dualities and topological manipulations of gauging 2-form symmetries and stacking counterterms, the construction of non-invertible duality defects in 6D exactly follows from our general discussion in Section \ref{sec:nonInvGeneral} .

For the rest of this section, we discuss concrete examples of 6D SCFTs: a irreducible $(2,0)$ example (the $D_4$ theory), a ``reducible'' $(2,0)$ example (the $A_4 \oplus A_4$ theory), together with some general comments on $(1,0)$ examples. We also briefly discuss some RG flows from 6D $(1,0)$ SCFTs to 6D $(2,0)$ SCFTs, along which we get non-invertible duality defects as emergent symmetries in the infrared.

\subsection{\texorpdfstring{$\mathcal{N}=(2,0)$ $D_4$}{N=(2,0) D4} Theory} \label{subsec:D4}
\begin{figure}
\centering
    \begin{tabular}{c}
    \begin{tikzpicture}[scale= .4]
	\node [style=none] (6) at (-10, 4) {$\mathcal{T}^{D_4}_{(1,1), (0,1)}$};
	\node [style=none] (7) at (0, 4) {$\mathcal{T}^{D_4}_{(0, 1), (1, 0)}$};
	\node [style=none] (8) at (-10, -4) {$\mathcal{T}^{D_4}_{(1, 1), (1, 0)}$};
	\node [style=none] (9) at (0, -4) {$\mathcal{T}^{D_4}_{(0, 1), (1, 1)}$};
	\node [style=none] (10) at (10, -4) {$\mathcal{T}^{D_4}_{(1, 0), (0,1)}$};
	\node [style=none] (11) at (10, 4) {$\mathcal{T}^{D_4}_{(1, 0), (1,1)}$};
    \node [style=none] (12) at (-10.5, 0) {{\color{red} $\tau$}};
    \node [style=none] (13) at (-0.5, 1) {{\color{red} $\tau$}};
    \node [style=none] (14) at (9.5, 0) {{\color{red} $\tau$}};
    \node [style=none] (15) at (-6.5, 2) {{\color{magenta} $\sigma$}};
    \node [style=none] (16) at (7.5, -1) {{\color{magenta} $\sigma$}};
    \node [style=none] (17) at (6, 3) {{\color{magenta} $\sigma$}};





	\draw [style=red arrows] (6) to (8);
	\draw [style=red arrows] (9) to (7);
	\draw [style=red arrows] (10) to (11);
	\draw [style=magenta arrows] (6) to (9);
	\draw [style=magenta arrows] (7) to (10);
	\draw [style=magenta arrows] (8) to (11);

    \node [style=none] (x1) at (12, 2) {};
    \node [style=none] (x2) at (18, 2) {};
    \draw [style=red arrows] (x1) to (x2);
    
    \node [style=none] (y1) at (12, -1) {};
    \node [style=none] (y2) at (18, -1) {};
    \draw [style=magenta arrows] (y1) to (y2);  

    \node [style=none,text=red] (x3) at (15, 1) {stacking SPT};
    \node [style=none,text=magenta] (y3) at (15, -2) {gauging $\bbZ_2^{(2)}$};
 
\end{tikzpicture}\\
    \begin{tikzpicture}[scale= .4]
	\node [style=none] (1) at (-10, -4){$\mathcal{T}^{D_4}_{(1,1), (1,0)}$};
	\node [style=none] (2) at (0, -4) {$\mathcal{T}^{D_4}_{(0, 1), (1, 1)}$};
 	\node [style=none] (3) at (10, -4) {$\mathcal{T}^{D_4}_{(1, 0), (0,1)}$};
	\node [style=none] (4) at (-10, 4) {$\mathcal{T}^{D_4}_{(1, 1), (0,1)}$};
	\node [style=none] (5) at (0, 4) {$\mathcal{T}^{D_4}_{(0, 1), (1, 0)}$};
	\node [style=none] (6) at (10, 4) {$\mathcal{T}^{D_4}_{(1, 0), (1, 1)}$};

    \draw [style=black arrow] (1) to (2);
    \draw [style=black arrow] (2) to (3);
    \draw [style=black arrow] (3) [bend left] to (1);
    \draw [style=black arrow] (4) to (5);
    \draw [style=black arrow] (5) to (6);
    \draw [style=black arrow] (6) [bend right] to (4);

    \draw [style=blue arrows] (1) to (4);
    \draw [style=blue arrows] (2) to (6);
    \draw [style=blue arrows] (3) to (5);

    \node [style=none] (7) at (12, 2) {};
    \node [style=none] (8) at (18, 2) {};
    \draw [style=blue arrows] (7) to (8);
    
    \node [style=none] (9) at (12, -1) {};
    \node [style=none] (10) at (18, -1) {};
    \draw [style=black arrow] (9) to (10);  

    \node [style=none,text=blue] (11) at (15, 1) {an order 2 GS};
    \node [style=none] (11) at (15, -2) {an order 3 GS};
    
\end{tikzpicture}
    \end{tabular}
\caption{The topological manipulations and Green--Schwarz automorphisms/dualities of the relative 6d $(2,0)$ SCFT of type $D_4$. TOP: gauging 2-form symmetries in magenta, and stacking a counterterm in red. BOTTOM: an order $2$ Green--Schwarz duality fixing $L_{SO} = \langle (1,1) \rangle$ while exchanging $L_{Ss} = \langle (1,0) \rangle$ with $L_{Sc} = \langle (0,1) \rangle$ is represented by blue arrows, and an order $3$ Green--Schwarz duality cyclically permuting $(L_{SO} = \langle(1,1)\rangle, L_{Sc} = \langle(0,1)\rangle, L_{Ss}=\langle(1,0)\rangle)$ is represented by black arrows.  After combining these two sets of topological manipulations into a single diagram, any closed loop should be interpreted as a duality defect, which is non-invertible if an odd number of $\sigma$ operations are involved.}
\label{fig:D4}
\end{figure}
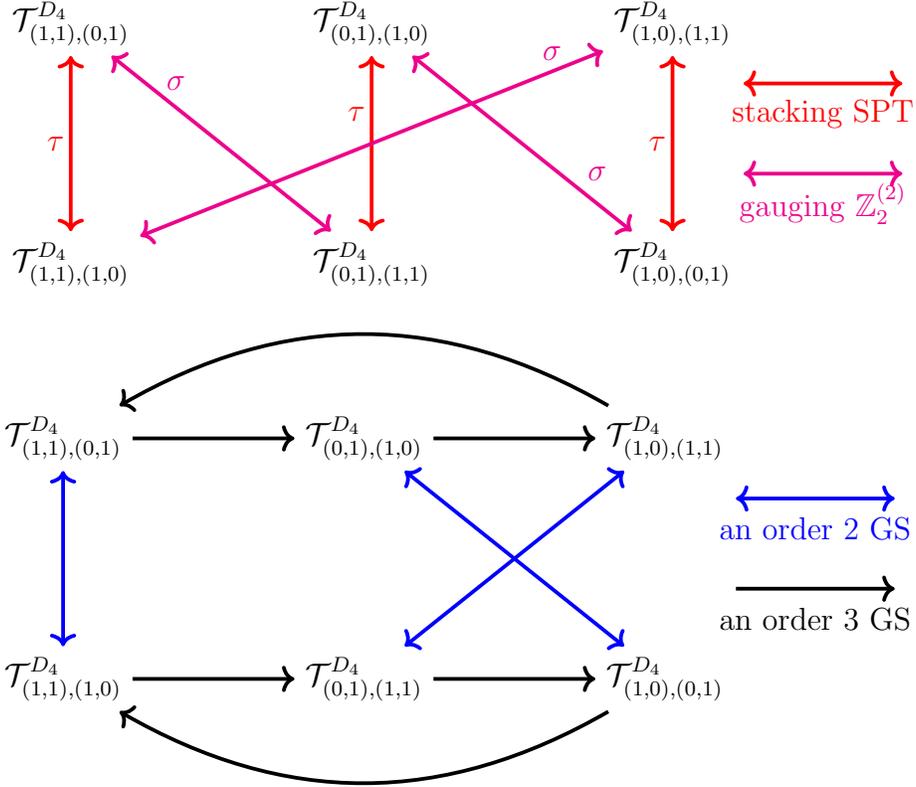

The first example we consider is the $D_4$ $(2,0)$ SCFT. The descirption of its tensor branch effective field theory is:
\begin{equation}
    \begin{array}{ccc}
      & 2 &  \\
    2 & 2 & 2 \,,
    \end{array}
\end{equation}
which translates into a Dirac pairing matrix which is the $\kso(8)$ Cartan matrix:
\begin{equation}
    K_{IJ} = \left(\begin{array}{cccc}
        -2 & 1 & 0 & 0 \\
         1 & -2& 1 & 1 \\
         0 & 1 &-2 & 0 \\
         0 & 1 & 0 &-2 
    \end{array} \right) \,.
\end{equation}

\paragraph{Polarization Pairs for the $D_4$ $(2,0)$ SCFT} The defect group of this theory is $\mathbb{D} = \mathbb{Z}_2 \oplus \mathbb{Z}_2$ with a quadratic form given by:
\begin{equation}\label{eq:quadratic form for D4}
    q = \left( \begin{array}{cc} 0 & \frac{1}{2} \\ \frac{1}{2} & 0 \end{array}\right) \,.
\end{equation}
Recall that for a Lagrangian subgroup, any pair of elements need to have integer bilinear pairing, or equivalently, any element should have a half-integer value of the quadratic form. Therefore, one has three possible choices of polarizations $L \subset \mathbb{D}$; we introduce the following compact notation to match with the conventions in the literature:
\begin{equation}
   \ell_{Ss} = (1,0), \ \ \ell_{Sc} = (0,1),\ \ \ell_{SO} = (1,1) \,.
\end{equation}

We remark that since all these subgroups are order $2$, the charge conjugations are all trivial and one also gets the full list of absolute theories by staying at the level of Lagrangian subgroups $L, \ovL$, which is the notation that was used in \cite{Gukov:2020btk}. As emphasized earlier, in more general cases then explicitly specifying the generators, as opposed to just the $(L, \ovl)$ is vital.

Conventionally, these three choices of polarizations are called the $Ss(8), Sc(8)$, and $SO(8)$ theories, with the understanding that these Lie groups no longer label the character lattice of gauge symmetry charges, but rather label the character lattice of string charges.\footnote{The character lattice of Lie group $\Lambda_{ch}$ is an intermediate lattice between the root lattice $\Lambda_{r}$ and the weight lattice $\Lambda_{w}$: $\Lambda_{r} \subset \Lambda_{ch} \subset \Lambda_{w}$. It captures the global form of the Lie group.}

Moreover, as we have explained in Section \ref{sec:nonInvGeneral}, the data of SPT phases on top of a 6D SCFT can be labeled by choosing the second element $\ovl$ in the polarization pair.\footnote{In this context, the SPT phase is fermionic and is given by the Arf-–Kervaire invariant. See, e.g.,\cite{Hsin:2021qiy}.} When the $\ell \subset \{\ell_{Ss},\ell_{Sc},\ell_{SO}\}$ has been chosen, $\ovl$ can be one of the remaining two subgroups. Therefore, after incorporating SPT data, we actually have six possible combinations:
\begin{equation}\label{eq:counter terms in D4}
    \calT^{D_4}_{(\ell_{Ss}, \ovl_{Sc})}, \ \ \calT^{D_4}_{(\ell_{Ss}, \ovl_{SO})}, \ \ \calT^{D_4}_{(\ell_{Sc}, \ovl_{Ss})}, \ \ \calT^{D_4}_{(\ell_{Sc}, \ovl_{SO})}, \ \ \calT^{D_4}_{(\ell_{SO}, \ovl_{Ss})}, \ \ \calT^{D_4}_{(\ell_{SO}, \ovl_{Sc})} \,,
\end{equation}
or when written explicitly in terms of polarization pairs:
\begin{equation}\label{eq:polar pairs in D4}
    \calT^{D_4}_{(1,0), (0,1)}, \ \ \calT^{D_4}_{(1,0), (1,1)}, \ \ \calT^{D_4}_{(0,1), (1,0)}, \ \ \calT^{D_4}_{(0,1), (1,1)}, \ \ \calT^{D_4}_{(1,1), (1,0)}, \ \ \calT^{D_4}_{(1,1), (0,1)} \,.
\end{equation}
See Figure \ref{fig:D4} for all allowed topological actions among these theories. Gauging amounts to flipping $\ell$ and $\ovl$, which we denote in green arrows.

The $S_3\cong \mathbb{Z}_2\ltimes \mathbb{Z}_3$ Green--Schwarz duality can be generated by an order $3$ element which cyclically permutes $(\ell_{SO},\ell_{Sc},\ell_{Ss})$ and any choice of an order $2$ element. Here we make a choice so that the latter switches $Ss$ and $SO$ but leaves $Sc$ invariant. 

\paragraph{Construction of Duality Defects} Next, we construct the duality defect: finding all possible closed chains of topological operations so that we go back to the same absolute theory. We give two concrete examples of non-invertible duality defects:
\begin{itemize}
\item The simplest example can be given as follows. We gauge the symmetry in half the spacetime so that we switch $\ell$ with $\overline{\ell}$, and then perform a Green--Schwarz duality to switch back $\ell$ and $\overline{\ell}$. As can be checked explicitly, there is always one particular order $2$ element inside $S_3$ that exchanges $\ell, \overline{\ell}$ while leaving the third $\mathbb{Z}_2$ subgroup invariant.

In the language of interfaces, by stacking with a half-space gauging interface and a Green--Schwarz duality interface, we can construct a topological operator
\begin{equation}\label{eq:non-invertible of D4}
    \calN(M_{5}) = \sigma(M_5) \cdot \calI_{a(\ell, \ovl)}(M_{5}) \,,
\end{equation}
with $a(\ell, \overline{\ell}) \in S_3$ an order $2$ element, such that $\calN(M_{5})$ implements the duality transformation in an \textit{absolute} $D_4$ theory with polarization $(\ell, \overline{\ell})$. The relevant fusion rules are given by:
\begin{align}
      \mathcal{N}(M_5) \times \overline{\mathcal{N}}(M_5) &= \sum_{S \in H_3(M_6; \mathbb{Z}_2)} \calU(M_3), \ \ \mathcal{N}(M_5) \times \calU(M_3) = \mathcal{N}(M_5) \,, 
\end{align}
where $\calU(M_3)$ is the symmetry operator for the 2-form symmetry.\footnote{From a string theory construction of 6D SCFTs, a more general construction of the 3d topological symmetry operator is to wrap a D3 brane on boundary 1-cycle, as concretely constructed in \cite{Heckman:2022muc}. Namely, this operator itself is a non-invertible one such that $\calU(M_3) \calU(M_3)^\dagger \neq 1$, and the leading term in its worldvolume TFT reduces to an invertible 2-form symmetry operator.}
\item As a second example, one can start from the $\calT^{D_4}_{(\ell_{Sc}, \ovl_{SO})} = \calT^{D_4}_{(0,1), (1,1)}$ theory, stack a counterterm to get the $\calT^{D_4}_{(\ell_{Sc}, \ovl_{Ss})} = \calT^{D_4}_{(0,1), (1,0)}$ theory, then gauge the 2-form symmetry to get the $\calT^{D_4}_{(\ell_{Ss}, \ovl_{Sc})} = \calT^{D_4}_{(1,0), (0,1)}$ theory, and finally implement an order $3$ Green--Schwarz duality $a_3$ to recover the $\calT^{D_4}_{(\ell_{Sc}, \ovl_{SO})} = \calT^{D_4}_{(0,1), (1,1)}$ theory:
\be
    \calT^{D_4}_{(0,1), (1,1)} \overset{\tau}{\longrightarrow} \calT^{D_4}_{(0,1), (1,0)} \overset{\sigma}{\longrightarrow} \calT^{D_4}_{(1,0), (0,1)} \overset{a_3}{\longrightarrow} \calT^{D_4}_{(0,1), (1,1)} \,.
\ee
Stacking the interfaces of these three operations builds a order-3 non-invertible duality defect, which is also commonly referred to as a triality defect.
\end{itemize}
We have thus demonstrated that there are indeed various constructions of invertible duality defects in such 6D SCFTs, in which all the topological manipulations play a role. One could also use topological operations other than an odd number of half-space gaugings of the $2$-form symmetry to build invertible duality defects. 

\subsection{\texorpdfstring{$\mathcal{N}=(2,0)$ $A_4\oplus A_4$}{N=(2,0) A4+A4} Theory} \label{subsec:A4A4}

Having studied an irreducible theory of $D_4$ type, we now give another example of a reducible 6D $(2,0)$ relative theory (which gives irreducible absolute theories), where we also identify non-invertible duality defects.\footnote{Usually an irreducible (relative) SCFT means that the theory has only one stress-tensor; an irreducible (relative) theory then has multiple stress-tensors which corresponds to having decoupled local operator sectors. A reducible relative theory may lead to an irreducible absolute theory in the sense that the extended operator spectrum may non-trivially connects the a priori decoupled relative sectors.}

Concretely, we examine the 6D $(2,0)$ theory $A_4 \oplus A_4$ as a direct sum of two relative 6D SCFTs
\begin{equation}
     2 \ \ 2 \ \ 2 \ \ 2\ \ \oplus\ \ 2 \ \ 2 \ \ 2 \ \ 2 \,,
\end{equation}
whose Dirac pairing matrix is the direct sum of two $\mathfrak{su}(5)$ Cartan matrices. As a relative theory, this theory is a direct sum of two $A_4$ $(2,0)$ theories. 

To identify polarizations and topological boundary conditions, we need to write down the pairing. We denote a generic element as $(a, b) \in \bbD = \mathbb{Z}_5 \times \mathbb{Z}_5$, then the quadratic form $q((a, b))$ inherited from the bilinear pairing on the $A_4 \oplus A_4$ weight lattice reads:
\begin{equation}
    q((a, b)) = \frac{2}{5}a^2 + \frac{2}{5}b^2 \in \bbQ/\bbZ \,.
\end{equation}
In this way, the generator of either $\mathbb{Z}_5$ subgroup associated with either $A_4$ factor has a non-trivial $q((a,b))$, and therefore neither $\bbZ_5$ is an isotropic subgroup of $\mathbb{D}$. Nonetheless, there are two possible choices for Lagrangian subgroups given by:
\begin{equation}
    \langle (1, 2) \rangle \quad \text{or} \quad \langle (1,-2) \rangle \quad  \subset \bbD = \bbZ_5 \times \bbZ_5 \,,
\end{equation}
since the generators $(1, \pm 2)$ of either $\bbZ_5$ has
\begin{equation}
    q((1, \pm 2)) = \frac{2}{5} (1^2 + (\pm 2)^2) \equiv 0 \mod \mathbb{Z} \,.
\end{equation}

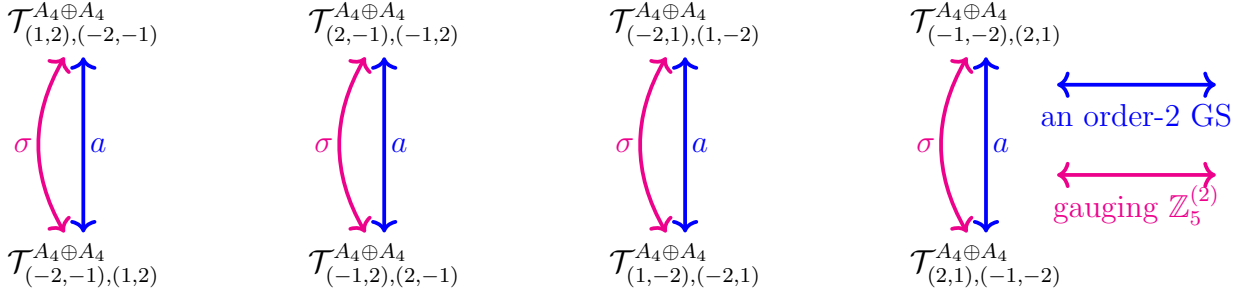
\begin{figure}
\centering
    \begin{tabular}{c}
    \begin{tikzpicture}[scale= .4]

 	\node [style=none] (4) at (-20, -4) {$\mathcal{T}^{A_4 \oplus A_4}_{(-2, -1), (1,2)}$};
	\node [style=none] (5) at (-20, 4) {$\mathcal{T}^{A_4 \oplus A_4}_{(1, 2), (-2, -1)}$};
	\node [style=none] (6) at (-10, 4) {$\mathcal{T}^{A_4 \oplus A_4}_{(2,-1), (-1, 2)}$};
	\node [style=none] (7) at (0, 4) {$\mathcal{T}^{A_4 \oplus A_4}_{(-2, 1), (1, -2)}$};
	\node [style=none] (8) at (-10, -4) {$\mathcal{T}^{A_4 \oplus A_4}_{(-1, 2), (2, -1)}$};
	\node [style=none] (9) at (0, -4) {$\mathcal{T}^{A_4 \oplus A_4}_{(1, -2), (-2, 1)}$};
	\node [style=none] (10) at (10, -4) {$\mathcal{T}^{A_4 \oplus A_4}_{(2, 1), (-1, -2)}$};
	\node [style=none] (11) at (10, 4) {$\mathcal{T}^{A_4 \oplus A_4}_{(-1, -2), (2, 1)}$};

    \node [style=none] (12-) at (-19.5, 0) {{\color{blue} $a$}};
    \node [style=none] (12) at (-9.5, 0) {{\color{blue} $a$}};
    \node [style=none] (13) at (0.5, 0) {{\color{blue} $a$}};
    \node [style=none] (14) at (10.5, 0) {{\color{blue} $a$}};
    \node [style=none] (14-) at (-22, 0) {{\color{magenta} $\sigma$}};
    \node [style=none] (15) at (-12, 0) {{\color{magenta} $\sigma$}};
    \node [style=none] (16) at (-2, 0) {{\color{magenta} $\sigma$}};
    \node [style=none] (17) at (8, 0) {{\color{magenta} $\sigma$}};





	\draw [style=blue arrows] (4) to (5);
        \draw [style=blue arrows] (6) to (8);
	\draw [style=blue arrows] (9) to (7);
	\draw [style=blue arrows] (10) to (11);
 	\draw [style=magenta arrows] (5) [bend right] to (4);
	\draw [style=magenta arrows] (6) [bend right] to (8);
	\draw [style=magenta arrows] (7) [bend right] to (9);
	\draw [style=magenta arrows] (11) [bend right] to (10);

    \node [style=none] (x1) at (12, 2) {};
    \node [style=none] (x2) at (18, 2) {};
    \draw [style=blue arrows] (x1) to (x2);
    
    \node [style=none] (y1) at (12, -1) {};
    \node [style=none] (y2) at (18, -1) {};
    \draw [style=magenta arrows] (y1) to (y2);  

    \node [style=none,text=blue] (x3) at (15, 1) {an order-2 GS};
    \node [style=none,text=magenta] (y3) at (15, -2) {gauging $\bbZ_5^{(2)}$};
 
\end{tikzpicture}\\
    \end{tabular}
\caption{All absolute theories descending from the $A_4 \oplus A_4$ $\calN = (2,0)$ relative theory. Here, each vertical pair of theories can be connected by gauging 2-form symmetry (in magenta arrows) and implementing Green--Schwarz duality (in blue arrows).}
\label{fig:2A4}
\end{figure}

Now, each of the $\bbZ_5$ subgroups has four non-trivial generators $n_1(1, 2)$ and $n_2(1,-2)$. By imposing the pairing condition $\langle \ell, \ovl\rangle  = 1$, we get that $n_1 n_2 \equiv 3 \mod 5$. By redefining the generators, we can get the full list of polarization pairs as in Figure \ref{fig:2A4}. There, each vertical pair of theories is connected by both gauging the $\mathbb{Z}_5$ 2-form symmetry, $\sigma$ and Green--Schwarz duality, $a$.

We notice that $\bbZ_5$ does not contain a $\bbZ_2$ factor, so we could not possibly form a non-trivial counterterm by a $\bbZ_5$-valued field. This is in perfect agreement with the fact that for the same $\ell$ we always have a unique choice of $\ovl$, following the general logic of Appendix \ref{app:absorb counterterm}. Therefore, the only topological manipulation that we have is gauging the 2-form symmetry. As we can see, gauging the 2-form symmetry will exchange the pair of theories in any individual column of \ref{fig:2A4}. On the other hand, the only GS automorphism which is a global symmetry exchanges the pair of $(2,0)$ theories, which therefore also exchanges the pair of $\bbZ_5$ generators in the polarization pair. Therefore, each operation of gauging 2-form symmetries is exchanged by the only order $2$ global symmetry element $a$ of the GS duality (out of the full Green--Schwarz automorphism which is $(S_5 \times S_5) \rtimes \mathbb{Z}_2$). Each absolute theory descending from the $A_4 \oplus A_4$ relative $(2,0)$ theory only admits one way of constructing the non-invertible duality defect, corresponding to the chain of topological manipulation
\be
    (\ell, \ovl) \overset{\sigma}{\longrightarrow} (\ovl, \ell) \xrightarrow{a} (\ell, \ovl) \,,
\ee
and thus the non-invertible duality defect
\begin{equation}
    U(M_5) = \sigma_{a}(M_5) \cdots \mathcal{I}_a(M_5) \,.
\end{equation}
By computing $U(M_5) U^\dagger(M_5)$, one can get a condensation operator of 2-form symmetry defects. In this way, one can see that $U(M_5)$ is indeed a non-invertible condensation defect.

For $A_{N-1} \times A_{N-1}$ theory with general $N$, the existence of polarization of the above type is discussed extensively in \cite{GarciaEtxebarria:2019caf,Gukov:2020btk}. If the background field of the two $\bbZ_N$ factors is given by $C_1, C_2$, then one type of topological boundary condition is imposed by the boundary term proportional to $\int_{M_6} r C_1 \cup C_2$, so that the boundary condition is given by:
\be
    C_1 = r C_2,\ \ C_2 = - r C_1 \,,
\ee
which together implies that $r^2 \equiv -1 \mod \bbZ_N$ (e.g., $r = \pm 2$ for $N = 5$, but $r$ does not exist for $N = 3$). Therefore, only for some $N$, there exists a pair of boundary conditions of the above diagonal type with $L = \langle (1, r_1) \rangle, \ovL = \langle (1, r_2) \rangle$ such that $r_2 = - r_1$. 

\subsection{\texorpdfstring{$\mathcal{N}=(1,0)$}{N=(1,0)} Case}

Now we examine the more general family of 6D SCFTs, with $\mathcal{N} = (1,0)$ supersymmetry. Such theories are constructed via F-theory on elliptically-fibered Calabi--Yau threefolds over a non-compact base \cite{Heckman:2013pva,Heckman:2015bfa}. Such theories generalize the $(2,0)$ family in two ways: the base configuration can be more general, and non-trivial degenerations of the elliptic fibers are allowed. The punchline is that $(1,0)$ theories mostly exhibit similar behavior as the $(2,0)$ theories, in terms of their possible polarization pairs and thus duality defects.

\paragraph{Bases for $(1,0)$ SCFTs} To begin with, we remark that all possible finite Abelian groups can be seen as the defect group of some (possibly reducible) 6D $(2,0)$ SCFT, so one cannot get new defect groups by considering $(1,0)$ theories. In addition, most of the time, even the quadratic pairing $q(\mu)$ for elements of a $(1,0)$ defect group $\mu \in \bbD$ coincides with that of certain $(2,0)$ theories. For example, an $A_{N-1}$ $(2,0)$ theory has a defect group $\mathbb{Z}_N$, from which one can already construct all finite simple Abelian groups. In addition, a $(1,0)$ SCFT with a $-N$ curve in the base has a quadratic form with spin $\frac{N-1}{2N}$ (so that the bilinear pairing has coefficient $-\frac{1}{N}$), which is isomorphic to the quadratic form on the $\bbZ_n$ center of an $A_{N-1}$ $(2,0)$ theory with quadratic form evaluated to $\frac{N-1}{2N}$ for $N$ odd. 
To exemplify this, we can consider the following $(1,0)$ theory consisting of two irreducible theories (where each tensor multiplet paired with a $\kf_4$ vector multiplet):
\be
    \overset{\mathfrak{f}_4}{5} \oplus \overset{\mathfrak{f}_4}{5} \,,
\ee
which has the following Dirac pairing matrix is:
\be
    \left( \begin{array}{cc} 5 & 0 \\ 0 & 5 \end{array} \right) \,.
\ee
The defect group is $\bbZ_{5} \oplus \bbZ_{5}$ with the associated quadratic form: $q((a, b)) = \frac{2}{5}(a^2 + b^2)$ for $(a, b) \in \bbD$. We notice that this is precisely the same as the defect group and quadratic form that appeared when we discussed the duality defects in the $A_4 \oplus A_4$ $(2,0)$ theory. Thus, this particular 6d $(1,0)$ SCFTs has a similar non-invertible symmetry structure to that depicted in Figure \ref{fig:2A4}.

\paragraph{Including Gauge Algebras} Another possibility is to pair the tensor multiplet with a vector multiplet, breaking the supersymmetry from $\calN = (2,0)$ down to $\calN = (1,0)$. Depending on the details of the gauge symmetry, the inclusion of gauge algebras may make the tensor multiplets to be no longer no equal footing, and thus reduce the admissible set of Green--Schwarz dualities down to those that preserve the structure of the gauge algebras. On the other hand, the data about 2-form symmetries, defect groups, polarization pairs (whenever applicable) are completely unaffected by this gauge algebra data.

For example, consider a $(1,0)$ theory where each tensor multiplet pairs with either a $\ksu(4)$ gauge theory or a $\ksu(8)$ gauge theory:
\begin{equation}
\begin{array}{cccc}
      &  & \overset{\mathfrak{su} (4)}{2}  &  \\
     {[SU(8)]} & \overset{\mathfrak{su} (8)}{2} & \overset{\mathfrak{su} (8)}{2} &  \overset{\mathfrak{su} (4)}{2} \,.
\end{array}
\end{equation}
Then the defect group and the bilinear pairing are identical to that of a $D_4$ $(2,0)$ theory. However, the Green--Schwarz automorphism is reduced from $S_3$ to $\mathbb{Z}_2$, the one exchanging the two external tensors paired with $\ksu(4)$ gauge algebras, but fixing the external tensor paired with the $\mathfrak{su}(8)$ gauge algebra. Then, inheriting from the $D_4$ $(2,0)$ theory analyzed above, in the following two polarizations (under a specific choice of basis of the $\bbD = \bbZ_2 \times \bbZ_2$):
\be
    \calT^{D_4}_{(\ell_{Ss}, \ovl_{Sc})} = \calT^{D_4}_{(1,0), (0,1)}, \quad \calT^{D_4}_{(\ell_{Sc}, \ovl_{Ss})} = \calT^{D_4}_{(0,1), (1,0)} \,,
\ee
there is still some remnant of the $\calN = (2,0)$ non-invertible duality defects associated with the preserved $\mathbb{Z}_2$ Green--Schwarz automorphism. In contrast, in the remaining four theories with either $\ell = \ell_{SO}$ or $\ovl = \ell_{SO}$, the non-invertible duality defect no longer exists.

We observe in passing that if we consider a Higgs branch renormalization group (RG) flow from this $(1,0)$ theory to the $D_4$ $(2,0)$ theory breaking all the paired gauge groups:
\be
\begin{array}{ccccc}
      &  & \overset{\mathfrak{su} (4)}{2}  &   \\
     {[SU(8)]} & \overset{\mathfrak{su} (8)}{2} & \overset{\mathfrak{su} (8)}{2} &  \overset{\mathfrak{su} (4)}{2}  \\
\end{array}
\xrightarrow{\text{Higgs Branch flow}}
\begin{array}{ccc}
      & 2 &  \\
    2 & 2 & 2 \,,
\end{array}
\ee
then the full group of $S_3$-valued Green--Schwarz dualities will get restored. Along such an RG flow, the original set of non-invertible duality defects for $(2,0)$ theories of type $D_4$ would appear in the infrared as an emergent global symmetry.\footnote{See \cite{Damia:2023ses} for analysis of non-invertible symmetries along RG flows in 4D. See also \cite{Elvang:2012st,Heckman:2015ola,Heckman:2016ssk,Heckman:2018pqx,Hassler:2019eso,Heckman:2021nwg,Baume:2021qho,Fazzi:2022yca,Distler:2022kjb} for some more studies on RG flows in 6D. We leave it for future work to understand the emergence of generalized symmetries along Higgs branch RG flows between 6d SCFTs.}

\section{Towards a String Theory Embedding} \label{sec:symTFTs}

In previous sections, we discuss polarization pairs and build non-invertible duality defects via them for QFTs in diverse dimensions. Many of our examples are QFTs which admit engineering via string theory, e.g., $\mathcal{N}=4$ SYM theories and 6D SCFTs. Furthermore, dualities of these theories usually admit top down interpretations as specific isometries of the extra-dimensional geometry.\footnote{String theory gives rise to geometric interpretations for both exact dualities and infrared dualities. Here we focus on exact dualities, e.g., S-duality of $\mathcal{N}=4$ SYM and Green--Schwarz duality of 6D SCFTs. For a geometric interpretation of infrared dualities in QFTs in diverse dimensions, see, e.g., \cite{Franco:2017lpa,Franco:2020ijt,Franco:2021vxq}.} Therefore, it is natural to ask for a stringy perspective for polarizations and non-invertible duality defects, especially in 6D SCFTs which largely arise via a string theory realization (see, e.g., \cite{Heckman:2018jxk} for a review). 

With this spirit, in this section, we set our stage at 6D $\mathcal{N}=(2,0)$ SCFTs, derive their 7D symmetry TFTs and discuss their polarizations and non-invertible duality defects from the Type IIB string theory point of view. The symmetry TFTs are computed from dimensional reduction of the Type IIB topological sector. This includes quadratic terms for the 2-form symmetry background fields of 6D SCFTs, which encode the information of the intermediate defect group and the Dirac pairing, as intensively studied in the previous section, as well as additional fields and interaction terms, encoding other symmetries and their mixed anomalies, coming from the cubic topological coupling in Type IIB supergravity. Based on this string theory construction, we discuss the branes behind charged and symmetry operators for the various global symmetries and take the first steps towards a geometric interpretation of non-invertible duality defects.\footnote{We refer the reader to \cite{Heckman:2022xgu} for a similar top-down discussion of the 4D $\mathcal{N}=4$ SYM theories.}

\subsection{Symmetry TFTs for \texorpdfstring{$\mathcal{N}=(2,0)$}{N=(2,0)} SCFTs from Type IIB}

We will now compute symmetry TFTs for 6D $\mathcal{N}=(2,0)$ SCFTs via Type IIB on an orbifold singularity $X=\mathbb{C}^2/\Gamma$, with $\Gamma$ is a finite subgroup of $SU(2)$. The computation is realized by reducing the topological sector of Type IIB string theory on the asymptotic boundary $\partial(\mathbb{C}^2/\Gamma)=S^3/\Gamma$ via treating the various Type IIB supergravity fluxes as elements in differential cohomology classes (see, e.g., \cite{Freed:2006ya,Freed:2006yc})
\begin{equation}\label{eq:cohomology class of s3orbifold}
    H^*(S^3/\Gamma)=\{ \mathbb{Z},0,\Gamma^\text{ab},\mathbb{Z} \} \,,
\end{equation}
where $\Gamma^\text{ab}$ is the Abelianization of $\Gamma$:
\begin{equation}
\Gamma^{ab}=\Gamma/[\Gamma,\Gamma] \,.
\end{equation}
The reduction leads to a 7D theory known as the symmetry TFT, as explained in \cite{Apruzzi:2021nmk,Heckman:2017uxe}. 

\paragraph{Quadratic terms} We have seen in previous sections that 7D quadratic CS terms play an essential role in characterizing 2-form symmetries as well as non-invertible symmetries of 6D SCFTs. A natural question is whether these terms can be reproduced from the top down. This is indeed the case. The self-dual RR 4-form gauge field $C_4$ and its field strength $F_5$ in Type IIB string theory should be regarded as the boundary mode of a 11D Chern--Simons theory via the anomaly inflow construction \cite{GarciaEtxebarria:2019caf} (see also \cite{Bah:2020uev,Hsieh:2020jpj,GarciaEtxebarria:toappear}) 
\begin{equation}
    S^{\text{11D}}_{\text{CS}}=\frac{1}{2}\int_{N_8\times S^3/\Gamma}\breve{F}_6\star \breve{F}_6 \,,
\end{equation}
where $\breve{F}_6$ is the differential cohomology uplift of the Type IIB flux $F_5$, and $N_8$ satisfies $\partial N_8=M_6\times \mathbb{R}_{\ge}$, i.e., a bulk manifold whose boundary is the spacetime where the 7D symmetry TFT lives. We expand $\breve{F}_6$ via the cohomology classes of $S^3/\Gamma$
\begin{equation}
    \breve{F}_6=\breve{f}_6\star \breve{1}+\sum_i\breve{E}_4^{(i)}\star \breve{t}_{2(i)}+\breve{f}_3\star\breve{\text{v}}\text{ol} \,,
\end{equation}
where $\breve{1}$ and $\breve{\text{v}}\text{ol}$ correspond to the free part $\mathbb{Z}$ of cohomology classes, whereas $\breve{t}_{2(i)}$ correspond to the torsional part $\Gamma^\text{ab}$ and $i$ runs over its generators. Note that the asymptotic boundary $S^3/\Gamma$ has formally infinite volume, so any fields arising as coefficients of $\breve{\text{v}}\text{ol}$ are non-dynamical \cite{Heckman:2022muc, Heckman:2022xgu}. Thus, $\breve{f}_3=0$. The self-dual property of the D3-brane flux then implies that $\breve{f}_6=0$.

Integrating over $S^3/\Gamma$, we end up with the TFT terms 
\begin{equation}\label{eq:7D BF term from IIB}
\begin{split}
S^{\text{7D}}_{\text{quadratic}}&=\frac{1}{2}\sum_{i,j}\int_{S^3/\Gamma}\breve{t}_{2(i)}\star\breve{t}_{2(j)}\int_{N_8}  \breve{E}_4^{(i)}\star \breve{E}_4^{(j)} \\
&\equiv \int_{M_6\times \mathbb{R}_{\ge 0}}\frac{1}{2}\sum_{i,j}\mathcal{C}_{ij}\breve{E}_3^{(i)}\star \delta \breve{E}_3^{(j)},
\end{split}
\end{equation}
where at the second step we denote the integral over $S^3/\Gamma$ as 
\begin{equation}\label{eq:coefficient as linking number}
    \mathcal{C}_{ij}\equiv \int_{S^3/\Gamma}\breve{t}_{2(i)}\star\breve{t}_{2(j)},
\end{equation}
which is determined by linking numbers between torsional cycles of $S^3/\Gamma$ \cite{GarciaEtxebarria:2019caf}. We also assume $H^4(M_6\times \mathbb{R}_{\ge 0})=0$ so that $\breve{E}_4^{(i)}$ on $N_8$ is  trivialized to $\breve{E}_3^{(i)}$ on $M_6 \times \bbR_{\geq 0}$. The resulting action exactly consists of quadratic terms for the 2-form symmetry background fields 
\begin{equation}\label{eq:fields for 2-form discrete symmetries}
    \breve{E}_3^{(i)}\leftrightarrow G^{(2)}_{(i)},
\end{equation}
which encodes the information of the intermediate defect group $\mathbb{D}^{(2)}=\Gamma^\text{ab}$ and its pairing for discrete 2-form symmetries, which we have intensively studied in previous sections.

\paragraph{Cubic terms}
In addition to the 11D Chern--Simons term, Type IIB supergravity itself includes a cubic topological term
\begin{equation}\label{eq:IIB Topological term}
    -\int_{M_6\times \mathbb{C}^2/\Gamma}C_4\wedge dB_2\wedge dC_2 \rightarrow -\int_{M_6\times \mathbb{C}^2/\Gamma}\breve{F}_5\star \breve{H}_3 \star \breve{G}_3,
\end{equation}
where the Type IIB fluxes are again promoted to differential cohomology elements. Based on (\ref{eq:cohomology class of s3orbifold}), we expand fluxes as 
\begin{equation}
\begin{split}
    &\breve{F}_5=\breve{f}_5\star \breve{1}+\sum_i\breve{E}_3^{(i)}\star \breve{t}_{2(i)}+\breve{f}_2\star \breve{\text{v}}\text{ol},\\
    &\breve{H}_3=\breve{h}_3\star \breve{1}+\sum_{i}\breve{B}_1^{(i)}\star \breve{t}_{2(i)}+\breve{h}_0\star \breve{\text{v}}\text{ol},\\
    &\breve{G}_3=\breve{g}_3\star \breve{1}+\sum_{i}\breve{C}_1^{(i)}\star \breve{t}_{2(i)}+\breve{g}_0\star \breve{\text{v}}\text{ol}.
\end{split}
\end{equation}
Again we treat fields associated with cycles with infinite volumes as non-dynamical, thus
$\breve{f}_5=\breve{f}_2=\breve{h}_0=\breve{g}_0=0.$

The surviving coefficients of $\breve{1}$ correspond to field strengths of background fields for continuous $U(1)$ 1-form symmetries:
\begin{equation}\label{eq:fields for continuous symmetries}
 \breve{h}_3\leftrightarrow U(1)^{(1)}_b, ~\breve{g}_3\leftrightarrow U(1)^{(1)}_c.
\end{equation}
The coefficients of the torsional generator $\breve{t}_{2(i)}$ are background gauge fields for discrete symmetries:
\begin{equation}\label{eq:fields for discrete symmetries}
    \breve{E}_3^{(i)}\leftrightarrow G^{(2)}_{(i)}, ~\breve{B}_1^{(i)}\leftrightarrow G^{(0)}_{(i), b}, ~\breve{C}_1^{(i)}\leftrightarrow G^{(0)}_{(i), c},
\end{equation}
where $\breve{E}_3^{(i)}\leftrightarrow G^{(2)}_{(i)}$ is exactly the 2-form symmetry background fields produced in equation \eqref{eq:fields for 2-form discrete symmetries}. $G^{(p)}_{(i)}$ denotes $p$-form discrete symmetry group corresponding to the $i$-th generator of $\Gamma^{ab}$, and in particular, $G^{(0)}_{(i), b}, G^{(0)}_{(i), c}$ are two different sets of 0-form symmetries associated with the same set of cycles but generated via different brane fluxes.

The IIB topological term (\ref{eq:IIB Topological term}) are now expanded as 
\begin{equation}
\begin{split}
    &-\int_{M_6\times \mathbb{C}^2/\Gamma}\breve{F}_5\star \breve{H}_3 \star \breve{G}_3\\
    &=-\sum_{i,j}\int_{S^3/\Gamma}\breve{t}_{2(i)}\star\breve{t}_{2(j)}\int_{M_6\times \mathbb{R}\ge 0}\left( \breve{E}_3^{(i)}\star\breve{B}_1^{(j)}\star\breve{g}_3+\breve{E}_3^{(i)}\star\breve{h}_3\star\breve{C}_1^{(j)}\right)
\end{split}
\end{equation}
Performing the integral over the internal geometry $S^3/\Gamma$, we derive 
\begin{equation}\label{eq:7D symTFT from IIB}
\begin{split}
S^{\text{7D}}_{\text{cubic}}&=-\int_{M_6\times \mathbb{R}_{\ge 0}} \sum_{i,j}\mathcal{C}_{ij}\left( \breve{E}_3^{(i)}\star\breve{B}_1^{(j)}\star\breve{g}_3+\breve{E}_3^{(i)}\star\breve{h}_3\star\breve{C}_1^{(j)}\right). 
\end{split}
\end{equation}
It is easy to see each term encodes a mixed anomaly between the discrete 2-form symmetry $G^{(2)}_{(i)}$, a discrete 0-form symmetry (either $G^{(0)}_{(i), b}$ or $G^{(0)}_{(i), c}$) and a $U(1)$ 1-form symmetry.

Combining quadratic terms in (\ref{eq:7D BF term from IIB}) and cubic terms in (\ref{eq:7D symTFT from IIB}), we can now write down the full symmetry TFT for 6D $\mathcal{N}=(2,0)$ theories as 
\begin{equation}\label{eq:full (2,0) TFT}
\begin{split}
    S_{\text{symTFT}}=\int_{M_6\times \mathbb{R}_{\ge 0}}\bigg\{ \frac{1}{2}\sum_{i,j}\mathcal{C}_{ij}E_3^{(i)}\cup \delta E_3^{(j)}-\sum_{i,j}\mathcal{C}_{ij}\left( E_3^{(i)}\cup B_1^{(j)}\cup g_3+E_3^{(i)}\cup h_3\cup C_1^{(j)} \right) \bigg\},
\end{split}
\end{equation}
where all fields are now expressed as elements in ordinary cohomology. The correspondence between fields and global symmetries has already been given in (\ref{eq:fields for continuous symmetries}) and (\ref{eq:fields for discrete symmetries}). 

Before discussing the string theory objects underlying charged operators and symmetry operators, we would like to remark that  $U(1)$ 1-form symmetries, denoted by field strengths $g_3$ and $h_3$, are not coupled to 6D SCFTs. In fact, it is shown in \cite{Cordova:2020tij} that there is no continuous 1-form symmetry in any 6D SCFT because non-trivial conserved 2-form current cannot exist in any unitary superconformal multiplet. From the symmetry TFT computation above, it is easy to see that $g_3$ and $h_3$ exist even when reducing the IIB topological term on $S^3$, while IIB string theory on $\mathbb{C}^2$ does not lead to 6D SCFTs. Therefore, these two classes of $U(1)$ 1-form symmetries (namely $U(1)^{(1)}_b$ and $U(1)^{(1)}_c$) do not belong to 6D SCFTs but rather are included in decoupled sectors.

\subsection{Branes for Symmetry Operators and Polarization Pairs}

Let us now discuss the brane construction for charged and symmetry operators of the discrete 2-form symmetries $G^{(2)}_{(i)}$ and two classes of 0-form symmetries $G^{(0)}_{(i), b}$, $G^{(0)}_{(i), c}$.  According to \cite{Heckman:2022muc} (see also \cite{Heckman:2022xgu, GarciaEtxebarria:2022vzq,Apruzzi:2022rei}), the charged heavy defects are built via branes along $\mathbb{R}_{\ge 0}$ direction, while the symmetry operators are built via branes wrapping cycles ``at infinity", which in the current context reads in Table \ref{tab:branes behind symmetries}. 

\begin{table}[H]
    \centering
    \begin{tabular}{|c|c|c|c|}
\hline
Fields & Global symmetries & Charged operators & Symmetry operators \\
     \hline
      $E_3^{(i)}$ & $G^{(2)}_{(i)}$ & D3-branes wrapping $\mathbb{R}_{\ge 0}\times \gamma_{1(i)}$&D3-branes wrapping $\tilde{\gamma}^{(i)}_1$\\
      \hline
      $B_1^{(i)}$ & $G^{(0)}_{(i), b}$ & F1-string wrapping $\mathbb{R}_{\ge 0} \times \gamma_{1(i)}$ & NS5-branes wrapping $\tilde{\gamma}_1^{(i)}$\\
      \hline
      $C_1^{(i)}$ & $G^{(0)}_{(i), c}$ & D1-string wrapping $\mathbb{R}_{\ge 0} \times \gamma_{1(i)}$ & D5-branes wrapping $\tilde{\gamma}_1^{(i)}$\\
      \hline
\end{tabular}
    \caption{Fields in the 7D symmetry TFT and their corresponding global symmetries in 6D SCFTs. The charged and symmetry operators composed of various types of branes are also presented. $\gamma_{1(i)}$ and $\tilde{\gamma}^{1(i)}$ denote torsional 1-cycles linking to each other in $S^3/\Gamma$.}
    \label{tab:branes behind symmetries}
\end{table}

Instead of dealing with generic $S^3/\Gamma$, we focus on the simple example of $D_4$ theory discussed in Section \ref{sec:6DSCFTs} to illustrate our idea of geometric realization for non-invertible duality defects. The defect group for the relative QFT is given by 
\begin{equation}
    \mathbb{D}\equiv \Gamma^{ab}=\bbZ_2^{(x)}\times \bbZ_2^{(y)}
\end{equation}
where $x$ and $y$ are used to distinguish the two Lagrangian subgroups of $\mathbb{D}$. The coefficient $\mathcal{C}_{ij}$ in the symmetry TFT is given by the linking pair of $S^3/D_4$,
\begin{equation}
    \frac{1}{2}\begin{pmatrix} 
    2&1\\
    1&2
    \end{pmatrix}
\end{equation}
from which we derive the quadratic CS term as
\begin{equation}
S_{\text{quadratic}}=\int_{M_6\times \mathbb{R}_{\ge 0}} \frac{1}{2}E_3^{(x)}\cup \delta E_3^{(y)}.
\end{equation}
This is exactly the 7D $\mathbb{Z}_2$ gauge theory matching the defect group and its Dirac pairing given by the quadratic form in (\ref{eq:quadratic form for D4}).

\paragraph{Polarizations pairs from branes patterns ``at infinity."} Recall that absolute theories corresponds to Lagrangian subgroups of the defect group, and for relative theories with split defect group one further needs to specify a polarization pair. From the symmetry TFT perspective, such algebraic data encodes the topological boundary conditions. This statement is now nicely inherited from the string theory as the non-commutativity of RR fluxes, which give rise to different wrapping patterns of D3-branes on topolocial cycles ``at infinity". 

For example, pick the polarization pair 
\begin{equation}
    \calT^{D_4}_{(\ell_{Sc}, \ovl_{Ss})}=\calT^{D_4}_{(0,1),(1,0)},
\end{equation}
which is introduced in (\ref{eq:counter terms in D4}) and (\ref{eq:polar pairs in D4}). This corresponds to choosing the 
Lagrangian subgroup $L=\mathbb{Z}_2^{(y)}$, which is now a gauge symmetry. The associated boundary condition is Neumann for $E_3^{(y)}$ and the Dirichlet boundary condition for $E_3^{(x)}$, without counterterm, in the symmetry TFT. From the string theory perspective, this polarization pair is realized as an asymptotic boundary condition for RR fluxes ``at infinity" (i.e., infinitely far away from the singularity where the 6D SCFT is engineered), which translates in the following certain patterns for D3-branes wrapping torsional 1-cycles:
\begin{equation}
    \begin{split}
        \text{D3-branes wrapping}~\gamma_{1(y)}~\text{``at infinity" fluctuating}&\longleftrightarrow\text{symmetry operators generating}~\mathbb{Z}_2^{(x)}\\
        \text{D3-branes wrapping}~\gamma_{1(x)}~\text{terminating ``at infinity"} &\longleftrightarrow \text{charged operators under}~\mathbb{Z}_2^{(x)}
    \end{split}
\end{equation}

\subsection{Comments on a Geometric Interpretation of Duality Defects} 

We briefly comment on the geometric interpretation of non-invertible duality defects. Before starting, we remark that a similar idea has been illustrated in \cite{Heckman:2022xgu} in the context of 4D SCFTs.

For QFTs built from string theory via geometric engineering, discrete global symmetries, including duality symmetries, are inherited from discrete isometries of the internal geometry. In the context of 6D SCFTs, as we mentioned in Section \ref{sec:6DSCFTs}, dualities symmetries come from Green--Schwarz automorphisms, which are inherited from discrete isometries of the complex 2-dimensional F-theory base space. Back to our current stage on $\mathcal{N}=(2,0)$ SCFTs from IIB string theory, this reduces to the discrete isometries of the orbifold $\mathbb{C}^2/\Gamma$.

\begin{figure}
    \centering
    \includegraphics[width=5cm]{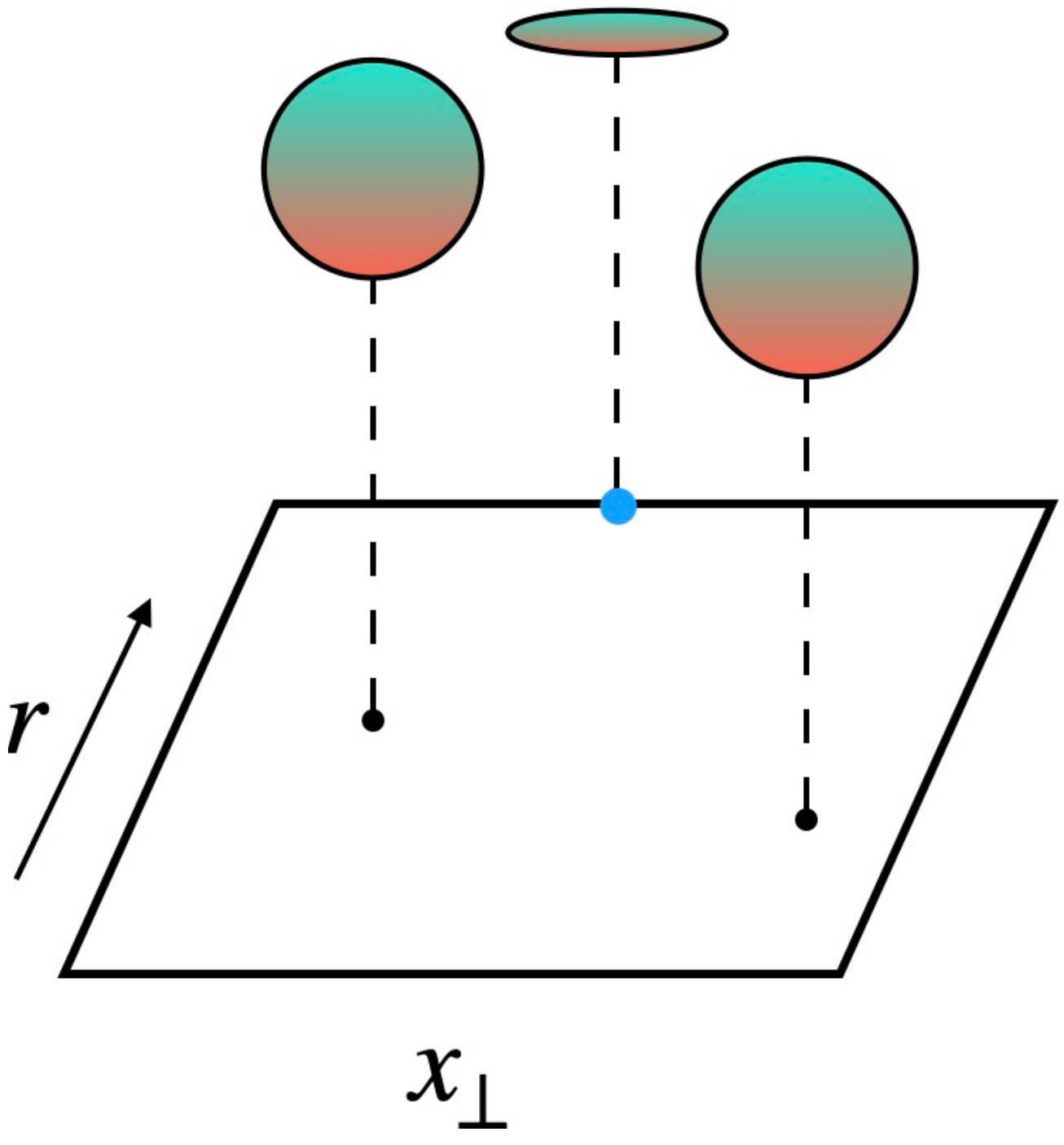}
    \caption{$S^3/\Gamma$ fibered over the 2-dimensional base space $(x_\perp,r)$. We use the colorful ball to denote the 3-dimensional fiber and a squashed ball to denote the degeneration of some cycles. The fiber degenerates at $x_\perp=0,r=\infty$, whose topological effect on 6D SCFTs engineered at $r=0$ is introducing the non-invertible duality defect.}
    \label{fig:lensfiber}
\end{figure}

In order to build non-invertible duality symmetries, instead of treating $M_6\times \mathbb{C}^2/\Gamma$ as a direct product, we consider the asymptotic boundary $S^3/\Gamma$ non-trivially fibered over a complex 2-dimensional base. The base space is composed by the transverse direction $x_\perp$ of the 5D duality defect in $M_6$ and the radial direction $\mathbb{R}_{\ge 0}$ of $\mathbb{C}^2/\Gamma$, whose coordinate can then be written as $(x_\perp, r)$. Similarly to the spirit of building topological operators via branes ``at infinity", we consider \emph{fiber degeneration ``at infinity"}. Namely, the $S^3/\Gamma$ fiber degenerates at $x_\perp=0, r=\infty$, so that it realizes a non-trivial transition from $x_\perp<0$ to $x_\perp>0$ at $r=\infty$. More concretely, we require this transition to be inherited from the discrete isometry transforming different torsional 1-cycles of $S^3/\Gamma$. Since this fiber degeneration is an asymptotic profile that is infinitely far away from ADE singularities, it only has topological effects on the 6D SCFT localized at $r=0$ as we want. See Figure \ref{fig:lensfiber} for a rough depiction. 

Let us revisit our $D_4$ example to illustrate our idea further. As before, we denote torsional 1-cycles corresponding to the two generators of the $\mathbb{Z}_2^{(x)}\times \mathbb{Z}_2^{(y)}$ defect group as $\gamma_{1(x)}$ and $\gamma_{1(y)}$. Choosing the polarization pair $\calT^{D_4}_{(\ell_{Sc}, \ovl_{Ss})}=\calT^{D_4}_{(0,1),(1,0)}$, then according to our previous discussion, $\gamma_{1(x)}$ is the cycle wrapping which D3-branes are allowed to end ``at infinity." Now introduce $S^3/D_4$ fiber degeneration at $x_\perp=0, r=\infty$ so that there is a non-trivial transition from $x_\perp < 0$ to $x_\perp >0$ based on the isometry exchanging $\gamma_{1(x)}$ and $\gamma_{1(y)}$. This gives rise to the non-invertible duality symmetry of $\calT^{D_4}_{(\ell_{Sc}, \ovl_{Ss})}$ theory as shown in (\ref{eq:non-invertible of D4}) in Section \ref{sec:6DSCFTs}. 

\begin{figure}
    \centering
    \includegraphics[width=14cm]{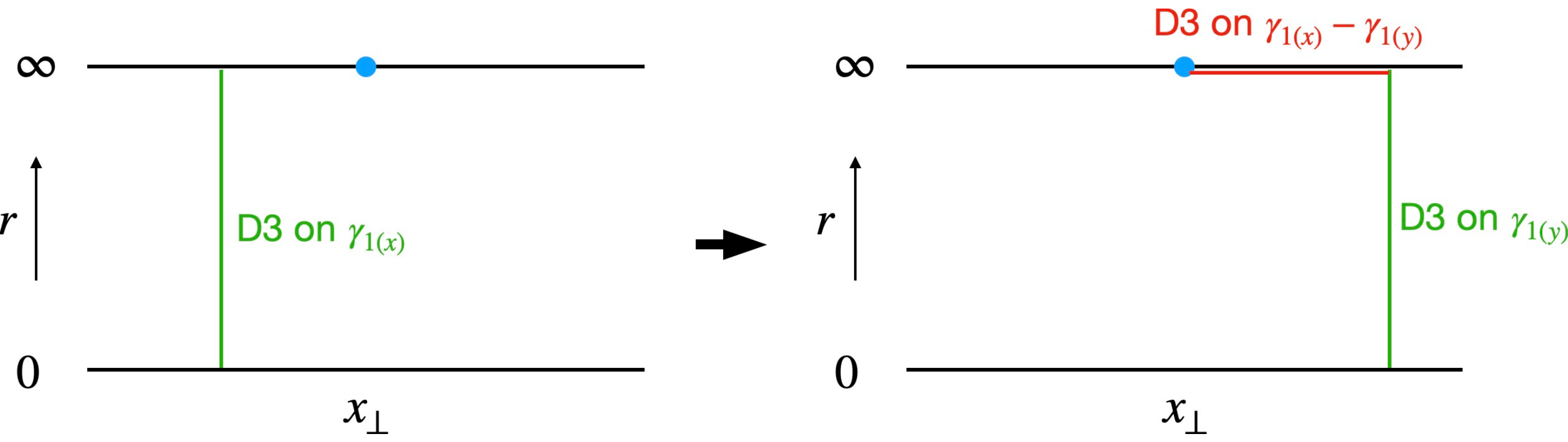}
    \caption{The transition of cycles wrapped by D3-branes gives rise to the action of non-invertible duality defects on 2-form symmetry charged operators. The blue point denotes the position where the $S^3/\Gamma$ fiber degenerates. The red line denotes D3-branes wrapping 1-cycles ``at infinity", which gives rise to symmetry operators $\eta$ aligned with Figure \ref{fig:6D transition}.}
    \label{fig:D3transition}
\end{figure}

\begin{figure}
    \centering
    \includegraphics[width=12.493cm]{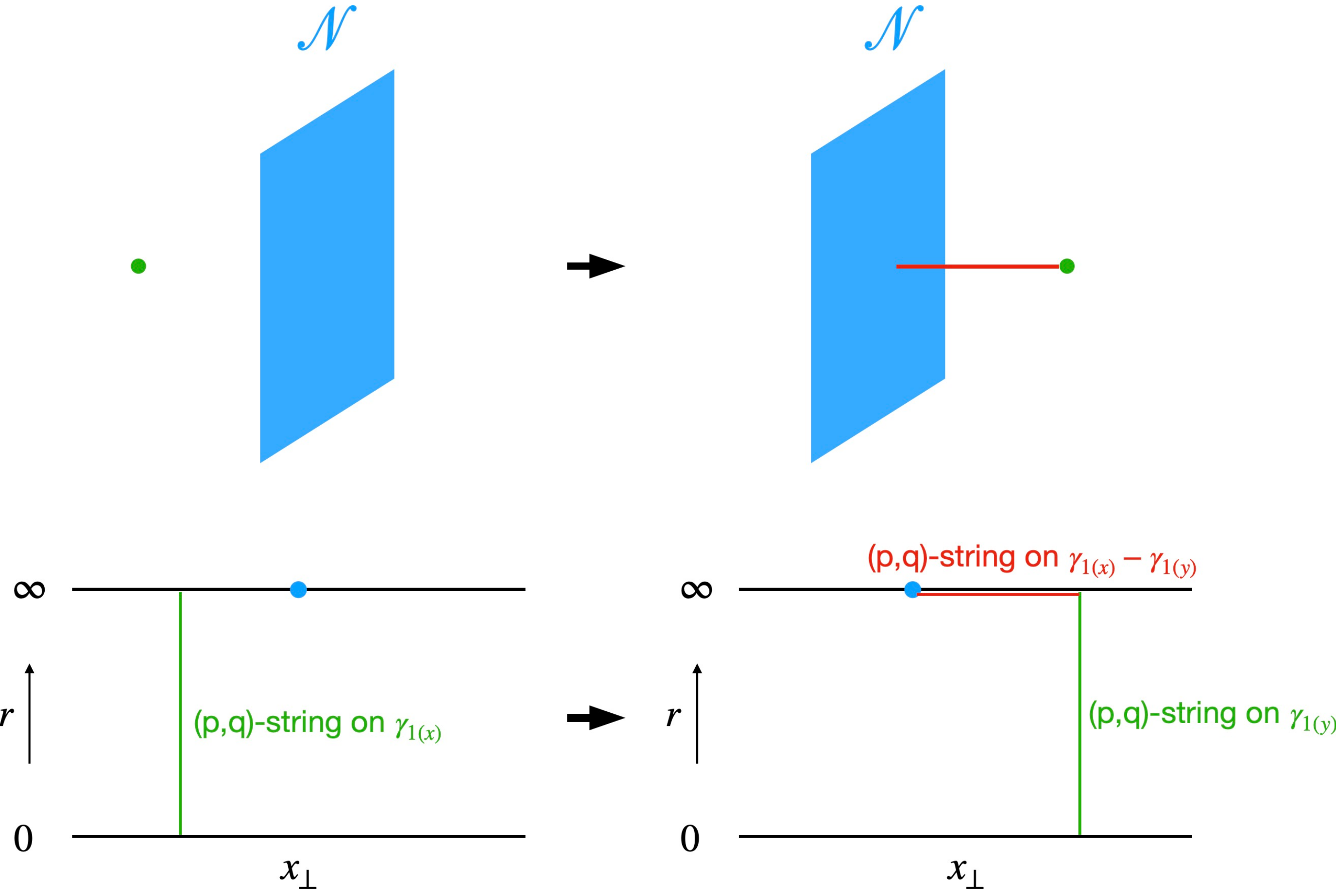}
    \caption{Local operators charged under $\mathbf{G}_{(i), b}$ and $\mathbf{G}_{(i), c}$ 0-form symmetries also admit non-trivial action by the non-invertible duality defect $\mathcal{N}$. The string theory origin is the transition of cycles wrapped by (p,q)-strings, associated with a certain combination of $\mathbf{G}_{(i), b}$ and $\mathbf{G}_{(i), c}$ symmetries.}
    \label{fig:stringtransition}
\end{figure}

The action of the non-invertible duality defect on the 2-form symmetry charged operators, discussed in Figure \ref{fig:6D transition}, is nicely encoded in the non-trivial transition of D3-branes crossing through the fiber degeneration, as shown in Figure \ref{fig:D3transition}.\footnote{We thank Inaki Garcia Etxebarria and Max Hubner for valuable discussions on this point.} This is very similar to the Hanany-Witten transition \cite{Hanany:1996ie} interpretation for non-invertible symmetries in 4D \cite{Apruzzi:2022rei,Heckman:2022xgu}. \footnote{This geometric transition can be thought of as sharing the same origin with the Hanany-Witten transition via various dualities in string theory. See, e.g., \cite{Kimura:2016yqa} for related discussions.} 
Interestingly, since this transition is based on exchanging torsional 1-cycles, F1- and D1-strings wrapping on cones over $\gamma_{1(i)}$ also transforms under it. This gives rise to a non-trivial action of the non-invertible duality defect on the charged operators under 0-form symmetries $\mathbf{B}_{(i)}$ and $\mathbf{C}_{(i)}$. See Figure \ref{fig:stringtransition} for a schematic depiction. This phenomenon of non-invertible defect acting on both 0-form and higher-form charged operators has been observed in 4D SCFTs from the string theory point of view in \cite{Heckman:2022xgu}.

Our discussion of the geometric realization of non-invertible duality defects is at a qualitative level. It would be interesting to have an algebraic description of the fiber degeneration ``at infinity" setup, similar to the holomorphic Weierstrass model considered in \cite{Heckman:2022xgu}, which we leave for future work. 

\section{Discussion} \label{sec:conclusion}

In this paper, we studied $2k$ dimensional QFTs, involving self-dual gauge fields, whose $(k-1)$-form symmetry can be gauged using the formalism of relative and absolute QFTs. Building upon previous studies of polarizations such as \cite{Gukov:2020btk}, we established \textit{polarization pairs} as an algebraic characterization of polarizations and the SPT phase data of such a $2k$ dimensional QFT by directly relating a polarization pair to the partition function of such a QFT. We explained how the operations of gauging $(k-1)$-form symmetries, stacking counterterms, and implementing charge lattice isomorphism are all incorporated into the polarization pair. We then combined these operations to give general constructions of duality defects in such even-dimensional QFTs. To demonstrate the generality of our formalism, we first revisited the well-studied case of 4D $\calN = 4$ SYM. Here, we show that the construction of non-invertible defects in \cite{Kaidi:2022uux} can be reformulated by polarization pairs in full generality. 

We then moved on to the novel case of 6D QFTs. There, by focusing on examples of 6D SCFTs, we are again able to identify non-invertible duality defects by combining the gauging of 2-form symmetries, stacking counterterms, and implementing discrete automorphisms of the charge lattice (known as Green--Schwarz automorphisms \cite{Apruzzi:2017iqe}). We gave some examples of non-invertible duality defects in 6D $(2,0)$ and $(1,0)$ SCFTs. Finally, we discussed the symmetry TFT for 6D SCFTs from Type IIB string theory on ADE orbifolds. In this way, we not only reproduced the 7D Chern--Simons terms relevant for the intermediate defect group and the non-invertible symmetries but also observed background fields for global symmetries which have non-trivial interplay with the non-invertible duality defects.

Our work suggests various natural directions for future investigation, especially in higher-dimensional theories where non-invertible symmetries have not been discussed overmuch in the literature. Below we comment on three possible directions.

\paragraph{Generic 4D Gauge Theory} In this paper, when we have considered explicit examples of 4D theories which realize non-invertible duality defects, we have focused on $\mathcal{N}=4$ super-Yang--Mills. For a 4D QFT with self-dual gauge fields, the analysis of the polarizarion pair put forth in this paper generically describes interfaces corresponding to gauging the one-form symmetry and stacking counterterms. Then, all that remains to be understood before being able to state the different types of non-invertible symmetries that exist in a given theory is the automorphism group of the charge lattice and whether this action extends to a full duality of the (relative) QFT. For $\mathcal{N}=4$ SYM this occurred only for certain values of the complexified coupling $\tau$. 

There are a variety of other classes of 4D gauge theories where the construction, whether top-down or bottom-up, provides a mechanism to study the possible dualities. Class $\mathcal{S}$ theories are obtained from compactifications of 6d $(2,0)$ SCFTs on punctured Riemann surfaces -- the duality action is determined from the Riemann surface. Non-invertible symmetries of class $\mathcal{S}$ theories have been studied from this geometric perspective in \cite{Bashmakov:2022jtl,Bashmakov:2022uek,Antinucci:2022cdi}. A recent exploration of a class of non-Lagrangian $\mathcal{N}=1$ and $\mathcal{N}=2$ SCFTs has been constructed via gaugings of Argyres--Douglas theories in \cite{Kang:2021lic,Kang:2021ccs,Kang:2022zsl,Kang:2022vab,Kang:2023dsa}. These theories often have identical central charges, and in many ways evince similar behavior to $\mathcal{N}=4$ SYM. Following from the techniques in this paper, the demonstration of non-invertible symmetries is evident once one has shown that the automorphism of the charge lattice extends to a duality of the local operator spectrum. It would be interesting to study whether the similarly to $\mathcal{N}=4$ SYM persists to this non-invertible sector. Some work in this regard has appeared in \cite{Carta:2023bqn}.

\paragraph{6D Little String Theories} In 6D, there is a class of non-gravitational yet non-local quantum theories called Little String Theories (LSTs) that were first identified in \cite{Seiberg:1997zk} and later given a geometric classification in \cite{Bhardwaj:2015oru} (see \cite{Aharony:1999ks} for an early review). Their geometric constructions almost come hand-in-hand with 6D SCFTs: instead of constructing 6D SCFTs putting F-theory on a local elliptically fibered Calabi--Yau threefold whose base $B$ contains a collection of complex curves with \textit{negative definite} intersection matrix $A_{ij} = \Sigma_i \cdot \Sigma_j < 0$, for LSTs, we only need to change one condition to $A_{ij} \leq 0$, namely the intersection pairing is negative semi-definite. For an irreducible LST, $A_{ij}$ will only have a single zero eigenvalue, whose eigenstate in the 2-form charge lattice encodes the charge of the little string that sets a mass scale for the LST.

One might wonder whether it is straightforward to construct non-invertible duality defects in 6D LSTs. The answer is, unfortunately, not. According to our formulation based on relative QFTs, the major obstacle is that the conception of relative versus absolute LSTs has not yet been established - since, in the first place, they are not QFTs at all. This conceptual obstacle comes hand-in-hand with a technical one - since the defect group of an LST can be given as $\bbD = \bbZ \oplus \bbD_{SCFT}$ \cite{DelZotto:2020sop}, where the pairing is trivial on $\bbZ$. Therefore, the whole notion of polarization (which builds upon a non-degenerate pairing) can only be safely discussed if we restrict our attention to $\bbD_{SCFT}$,  thus forcing us to ignore the little string physics and step back into the SCFTs discussed in this paper. 

\paragraph{8D QFTs} Following the path of increasing dimension, the next natural place to look for non-invertible symmetries is in 8D QFTs. Recently, 8D QFTs have been studied, in view of 8D compactifiations of string theory, in \cite{Cvetic:2020kuw,Font:2020rsk,Cvetic:2021sxm,Font:2021uyw,Cvetic:2021sjm,Cvetic:2022uuu,Garcia-Etxebarria:2017crf,Lee:2022spd}, however, the focus is typically on the defect group $\bbD \supset \bbD_1^{(e)} \oplus \bbD_5^{(m)}$ associated with the electric 1-form and magnetic 5-form symmetries. In contrast, self-dual 3-form symmetries, associated with the intermediate defect group discussed in this paper, have not received much attention. It would be interesting to write down specific examples of 8D QFTs that exhibit non-invertible duality defects, and to understand whether and in what manner such theories can be obtained from string theory.

\section*{Acknowledgements}
We thank J.~J.~Heckman for comments on a draft version of this paper. We further thank I.~Garcia~Etxebarria, J.~J.~Heckman, M.~H\"{u}bner, H.-T.~Lam, D.~Pei, M.~Porrati, and Y.~Wang for helpful discussions. XY and HYZ also thank J.~J.~Heckman, M.~H\"{u}bner, and E.~Torres for their collaboration on related projects. XY also thanks H.-T.~Lam and Y.~Wang for inspiring discussions in the early phase of the project during the Porratifest Symposium at NYU. CL and HYZ thank the Geometry of (S)QFT workshop at the Simons Center of Geometry and Physics for hospitality in the early phase of the project. XY thanks the UPenn Theory Group and Tsinghua String Theory Group for their hospitality during part of this work. CL acknowledges support from DESY (Hamburg, Germany), a member of the Helmholtz Association HGF. XY acknowledges support from the James Arthur Graduate Associate fellowship. The work of HYZ is supported by the Simons Foundation Collaboration grant $\#724069$ on ``Special Holonomy in Geometry, Analysis and Physics''.

\appendix

\section{\texorpdfstring{$(k-1)$}{(k-1)}-form Defect Group in \texorpdfstring{$2k$}{2k}-dimensional QFT} \label{apdx:relative}

In this appendix, we review the relevant background material on relative QFTs in even spacetime dimensions necessary for the construction of the polarization pair in Section \ref{sec:polarizationPair}. We emphasize the key role of the self-dual $(k-1)$-form gauge fields, and the related $(k-1)$-form discrete global symmetries together with their anomalies and possible gauging. 

\subsection{Dirac Pairing}\label{subapdx:Dirac}

In the $2k$ dimensional QFT, we consider the dynamical state of $k-1$ spatial dimensions so that its worldvolume can be coupled to a $k-1$ dimensional gauge field. The gauge field of this $(k-1)$-form is self-dual so that one can write down dual gauge field $A^{dual}_{k-1}$ of the same form degree and impose the Hodge duality condition:
\begin{equation}
    dA = \pm * dA^{dual}.
\end{equation}
Under this Dirac pairing, we can define the \textit{charge lattice} $\Lambda$ of dynamical states with spatial dimension $k-2$. This charge lattice is mathematically a $\mathbb{Z}$-module equipped with a $\mathbb{Z}$-valued bilinear pairing:
\begin{equation}
     \langle \cdot, \cdot\rangle: \Lambda \times \Lambda \rightarrow \mathbb{Z}
\end{equation}
Due to the exchanging property of differential forms, this pairing is symmetric for $k$ odd and antisymmetric for $k$ even. 

As an example, in four dimensions ($k=2$), we consider the anti-self-dual Dirac pairing $\langle (e, m), (e', m') \rangle:= e m' - e' m$ on the electromagnetic charge lattice. There the self-pairing of any single state is trivial, so the notion of electric-magnetic duality is necessary so that there is a magnetic state which pairs trivially with any given electric state. 

If we instead consider six dimensions ($k=3$), there is a self-dual Dirac pairing of string charges under various 2-form gauge fields. These strings are called ``dyonic strings", which generically have non-trivial self-pairings.\footnote{An example is the dynamical strings in 6D SCFTs, also elaborated in Section \ref{sec:6DSCFTs}. These strings are coupled to the anti-self-dual field comes from the tensor multiplet:
\begin{equation}
     (d A^{dual})_{3} = \pm * (dA)_{3}.
\end{equation}}

\subsection{Heavy Defects and the Defect Group} \label{subapdx:defectGroup}

In addition to the charge lattice of dynamical states $\Lambda$, there are also \textit{heavy defects} in the spectrum of extended objects of the QFT that carries charges valued in a more refined version of $\Lambda$. Such a ($\mathbb{Q}$-)refinement is given by $\Lambda^* \supset \Lambda$ and $\Lambda^* \subset \mathbb{Q} \otimes \Lambda$, which is defined as the set of element in $\Lambda \otimes \mathbb{Q}$ that have integer pairings with all elements in $\Lambda$ under $\langle \cdot, \cdot \rangle$. This definition is indeed justified by noticing that all defects must have integer Dirac pairing with the physical states in order for the QFT to admit a consistent quantization.

Such heavy defects can be physically viewed as some external heavy probes of a theory. Concepts of heavy defects carrying fractional charges in comparison to dynamical objects are well-known in 4D as the ``$N$-ality" of the fundamental Wilson lines of 4D $\mathfrak{su}(N)$ Yang-Mills theory, which can be seen as the worldline of a heavy probe quark in the fundamental representation. In 6D SCFTs engineered from string theory, such defects have also been carefully studied from the top-down perspective in \cite{DelZotto:2015isa}. 

Following 't Hooft screening arguments \cite{THOOFT19781,THOOFT1979141}, the defect group of a $2k$-dimensional QFT with respect to the $k-1$ dimensional charged objects can be defined by modding the group of heavy defects by the group dynamical defects:
\begin{equation}
    \mathbb{D} = \Lambda^*/\Lambda,
\end{equation}
which is a non-trivial finite Abelian group for generic $\Lambda$. A special type of charge lattice $\Lambda$ having trivial $\mathbb{D}$ (hence $\Lambda = \Lambda^*$) is called \textit{unimodular} lattices, which are also called self-dual lattices under its a bilinear pairing.

\subsection{\texorpdfstring{$(k-1)$}{k-1}-form Symmetries and Their Anomalies} \label{subapdx:k-1Anomaly}

After having understood the key role of defect group and its Lagrangian subgroup(s) in controlling the collection of physically allowed operators in an absolute quantum field theory,
we now systematically formulate our understanding of the $(k-1)$-form symmetries together with their counter-term data. 

Picking a Lagrangian subgroup $L\subset \mathbb{D}$ for a relative theory would give an absolute theory whose $(k-1)$-form symmetry is valued in $\mathbb{D}/L$. However, this $(k-1)$-form symmetry could still suffer from a 't Hooft anomaly associated with the 2-form symmetry $L^\vee$ precisely when the following short exact sequence does not split:
\begin{equation}
        1 \rightarrow L \rightarrow \mathbb{D} \rightarrow \mathbb{D}/L = L^\vee \rightarrow 1,
\end{equation}
namely, whenever a direct sum decomposition $\mathbb{D} = L \oplus \overline{L}$ (with $\overline{L} \cong L^\vee$ is an uplift of $L^\vee$ back into $\mathbb{D}$) is \textit{impossible}. Such a theory with a 't Hooft anomaly of the 2-form symmetry is called a \textit{projective} theory in \cite{Gukov:2020btk}, which is a specific type of absolute theory. Physically, such projective theory lives at the boundary of bulk $2k+1$-dimensional Chern-Simons theories. 

Returning to the anomaly-free case when the direct sum decomposition $\mathbb{D} = L \oplus \overline{L}$ is possible, we remark that such a decomposition might not be unique. Different direct sum decompositions correspond to different choices of local counterterms of this theory, and any pair of such resulting theories only differ by a Symmetry-Protected Topological (SPT) phase (also known as ``stacking a counterterm"). 

Therefore, we propose to rephrase the usual statement of 
\begin{equation}
\textit{a certain }\mathbb{D}/L\textit{ theory with a specific choice of SPT} \nonumber
\end{equation}
as 
\begin{equation}
\textit{a theory with defect group }\mathbb{D}\textit{ and a pair }(l, \overline{l})\textit{ generating Lagrangian subgroups }(\ell, \ovl), \nonumber
\end{equation}
with $\overline{L}$ a particular choice of \textit{uplift} of $L^\vee= \mathbb{D}/L$ to $\mathbb{D}$. Said differently, the information of an SPT choice is fixed via the choice of $\overline{L}$. A nice feature of this phrasing is that all choices of $\overline{L}$ would be placed on equal footing, so we no longer need to make the decision of which choice $\overline{L}$ corresponds to ``a theory without a counterterm". 

\subsection{Symmetry TFT for Polarization Pairs for non-cyclic \texorpdfstring{$(\ell, \ovl)$}{(L, barL)}} \label{subapdx:symTFTnoncyclic}

In this part, we elaborate on the definition of polarization pairs for QFTs with split defect groups in general, without imposing the restriction of $L \cong \ovL \cong \bbZ_N$ as done in the main text.

Consider the defect group $\mathbb{D}$ with $n$ generators $\mathbb{D}_I, I=1,\cdots m$, each of which we associate with an $k$-form gauge field $b_I$ valued in $\mathbb{D}_I$. The symmetry TFT then has an action in the generic form 
\begin{equation}
    S_{\text{symTFT}}[b_1,b_2,\cdots, b_m]=\frac{1}{2}\int_{M_{2k+1}}b_IQ_{IJ}\delta b_J
\end{equation}
where the matrix $Q_{IJ}$ is inherited from the quadratic pairing $q(\mu)$, defined around (\ref{eq:quadratic pairing on defect group}), on the defect group $\mathbb{D}$.

We then specify to split defect groups with $\bbD = L \oplus \ovL$, so $m = 2n$ that it has two identical sets of generators $b_i^x, b_i^y$ where $i \in 1, \dots, n$. In addition, $L, \ovL$ both begin Lagrangian subgroups of $\bbD$ means that the $xx$ components and $yy$ components of $Q_{IJ}$ vanish identically so we have,
\be
Q_{IJ} = \left(
\begin{array}{cc}
0 & Q^{xy}_{ij} \\
Q^{yx}_{ij} & 0
\end{array}\right)
\ee
and only $xy$ component remains. And the symmetry TFT now reads:
\be
    S_{\text{symTFT}}[b_1^x,b_2^x,\cdots, b_n^x, b_1^y, b_2^y, \dots, b_n^y] = \int_{M_{2k+1}} b_i^x Q^{xy}_{ij} \delta b_j^y, \ \ i, j = 1, \dots, n
\ee

The $2k+1$-dimensional symmetry TFT always has a dynamical boundary, where the $2k$-dimensional relative QFT lives. The existence of a Lagrangian subgroup $L\subset \mathbb{D}$ translates into the existence of a topological (gapped) boundary of the symmetry TFT, on which there is a set of well-defined topological boundary conditions. This includes Dirichlet boundary conditions (where $J$ labels a set of $2n$ generators, and $i$ labels $n$ linear combinations $\ovl_i$ of these generators that satisfy Dirichlet boundary conditions): 
\begin{equation}
   \delta(d_i(b_J)-(\ovl_i)_J D_J)|_{M_{2k}}=0,\ \ i = 1, \dots, n, \ J = 1, \dots, 2n,
\end{equation}
where $d_i(b_J),\ i = 1, \dots, m$ are $m$ linear combination of $\bbD$-valued fields that has Dirichlet boundary condition (hence the letter $d$), whose boundary value are given by $\ovl_J D_J$ (which can also be written as $\ovl^x_i D^x_i + \ovl^y_i D^y_i$) Here, for each individual $i$, we recover the cyclic story as in the main text.

We define the polarization pair for the general case as:
\be
    (\{\ell_1, \dots, \ell_n\}; \{\ovl_1, \dots, \ovl_n\}).
\ee
With this polarization pair, the boundary states can be denoted as 
\be
|\{\ell_1, \dots, \ell_n\}, \{\ovl_1, \dots, \ovl_n\}, B_1, \dots, B_n \rangle.
\ee
As our notation suggests, the orders within $\ell_1, \dots, \ell_n$ and the order with in $\ovl_1, \dots, \ovl_n$ are both irrelevant, since they are $n$ Dirichlet boundary conditions implemented by $\delta$ functions that commutes with each other.

One could also concatenate the general topological boundary with a dynamical boundary $|\calR\rangle$ to get the partition, whose description thus goes completely in parallel as the topological boundary states.

We proceed with giving some general remarks:
\begin{itemize}
    \item For the generic situation discussed in this appendix, there are many ways of implementing gauging of $(k-1)$-form symmetry and stacking counterterms, one associated with each cyclic subgroup of $\ovL$. Formulating the story in full details introduces more technical complications, but the main idea still follows from our main text.
    \begin{itemize}
        \item One could consider gauging not only gauging the cyclic subgroup associated with a single generator, but also gauging a ``diagonal" cyclic subgroup. Then, one should first implement a linear transformation as above such that this diagonal generator is an basis in the new set of basis, and then switch it with a ``dual" generator out of $\{\ovl_i\}$.
    \end{itemize}
    \item The permutation within $\{\ovl_1, \dots, \ovl_n\}$ is irrelevant, so is adding one element to another: $\ovl_j \rightarrow \ovl_j + \ovl_i, \ (i \neq j)$ while keeping any other $\ovl$s fixed. However, multiplying the generator of a single $\ovl$ is still brings us to another absolute theory via implementing an operation analogous to charge conjugation. 
\end{itemize}

\paragraph{4D $\calN = 4$ $\kso(8)$ Theory} To conclude this appendix, we give some discussions of how polarization pair describes all global structure of the 4D $\calN = 4$ $\kso(8)$ SYM theory, which is also covered in \cite{Kaidi:2022uux}.

Here, $\bbD = \bbZ_2^4$ with the following pairing \footnote{This pairing holds for $\kso(8k)$ in general, while it will take a different form for $\kso(8k+4)$, see e.g., \cite{GarciaEtxebarria:2019caf}.}:
\be
    \left(\begin{array}{cccc}
    0 & 0 & 0 & \frac{1}{2} \\ 0 & 0 & \frac{1}{2} & 0 \\ 0 & \frac{1}{2} & 0 & 0 \\ \frac{1}{2} & 0 & 0 & 0
    \end{array}\right),
\ee
such that $L \cong \ovL \cong \bbZ_2 \oplus \bbZ_2$. Any pair of non-trivial elements inside $\bbZ_2^4$ would give a $\bbZ_2^2$ subgroup, so modulo repetition there are ${\tiny \left( \begin{array}{cc} 15 \\ 2 \end{array}\right)}/3 = 35$ such $\bbZ_2^2$ subgroups of $\bbZ_4^2$. However, none of them are isomorphic. To look for isomorphic $\bbZ_2^2 \subset \bbZ_2^4$, we only need to examine the mutial pairing of the two generators $\mathbf{a}, \mathbf{b}$, which reads $(a_1 b_4 + a_2 b_3 + a_3 b_2 + a_4 b_1)/2$. One can check that for a fixed $a$, there are 8 out of 16 $b$ that will make this pairing integral. But after excluding $b = 0$ and $b = a$, this number becomes 6 out of 14. So the total number of Lagrangian subgroups of this $\bbZ_2^4$ becomes$\left( \begin{array}{cc} 15 \\ 2 \end{array}\right) \times \frac{1}{3} \times \frac{6}{14} = 15$.

Now $\ovL$ can be any one of these 15 Lagrangian subgroups. Our case is simple in that once we fix $L$, all possibilities of $\{\ell_1, \ell_2\}$ modulo linear relations are completely fixed, since all elements in $L$ have order-2. The last step is to also pick an $\ovL$ such that $\bbD = L \oplus \ovL$, which is equivalent to the statement that $L \cap \ovL = 0$. For any given $L$, it turns out that 8 out of the remaining $15 - 1$ Lagrangian subgroups can be chosen to be $\ovL$ that satisfy this direct sum condition, so there are $15 \times 8 = 120$ polarizations of the 4D $\calN = 4\ \kso(8)$ SYM theory. Reproducing the correct collection of SYM theories is a nice confirmation and illustration that a polarization pair is indeed capable of completely describing polarization for theories with $\bbD = L \oplus \ovL$ in general, even though the defect group is not of the form $\bbZ_N \times \bbZ_N$.

Now, associated with any particular $(\{\ell_1, \ell_2\}, \{\ovl_1, \ovl_2\})$, there is a $\bbZ_2 \times \bbZ_2$-valued 1-form symmetry which has three $\bbZ_2$ subgroups. So the generic possibility of stacking counterterms can be expressed as $\int_{M_4} \calP(B_2^{(1)}) +  \calP(B_2^{(2)}) +  \calP(B_2^{(1)} + B_2^{(2)})$ which, nonetheless, can still be accounted for by shifting $\{\ovl_1, \ovl_2\}$ via $\{\ell_1, \ell_2\}$. Similarly, gauging any $\bbZ_2$ subgroups of this $\bbZ_2 \times \bbZ_2$ can be accounted for by flipping a single pair of $\ell, \ovl$. We leave the full detail as an exercise for the interested reader.

\section{Changing \texorpdfstring{$\ovl$}{lbar} to Absorb Quadratic Counterterms} \label{app:absorb counterterm}
As we have seen in the main text and in Appendix \ref{subapdx:k-1Anomaly}, a nice feature of the polarization pair is that all choices of $\overline{L}$ are placed on equal footing, so we no longer need to make the decision of which choice $\overline{L}$ corresponds to ``a theory without a counterterm". In this Appendix, we further justify this statement from the symmetry TFT perspective.

We have formally defined the boundary state $|\ell, \ovl, B \rangle$ in a way that depends on the choice of $(\ell, \ovl)$ in the main text. The dependence of this basis on $\ovl$ can be examined via the consequence of basis change for a theory with, e.g., $\bbD = L \oplus \ovL = \bbZ_N \times \bbZ_N$:
\be
    (\ell, \ovl) \rightarrow (\ell, \ovl + \ell)
\ee
such that $\ovL' = \langle \ovl + \ell \rangle$. Now the new set of clock-shift operators are are
\be
    \{\Phi(A), \Phi(B')\}, \quad A \in H^k(M_{2k}, L),\ B' \in H^k(M_{2k}, \ovL').
\ee
Now the creation operators $\Phi(B') = \Phi(A + B)$ that we use to construct the entire set of basis vectors $\{|\ell, \ovl, B+A \rangle \}$ are different from $\{|\ell, \ovl, B \rangle \}$ which is constructed via $\Phi(B)$, so it is natural to expect differences between these two sets of basis. 

We note that the two set of basis still share the same clock operator $\Phi(A)$. Namely, for fixed $B = B_0$, $|\ell, \ovl, B_0+A_0 \rangle$ and $|\ell, \ovl, B_0 \rangle$ are eigenvectors in the same 1d eigenspace of $\Phi(A)$, so they should only differ by a phase. To then determine this phase, we need the following formula of $\Phi(A+B)$ \footnote{Such a choice of quadratic refinement has the effect of redefining the zero point of the background field, as illustrated in 6D in Section 4.4 of \cite{Gukov:2020btk}. We leave a more detailed exploration of such a possibility and its more physical implications in future work.}
\be
    \Phi(A+B) = \exp\left(-\frac{2\pi i}{2N} \int_{M_{2k}} A \cup B\right) \Phi(A)\Phi(B) = \exp\left(\frac{2\pi i}{2N} \int_{M_{2k}} A \cup B\right) \Phi(B)\Phi(A)
\ee
Namely, $\Phi(A+B)$ is neither $\Phi(A)\Phi(B)$ nor $\Phi(B)\Phi(A)$, but a phase-averaged version of them. With this formula, we can act $\Phi(A+B)$ on to $|\ell, \ovl', B'=0 \rangle$ to get (where in the second line, we used the fact that when $B' = 0$, the boundary state does not depend on $\ovl$):
\begin{align}
\begin{split}
|\ell, \ovl', B'=A_0+B_0 \rangle &= \Phi(A_0+B_0) |\ell, \ovl', B'=0 \rangle \\
& = \exp\left(\frac{2\pi i}{2N} \int_{M_{2k}} A_0 \cup B_0\right) \Phi(B_0)\Phi(A_0) |\ell, \ovl, B=0 \rangle \\
&= \exp\left(\frac{2\pi i}{2N} \int_{M_{2k}} A_0 \cup B_0\right) \Phi(B_0) |\ell, \ovl, B=0 \rangle \\
&= \exp\left(\frac{2\pi i}{2N} \int_{M_{2k}} A_0 \cup B_0\right) |\ell, \ovl, B=B_0 \rangle
\end{split}
\end{align}
By doing the above procedure $k$ times we can see the quadratic nature of the phase factor:
\begin{align}
\begin{split}
|\ell, \ovl', B'=k(A_0+B_0) \rangle = \exp\left(\frac{2\pi i}{2N} k^2 \int_{M_{2k}} A_0 \cup B_0\right) |\ell, \ovl, B=kB_0 \rangle
\end{split}
\end{align}
As we have seen, shifting $\ovl$ creates quadratic phase factors.

Exactly in the same fashion, we can also use the freedom of shifting $\ovl$ to cancel/absorb all quadratic phase factors. In particular, If we have a phase factor of any quadratic counterterm $\epsilon(B)$:
\be
    \exp\left(2\pi i \int_{M_{2k}} \epsilon(B)\right) |\ell, \ovl, B \rangle
\ee
then we can also absorb it by doing a shift of $\ovl \rightarrow \ovl + r\ell$ such that the following identity holds 
\be
    r \int_{M_{2k}} A_0 \cup B_0 = \int_{M_{2k}} \epsilon(B_0).
\ee
In this case, the counterterm is indeed absorbed by the following operation:
\be
    \exp\left(2\pi i \int_{M_{2k}} \epsilon(B)\right) |\ell, \ovl, B \rangle 
 = \rightarrow |\ell, \ovl + r\ell, B' \rangle.
\ee

\bibliographystyle{utphys}
\bibliography{Duality}

\end{document}